\documentclass[11pt]{article}

\usepackage{graphicx} 
\usepackage{booktabs} 
\usepackage{float}
\usepackage{epsfig}
\usepackage{cite}
\usepackage{amsmath, amssymb}
\usepackage[footnotesize]{caption}
\usepackage{slashed}
\usepackage{xcolor}
\usepackage[utf8]{inputenc}
\usepackage{tikz}
\usetikzlibrary{shapes.geometric}
\usetikzlibrary{arrows.meta,arrows}
\usetikzlibrary{quotes,angles}
\usepackage{multicol}
\usepackage{multirow}
\usepackage{array}
\usepackage[export]{adjustbox} 
\usepackage{subcaption}
\usepackage{hyperref} 
\usepackage{bm}
\usepackage[normalem]{ulem}
\usepackage{comment}

\numberwithin{equation}{section}

\newcommand{\be}{\begin{equation}}
\newcommand{\ee}{\end{equation}}
\newcommand{\beq}{\begin{eqnarray}}
\newcommand{\eeq}{\end{eqnarray}}

\begin{document}

\title{
\vspace*{-3.0cm}
\phantom{h} \hfill\mbox{\small }\vspace*{-1.1cm}
\vspace*{0.7cm}
\\[1cm]
\vspace{13mm}
\textbf{
$t \bar{t}$ production as a window to invisible new physics
\\[4mm]}}

\date{}
\author{
Rodrigo Capucha$^{1, 2\,}$\footnote{E-mail:
\texttt{rscapucha@fc.ul.pt}} ,
Jo\~ao Lopes$^{1, 2\,}$\footnote{E-mail:
\texttt{joaocarreira2000@hotmail.com}} ,
Jo\~ao Bravo Martins$^{1, 2\,}$\footnote{E-mail:
\texttt{joaobravomartins24@hotmail.com}} , \\
Ant\'{o}nio Onofre$^{3\,}$\footnote{E-mail:
\texttt{antonio.onofre@cern.ch}},
Rui Santos$^{1,2\,}$\footnote{E-mail:
  \texttt{rasantos@fc.ul.pt}} 
\\[5mm]
{\small\it
$^1$Departamento de F\'{\i}sica, Faculdade de Ci\^{e}ncias,} \\
{\small \it   Universidade de Lisboa, 1749-016 Lisboa, Portugal} \\[3mm]
{\small\it
$^2$Centro de F\'{\i}sica Te\'{o}rica e Computacional,
    Faculdade de Ci\^{e}ncias,} \\
{\small \it    Universidade de Lisboa, Campo Grande, Edif\'{\i}cio C8
  1749-016 Lisboa, Portugal} \\[3mm]
{\small\it
$^3$ Centro de F\'{\i}sica da Universidade do Minho e Universidade do Porto (CF-UM-UP),}\\ {\small\it Universidade do Minho, 4710-057 Braga, Portugal} \\[3mm]
}
\maketitle

\begin{abstract}
\noindent

We present a phenomenological study where we probe the sensitivity to invisible dark matter (DM) mediators produced in association with a $t\bar{t}$ pair at the Large Hadron Collider (LHC). Building on previous work focused on scalar mediators, we extend the analysis to include spin-1 mediators, $Y_1$, with both vector and axial-vector couplings to top quarks. The mediator mass is fixed to $5$~GeV. Signal samples of $pp \rightarrow t\bar{t}Y_i$ ($i = 0, 1$) are generated using a \texttt{MadGraph5\_aMC@NLO} simplified DM model. Only dileptonic final states of the $t\bar{t}$ system are considered, and the reconstruction is performed through a kinematic fit without explicitly reconstructing the invisible mediator. All relevant Standard Model backgrounds are included. We consider several exclusion scenarios to assess the sensitivity to the presence of a spin-1 mediator, as well as the ability to distinguish a pure vector or axial-vector mediator from alternative hypotheses with different spin and CP properties. We find that the analysis is sensitive to light spin-1 mediators and that CP-sensitive angular observables provide discrimination power between vector, axial-vector, scalar and pseudoscalar scenarios. These results highlight the potential of $t\bar{t}$ final states not only to search for invisible particles, but also to characterize their spin and parity properties in case of discovery.

\end{abstract}

\thispagestyle{empty}
\vfill
\newpage
\setcounter{page}{1}

\section{Introduction}
\hspace{\parindent}

Over the years, numerous particles have been proposed as dark matter (DM) candidates. However, none have yet been observed experimentally~\cite{Rosenberg:2000wb, Schumann:2019eaa}. Developing and refining methods to detect or exclude DM candidates remains, consequently, a central goal of dark matter searches.

In many models containing dark matter (DM) candidates, the Standard Model (SM) is extended by a dark sector whose particles interact only weakly with those of the SM.
In this study, we will explore one such model, where the dark sector particles interact with the SM through a mediator particle, $Y$. 
We focus on the $pp \to t\bar{t}Y$ production process and simulate the response of a typical Large Hadron Collider (LHC) detector, like ATLAS, to reconstruct the visible particles arising from the dileptonic decay channel of the $t\bar{t}$ pair.
Furthermore, we establish confidence level exclusion limits for the couplings of this mediator to the top quarks using signal and background distributions from two angular observables, $\Delta \phi_{\ell^+ \ell^-}$ and $b_4$. This paper follows a similar methodology to Ref.~\cite{Azevedo:2023xuc}, but applies it on spin-1 DM mediators ($J^{P}= 1^{\pm}$). 
The goal is to explore the experimental sensitivity to the CP-nature of the mediators and their presence alongside a $t\bar{t}$ pair, as well as its effectiveness at excluding the presence of spin-0 mediators (scalars and pseudoscalars).

This work is particularly timely in light of the recent observations by the ATLAS~\cite{ATLAS:2026nrx} and CMS~\cite{CMS:2025kzt} collaborations of an enhancement in the $t\bar{t}$ production cross section near threshold in high-precision LHC data, compatible with the formation of a quasi-bound toponium state. While SM threshold dynamics can account for a significant fraction of the observed enhancement, current measurements still leave room for additional contributions arising from beyond-the-Standard-Model (BSM) pseudoscalar mediators coupled to top quarks~\cite{Flacke:2025dwk}.

This paper is divided in the following sections: we present the simplified DM model, the relevant parameters and the angular observables we used, in Section~\ref{sec:TH}. We describe the event generation and detector simulation in Section~\ref{sec:generation} and, in Section~\ref{sec:matching}, discuss the event selection and kinematic reconstruction. Finally, we present our results in Section~\ref{sec:results} and the main conclusions in Section~\ref{sec:conclusion}.

\section{The DM Lagrangian \label{sec:TH}}
\hspace{\parindent} 

In this study, we use the simplified DM model \texttt{DMsimp}~\cite{Backovic:2015soa}, which includes spin-0 ($Y_0$) and spin-1 ($Y_1$) mediators, as well as a dark sector composed of Dirac fermions~\footnote{These particles are assumed to be singlets under the SM gauge group.}. In the following, we remain agnostic about the dark sector, focusing solely on the interactions between the DM mediator and the SM fields. More specifically, we will study only the interactions of the DM mediator with the top quarks using an inclusive analysis, without targeting any particular mediator decays.

In the case of the spin-0 mediator, we assume Yukawa couplings proportional to the mass of the corresponding SM fermion and, hence, dedicate ourselves exclusively to top quarks. The Lagrangian density can thus be simplified and written as follows:
\begin{equation}
    \mathcal{L}_{SM}^{Y_0} = \frac{y_{33}^t}{\sqrt{2}} \bar{t}(g_{u_{33}}^S + ig_{u_{33}}^P \gamma^5) t Y_0,
\label{eq:SMinteraction_spin0}
\end{equation}
where $g^{S/P}_{u_{33}}$ are the scalar/pseudoscalar couplings of the spin-0 DM mediator to top quarks. They are normalized to the SM top Yukawa coupling, $y^t_{33}=\sqrt{2}m_t/v$. The pure scalar scenario ($CP=+1$; $Y_{0^+}$) corresponds to $g^S_{u_{33}} = g^S_{\mathrm{SM}} = 1$ and $g^P_{u_{33}} = 0$, while the pure pseudoscalar case ($CP=-1$; $Y_{0^-}$) is obtained by setting $g^S_{u_{33}} = 0$ and $g^P_{u_{33}} = g^P_{\mathrm{SM}} = 1$. The choice $g^{S/P}_{SM} \equiv 1$ was suggested in \cite{Backovic:2015soa, Boveia:2016mrp} as a benchmark for the spin-0 mediator scenario to reproduce the SM predictions for $pp \rightarrow ht\bar{t} \rightarrow \tau^+  \tau^- t\bar{t}$~\footnote{Note that the value of these couplings is just a choice to reproduce SM-like cross sections from the event generator we used (\texttt{MadGraph5\_aMC@NLO}). Other values for these couplings could have been considered. \label{note01}}.
When both $g^{S/P}_{u_{33}} \neq 0$, the interaction has both CP-even and -odd components and is therefore CP-violating.

For the spin-1 mediator, the Lagrangian density is given by:
\begin{equation}
    \mathcal{L}_{SM}^{Y_1} = \overline{t}\gamma_\mu(g^V_{u_{33}} + g^A_{u_{33}}\gamma_5)t Y_1^\mu,
\label{eq:SMinteraction_spin1}
\end{equation}
where $g^{V/A}_{u_{33}}$ denote the vector/axial-vector couplings of the spin-1 DM mediator to top quarks. The pure vector scenario ($P=-1$; $Y_{1^-}$) corresponds to $g^V_{u_{33}} = g^V_{\mathrm{SM}} \equiv 0.25$ and $g^A_{u_{33}} = 0$, while the pure axial-vector case ($P=+1$; $Y_{1^+}$) is obtained by setting $g^V_{u_{33}} = 0$ and $g^A_{u_{33}} = g^A_{\mathrm{SM}} \equiv 0.25$. 
As for the spin-0 case, the choice $g^{V/A}_{SM} \equiv 0.25$ was suggested in \cite{Backovic:2015soa, Boveia:2016mrp} as a benchmark for the spin-1 mediator scenario to reproduce the SM predictions for $pp \rightarrow Zj(j) \rightarrow \tau^+  \tau^- j(j)$~\footref{note01}.

In this paper, as in Ref.~\cite{Azevedo:2023xuc}, we explore how angular distributions of the top quarks and their decay products can be used to probe the invisible sector, by looking into the expected changes of these observables in the presence of new particles. Several CP-sensitive observables have been proposed in the literature to probe the CP nature of the coupling between top quarks and the Higgs boson at the LHC and future colliders, mainly in the $t\bar{t}H$ channel~\cite{Bernreuther:1993hq, Gunion:1996xu, BhupalDev:2007ftb, Frederix:2011zi, Ellis:2013yxa, Khatibi:2014bsa, Demartin:2014fia, Kobakhidze:2014gqa, Bramante:2014gda, Boudjema:2015nda, He:2014xla, Santos:2015dja, Gritsan:2016hjl, Dolan:2016qvg, Goncalves:2016qhh,  Buckley:2015ctj, Mileo:2016mxg, Buckley:2015vsa, AmorDosSantos:2017ayi, Goncalves:2018agy, Azevedo:2017qiz, Li:2017dyz, Ferroglia:2019qjy, Faroughy:2019ird, Azevedo:2020vfw, Azevedo:2020fdl, Bortolato:2020zcg, Cao:2020hhb, Goncalves:2021dcu, Barman:2021yfh, Bahl:2020wee, Bahl:2021dnc, Azevedo:2022jnd, ATLAS:2023cbt}. These observables are sensitive to the structure of the couplings and can also be used to discriminate scalar boson signals from irreducible backgrounds at the LHC over a wide mass range (see, e.g., Refs.~\cite{Azevedo:2020fdl, Azevedo:2020vfw}). 

We will use as benchmark observables the azimuthal angle difference between the charged leptons resulting from the dileptonic decay of the $t\bar{t}$ pair, $\Delta \phi_{\ell^-\ell^+}$, and the $b_4$ variable evaluated in the laboratory frame (LAB), $b_4 = p^z_t . p^z_{\bar{t}} / (|\vec{p}_{t}| . |\vec{p}_{\bar{t}}|)$,
where the $z$-direction corresponds to the beam direction, and $\vec{p}_{t(\bar{t})}$ and $p^z_{t(\bar{t})}$ correspond to the total and z-component of the top (anti-top) quark momentum measured in the LAB frame, respectively. It should be noted that the $b_4$ variable can also be expressed as $b_4=\cos{\theta_t} \times \cos{\theta_{\bar{t}}}$,  explicitly showing the dependency of $b_4$ on the $t$ and $\bar{t}$ polar angles in the LAB frame ($\theta_t$ and $\theta_{\bar{t}}$, respectively) with respect to the $z$-direction. The evaluation of these observables requires the reconstruction of the $t\bar{t}$ system and its decay products, which will be discussed in the following sections.

\begin{figure}[h!]
	\begin{center}
        \includegraphics[width=0.95\textwidth]{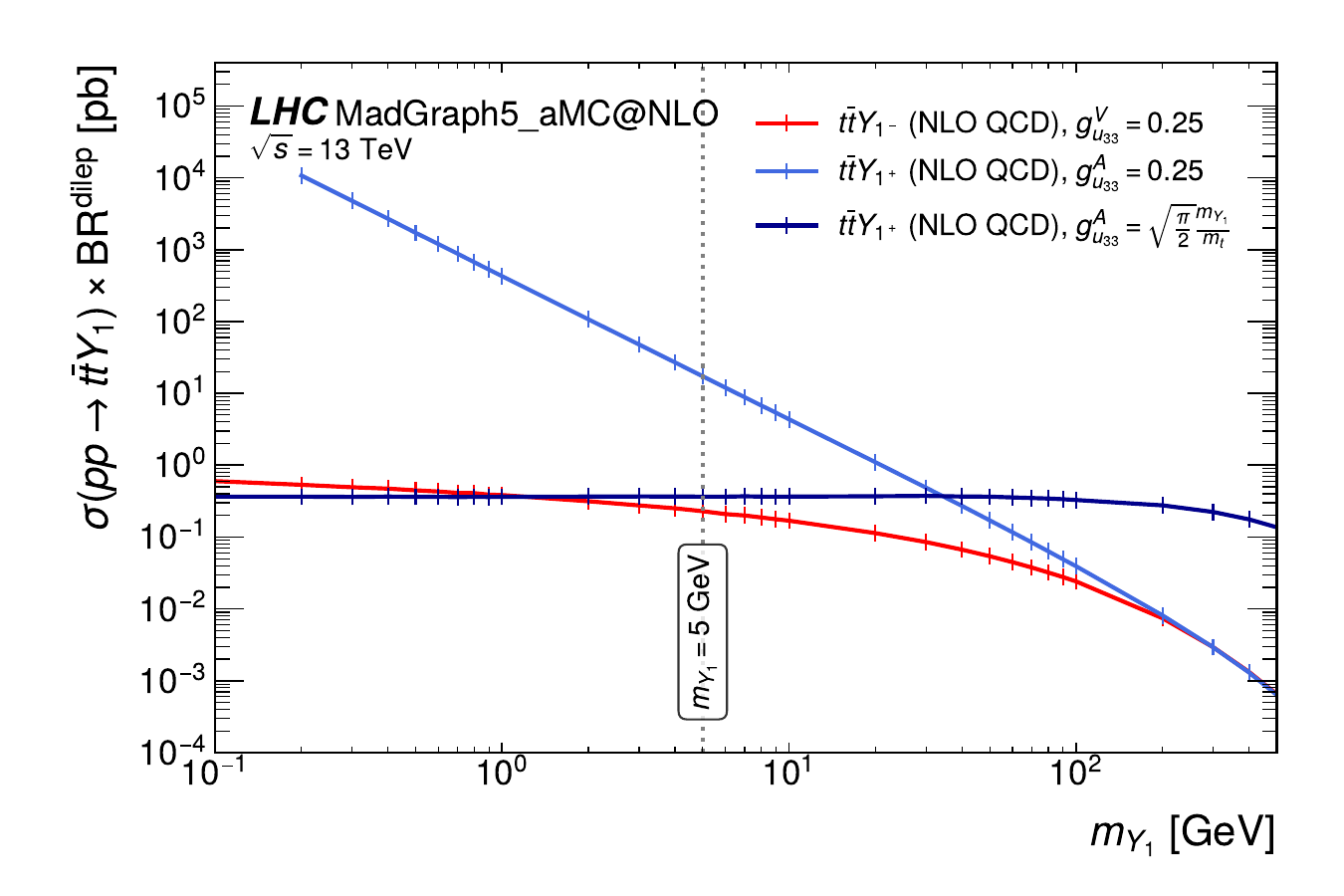}
		\caption{Total cross section for $pp \to t\bar{t}Y_1$, followed by $t\bar{t} \to b\ell^+\nu_\ell\bar{b}\ell^-\bar{\nu}_{\ell}$, as a function of the mediator mass, for a pure vector (red line) and a pure axial-vector (blue lines) mediators. While in the lighter blue line we set $g^A_{u_{33}} = 0.25$, in the darker one we set $g^A_{u_{33}} = \sqrt{\pi/2} \, \, m_{Y_1}/m_t$, the upper bound from perturbative unitarity (see Eq.~\ref{eq:uni1}). The cross sections were computed with \texttt{MadGraph5\_aMC@NLO}~\cite{Alwall:2011uj} at Next-to-Leading Order (NLO) in QCD, using the DM simplified model \texttt{DMsimp}~\cite{Backovic:2015soa}.}
		\label{fig:xsec_ttY1}
	\end{center}
\end{figure}

We conclude this section with a comparison between the total cross section for $pp \to t\bar{t}Y_1$ in the vector and axial-vector cases (see Figure~\ref{fig:xsec_ttY1}). It is clear that while the two cross sections converge to similar values for heavier mediators ($m_{Y_1} \gtrsim 100$~GeV), their behaviour differs significantly at lower mediator masses. In particular, as $m_{Y_1} \to 0$, the vector cross section increases very slowly, whereas the axial-vector cross section increases sharply, exhibiting an approximate $1/m_{Y_1}^2$ scaling. This different low-mass behaviour originates from the fact that the fermionic axial current, $J_A^\mu$, is not conserved, i.e., $\partial_\mu J_A^\mu \neq 0$, and therefore its contraction with the longitudinal polarization of the mediator does not vanish. To show this, note that summing over the mediator polarizations introduces the tensor structure
\begin{equation}
    \sum_\lambda \epsilon_\mu^{(\lambda)}(k)\,\epsilon_\nu^{(\lambda)\,*}(k) = -g_{\mu\nu}+\frac{k_\mu k_\nu}{m_{Y_1}^2} \, ,
\end{equation}
where $k_\mu$ is the momentum of the mediator. Therefore, ($1/m_{Y_1}^2$)-enhanced cross sections are associated with the longitudinal mode of $Y_1$. For an axial-vector coupling, $J_A^\mu = \bar u(p)\gamma^\mu\gamma^5 v(q)$. Making use of $k = p + q$, where $p$ and $q$ are the momenta of the top quarks, one finds
\begin{equation}
    k_\mu J_A^\mu = \bar u(p)(\slashed p+\slashed q)\gamma^5 v(q) = 2m_t\,\bar u(p)\gamma^5 v(q) \, ,
\end{equation}
where the Dirac equations have been used. As a result, the longitudinal contribution to the amplitude behaves as
\begin{equation}
    \mathcal M_A^{(L)} \sim \frac{g^A_{u_{33}} \, m_t}{m_{Y_1}},
\end{equation}
so that the squared amplitude, and therefore the cross section, exhibit the low-mass behaviour $\sigma_{t\bar{t}Y_{1^+}} \propto 1/m_{Y_1}^2$. For a vector coupling, the fermionic current is given by $J_V^\mu = \bar u(p)\gamma^\mu v(q)$, and
\begin{equation}
    k_\mu J_V^\mu = \bar u(p)(\slashed p+\slashed q)v(q)=0 \, .
\end{equation}
Hence, the longitudinal contribution vanishes in the vector case, and no $1/m_{Y_1}^2$ enhancement arises. 

In order for the coupling between the longitudinal component of $Y_1$ and the top quarks, or more generally any fermion $f$ with axial couplings $g^A_f \neq 0$, to remain perturbative, the following combination of parameters must satisfy the bound~\cite{Kahlhoefer:2015bea}
\begin{equation}
    g^A_f \lesssim \sqrt{\frac{\pi}{2}} \frac{m_{Y_1}}{m_f} \, .
    \label{eq:uni1}
\end{equation}
For purely vector couplings, there is no analogous constraint. This highlights the existence of a non-trivial relation between the mediator mass, the fermion mass and the axial coupling, which prevents an arbitrary choice of parameters. As long as Eq.~\ref{eq:uni1} is satisfied, the axial cross section remains well-behaved, even for lighter mediators. In particular, the $1/m_{Y_1}^2$ enhancement observed in the axial case is effectively regulated once perturbative unitarity is imposed, since the allowed parameter space enforces a correlated scaling between $g^A_f$ and $m_{Y_1}$.

Perturbative unitarity further requires
\begin{equation}
    \sqrt{s} < \frac{\pi \, m^2_{Y_1}}{(g^A_{\text{DM}})^2 \, m_{\text{DM}}} \, ,
    \label{eq:uni2}
\end{equation}
where $g^A_{DM}$ denotes the axial coupling between the mediator and DM and $m_{\text{DM}}$ is the dark matter mass. At higher energies, new physics must appear to restore unitarity~\cite{Kahlhoefer:2015bea}. 

\section{Event Generation and Detector Simulation
\label{sec:generation}}
\hspace{\parindent} 

LHC-like signal and background events were generated with the Monte Carlo event generator \texttt{MadGraph5\_aMC@NLO}~\cite{Alwall:2011uj}, with a center-of-mass energy ($\sqrt s$) of 13~TeV. The DM simplified model, \texttt{DMsimp}~\cite{Backovic:2015soa}\footnote{Available in the \texttt{FeynRules} repository.}, was used to generate $pp\to t\bar tY_0$ signal events, at Leading Order (LO), and $pp\to t\bar tY_1$ signal events, at Next-to-Leading Order (NLO) in QCD. For the spin-0 scenario, event generation was performed at LO since \texttt{MadGraph5\_aMC@NLO} was unable to generate NLO event samples for very light mediator masses in the parameter region considered in this work. For the spin-1 case (in particular the vector case), however, NLO event generation was required, since at low mediator masses the LO cross section increases rapidly and starts to deviate significantly from the NLO prediction. The pure scalar, pseudoscalar, vector and axial-vector signals were generated by setting the respective couplings as mentioned in the previous section (following the Lagrangian density in Equations~\ref{eq:SMinteraction_spin0} for the spin-0~scenario and~\ref{eq:SMinteraction_spin1} for the spin-1 scenario). The mass of the DM mediator was set to the benchmark value $m_{Y_1}=5$~GeV, and the top quark mass to $m_t=172.5$~GeV. Note that for the axial-vector case, perturbative unitarity imposes that $g^A_{u_{33}} \lesssim 0.036$ (see Eq.~\ref{eq:uni1}), a significantly smaller value than the benchmark value used, $g^A_{u_{33}} = 0.25$. The region violating perturbative unitarity will be explicitly indicated in the exclusion plots presented in Section~\ref{sec:results}, allowing a direct comparison between the constraints from perturbative unitarity and those obtained in our analysis. 

Only the dileptonic channel of the $t\bar{t}$ pair decay ($t\bar{t}\rightarrow bW^+\bar{b}W^-\rightarrow b\ell^+\nu_\ell\bar{b}\ell^-\bar{\nu}_{\ell}$) was considered in our signal generation, with $\ell = e, \mu$. In order to preserve the inclusiveness of the analysis, the DM mediator was not allowed to decay. Nevertheless, as our analysis only targets the SM detectable particles, our results do not change if the mediator is allowed to decay to DM particles. 
The QCD order at which the dileptonic channel of each signal process was generated in \texttt{MadGraph5\_aMC@NLO}, as well as the corresponding number of generated events and cross section values, are displayed in Table~\ref{tab:cross-sections}. For comparison, the predicted cross section for the dileptonic channel in SM top quark pair production at the LHC is approximately 87.56~pb at $\sqrt{s}=13$ TeV~\cite{ParticleDataGroup:2024cfk}. 

\begin{table}[H]
\centering
\renewcommand{\arraystretch}{1.3}
\begin{tabular}{|c|c|c|c|}
\hline
\textbf{Process (dileptonic channel)}   & \textbf{Order} & \textbf{Number of Events} & \textbf{Cross section (pb)} \\ \hline
$t\bar{t}Y_{1^-}$ $(m_{Y_{1^-}}=5\text{ GeV; }g^V_{SM}=0.25)$  & NLO & 499936              & 0.2311                      \\ \hline
$t\bar{t}Y_{1^+}$ $(m_{Y_{1^+}}=5\text{ GeV; }g^A_{SM}=0.25)$  & NLO & 499913              & 17.27                       \\ \hline
$t\bar{t}Y_{0^+}$ $(m_{Y_{0^+}}=5\text{ GeV; }g^S_{SM}=1)$  & LO & 500000                  & 1.557                       \\ \hline
$t\bar{t}Y_{0^-}$ $(m_{Y_{0^-}}=5\text{ GeV; }g^P_{SM}=1)$  & LO & 500000                  & 0.02336                     \\ \hline
\end{tabular}
\caption{Order in QCD, number of generated events and cross section values of the dileptonic channel of each signal process generated in \texttt{MadGraph5\_aMC@NLO}.}
\label{tab:cross-sections}
\end{table}

The considered SM background processes were $t\bar{t}$ (plus up to 3~jets), $t\bar{t}V$ (plus up to 1 jet and $V=W,Z$), single top quark production ($s$-, $t$- and $Wt$-channels), $W/Z$ (plus up to 4~jets), $W$($Z$)$b\bar{b}$ (plus up to 2~jets), $WW, ZZ$ and $WZ$ diboson processes (plus up to 3 jets), $t\bar{t}H$ and $t\bar{t}b\bar{b}$. The $t\bar{t}H$ and $t\bar{t}b\bar{b}$ processes were generated at NLO, while the remaining backgrounds were generated at LO, using \texttt{MadGraph5\_aMC@NLO}.

Following event generation, where \texttt{MadSpin}~\cite{Artoisenet:2012st} was used to preserve spin correlations in the decays of heavy particles, and parton showering plus hadronization (performed with \texttt{PYTHIA}~\cite{Sjostrand:2006za}), all signal and background events were passed through a fast simulation of a typical LHC detector using \texttt{Delphes}~\cite{deFavereau:2013fsa}, with the default ATLAS detector parameter card. Further details on the event generation and detector simulation can be found in~\cite{Azevedo:2020vfw, Azevedo:2020fdl, JoaoLopes:2024MasterThesis}.

\section{Event Selection and Kinematic Reconstruction
\label{sec:matching}}
\hspace{\parindent} 

Events are selected by requiring the jets and leptons reconstructed by \texttt{Delphes} to satisfy $|\eta|<2.5$~\footnote{The pseudorapidity is defined as $\eta=-\ln[\tan(\theta/2)]$, where $\theta$ is the polar angle in the LAB frame with respect to the beam ($z$) direction.} and transverse momentum $p_T>20$~GeV. Only events with at least two jets and two isolated leptons of opposite charge are accepted. Furthermore, during the event reconstruction, described below, we require the invariant masses of the pairs $(b,\ell^+)$ and $(\bar{b},\ell^-)$ to be smaller than $m_t=172.5$~GeV. After the kinematic reconstruction, to avoid contamination from the $Z$+jets background, we require the invariant mass of the two-lepton system ($m_{\ell^+ \ell^-}$) to be at least 10~GeV away from the $Z$-boson mass ($m_Z$) by requiring $|m_{\ell^+ \ell^-}-m_Z|>10$~GeV. Finally, events are required to contain exactly two $b$-tagged jets. 

For the reconstruction of the $t\bar{t}$ system, we first assign the two highest-$p_T$ oppositely charged leptons as the ones coming from the $W^{\pm}$ bosons' decays. The next step is to assign a jet to the $b$ and $\bar{b}$ quarks originating from the $t\bar{t}$ decays. Since multiple jet assignments are possible, we employ methods implemented in the Toolkit for Multivariate Data Analysis, \texttt{TMVA}~\cite{hoecker2007tmva}, to identify the most likely pairing.
\begin{figure}[h]
    \centering
    \begin{subfigure}[b]{0.45\textwidth}
         \centering
         \adjincludegraphics[width=\textwidth,trim={0 {.5\height} {.66\width} 0},clip]{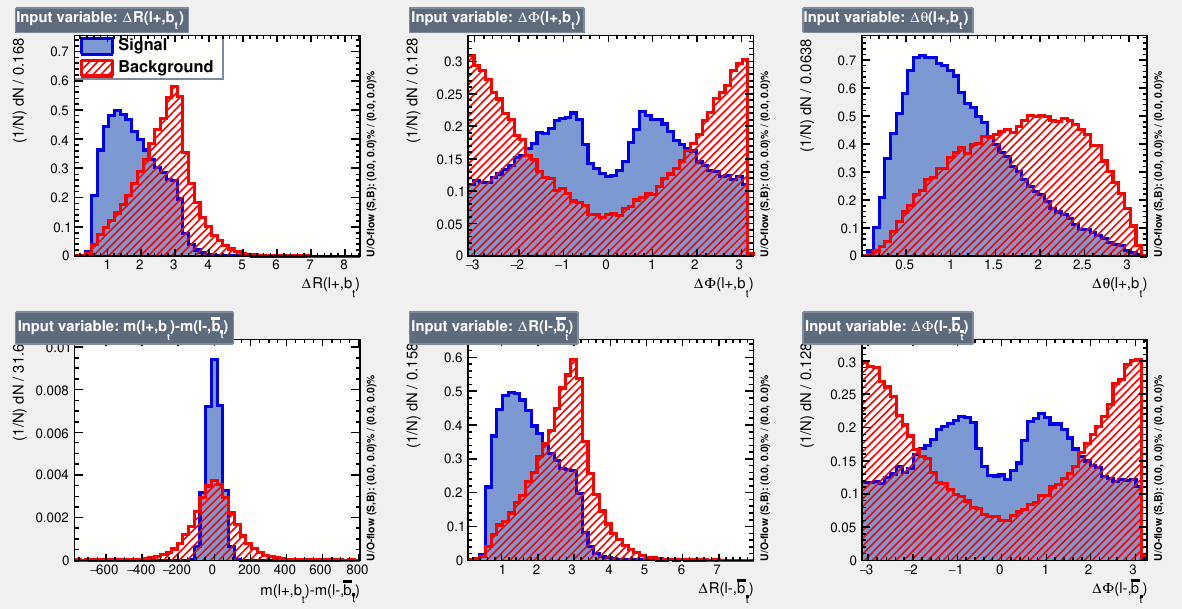}
         \caption{$\Delta R (\ell^+, b_t)$}
         \label{deltaR_TMVA}
     \end{subfigure}
     \hfill
    \begin{subfigure}[b]{0.45\textwidth}
         \centering
         \adjincludegraphics[width=\textwidth,trim={{.33\width} {.5\height} {.33\width} 0},clip]{Figures_Joao-Lopes/TMVA_m0.1_weights/variables_id_c1.pdf}
         \caption{$\Delta \phi (\ell^+, b_t)$}
         \label{deltaPhi_TMVA}
     \end{subfigure}
    \begin{subfigure}[b]{0.45\textwidth}
         \centering
         \adjincludegraphics[width=\textwidth,trim={{.66\width} {.5\height} 0 0},clip]{Figures_Joao-Lopes/TMVA_m0.1_weights/variables_id_c1.pdf}
         \caption{$\Delta \theta (\ell^+, b_t)$}
         \label{deltaTheta_TMVA}
     \end{subfigure}
     \hfill
    \begin{subfigure}[b]{0.45\textwidth}
         \centering
         \adjincludegraphics[width=\textwidth,trim={0 0 {.66\width} {.5\height}},clip]{Figures_Joao-Lopes/TMVA_m0.1_weights/variables_id_c1.pdf}
         \caption{$m (\ell^+, b_t) - m (\ell^-, \overline{b}_{\overline{t}})$}
         \label{invariantMass_TMVA}
     \end{subfigure}
    \caption{Normalized distributions of the \texttt{TMVA} input variables for \textit{correct} (labeled as ``Signal", in blue) and \textit{wrong} (labeled as ``Background", in red) combinations, from the $t\overline{t}Y_{1^-}$ signal events with $m_{Y_{1^-}}=0.1$ GeV. The input variables, $\Delta R, \Delta \phi, \Delta \theta$ and the invariant mass difference, are computed at parton-level. The $\Delta R$, $\Delta \phi$ and $\Delta \theta$ between the $\ell^+$ and the $b$-jet from the $t$ decay, $b_t$, are shown in \ref{deltaR_TMVA}, \ref{deltaPhi_TMVA} and \ref{deltaTheta_TMVA}, respectively. The invariant mass difference between the pairs $(\ell^+, b_t)$ and $(\ell^-, \bar{b}_{\bar{t}})$ is shown in \ref{invariantMass_TMVA}.}
    \label{fig: TMVA distributions}
\end{figure}
To this end, we use several parton-level distributions (which already include hadronization and shower effects), where {\it correct} and {\it wrong} combinations are compared. 
A {\it correct} combination occurs whenever the jets are correctly assigned to the $b$-quarks from both the $t$ and $\bar{t}$ decays. The {\it wrong} combination happens when these jets are swapped.
{\it Correct} (labelled as ``Signal", in blue) and {\it wrong} (labelled as ``Background"\footnote{Do not confuse with the SM background.}, in red) combinations are shown in Figure~\ref{fig: TMVA distributions}, for the $\Delta R$~\footnote{$\Delta R\equiv\sqrt{\Delta \Phi^2+\Delta \eta^2}$, where $\Delta \Phi \, (\Delta \eta)$ corresponds to the difference in the azimuthal angle (pseudorapidity) between two objects.}, $\Delta\Phi$ and $\Delta\theta$ between the $\ell^+$ lepton and the jet originating from the hadronization of the $b$-quark from the $t$ decay (denoted as $b_t$). 
Similar distributions were also obtained for the ($\ell^+, \bar{b}_{\bar{t}}$), ($\ell^-, b_t$), and ($\ell^-, \bar{b}_{\bar{t}}$) pairs, where $\bar{b}_{\bar{t}}$ originates from the $\bar{t}$ decay, in order to optimize the identification of the correct jet--lepton assignment. Clear differences between the {\it correct} and {\it wrong} combinations are visible in all the distributions shown in Figure~\ref{fig: TMVA distributions}. In addition, the invariant mass difference between the pairs $(\ell^+, b_t)$ and $(\ell^-, \bar{b}_{\bar{t}})$ is also used as an input variable. 
These distributions were obtained from parton-level information of $t\bar{t}Y_{1^-}$ events with $m_{Y_{1^-}}=0.1$ GeV in what we call the vector mediator analysis, and from $t\bar{t}Y_{1^+}$ events with $m_{Y_{1^+}}=5$ GeV for the axial-vector mediator analysis. For the vector mediator case, the choice $m_{Y_{1^-}}=0.1$ GeV follows previous studies~\cite{JoaoLopes:2024MasterThesis}~\footnote{For even smaller mediator masses, soft and collinear emissions of a nearly massless vector boson significantly increase the cross section. To avoid entering this infrared-sensitive regime, we restrict our analysis to mediator masses above $0.1$~GeV.}. We have verified that using $m_{Y_{1^-}}=5$ GeV instead leads to no significant differences in the resulting distributions or in the overall analysis. 

\begin{figure}[h!]
	\begin{center}
    \begin{subfigure}[b]{0.48\textwidth}
        \includegraphics[width=\textwidth]{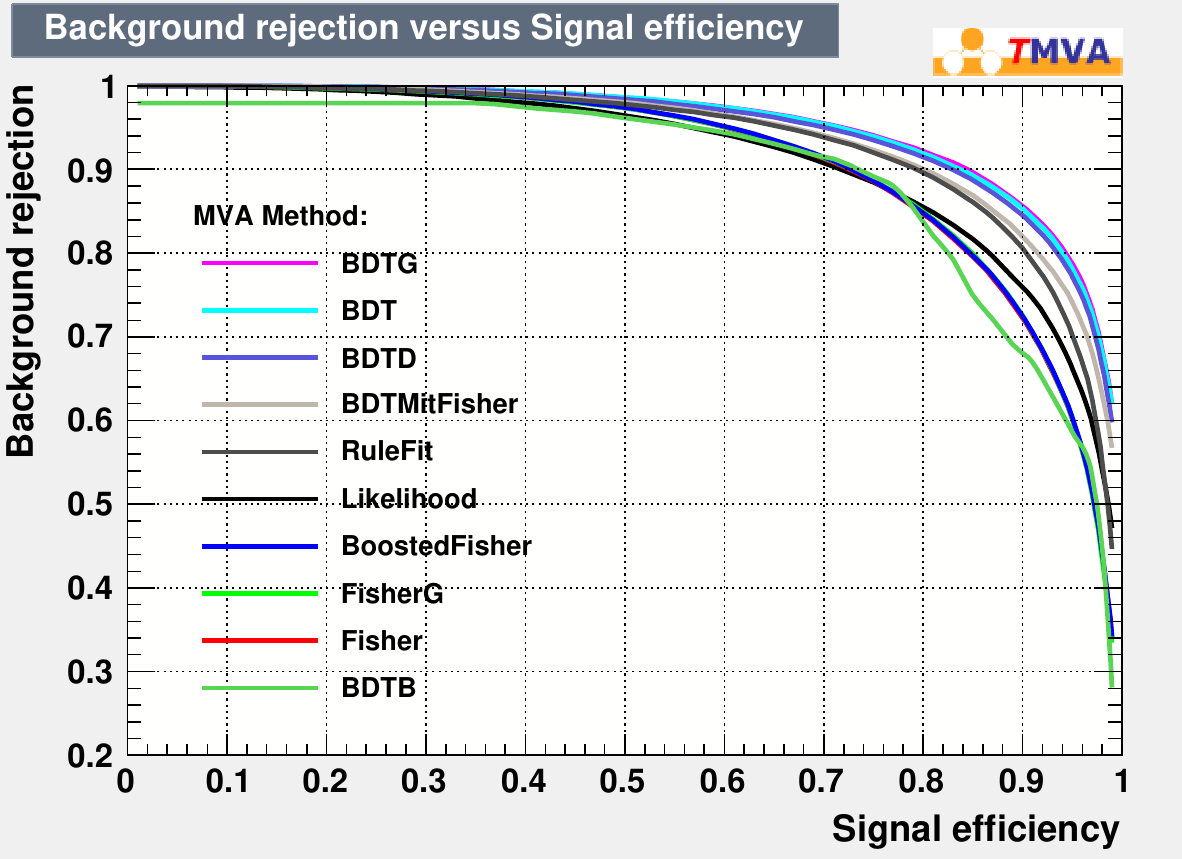}
        \caption{ROC Curve; $t\bar{t}Y_{1^-}$ with $m_{Y_{1^-}}=0.1$ GeV}
        \label{ROC_spin1plus}
    \end{subfigure}
    \begin{subfigure}[b]{0.48\textwidth}
        \includegraphics[width=\textwidth]{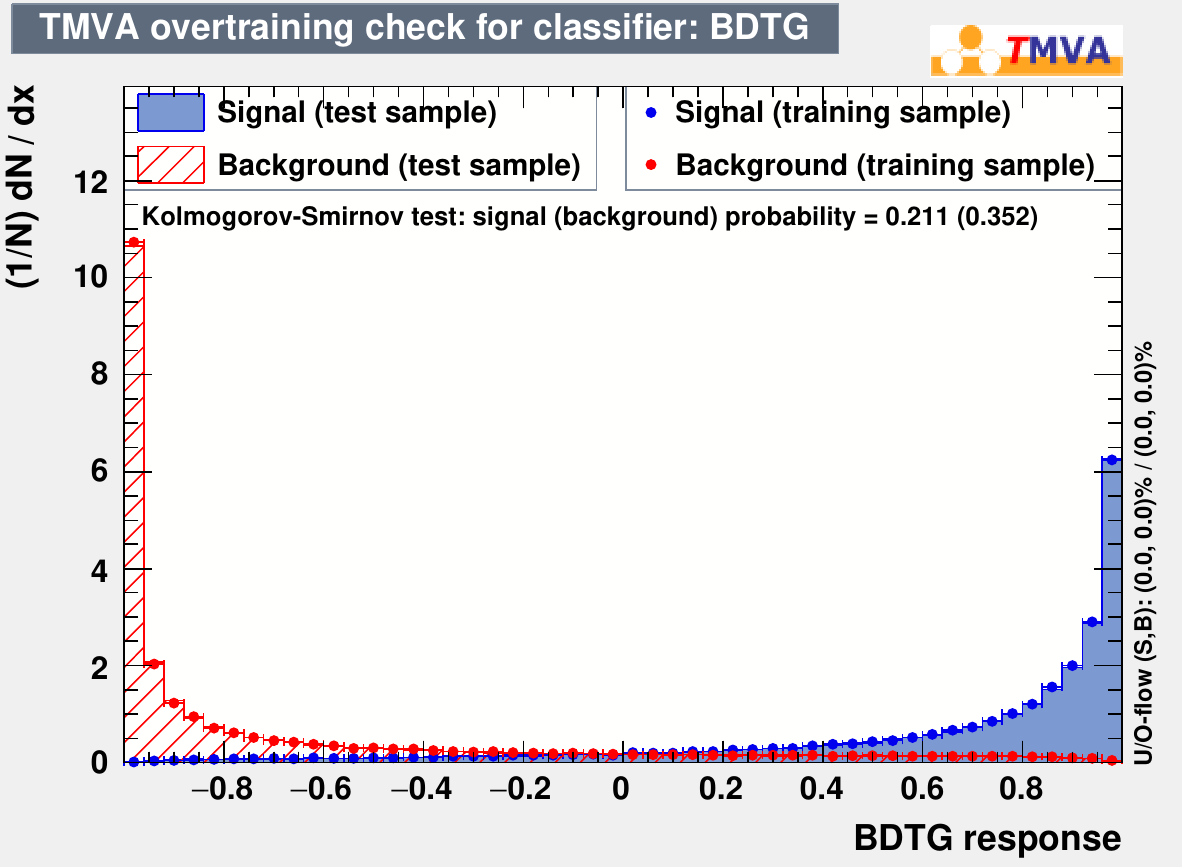}
        \caption{BDTG; $t\bar{t}Y_{1^-}$ with $m_{Y_{1^-}}=0.1$ GeV}
        \label{BDTG_spin1plus}
    \end{subfigure}
    \begin{subfigure}[b]{0.48\textwidth}
        \includegraphics[width=\textwidth]{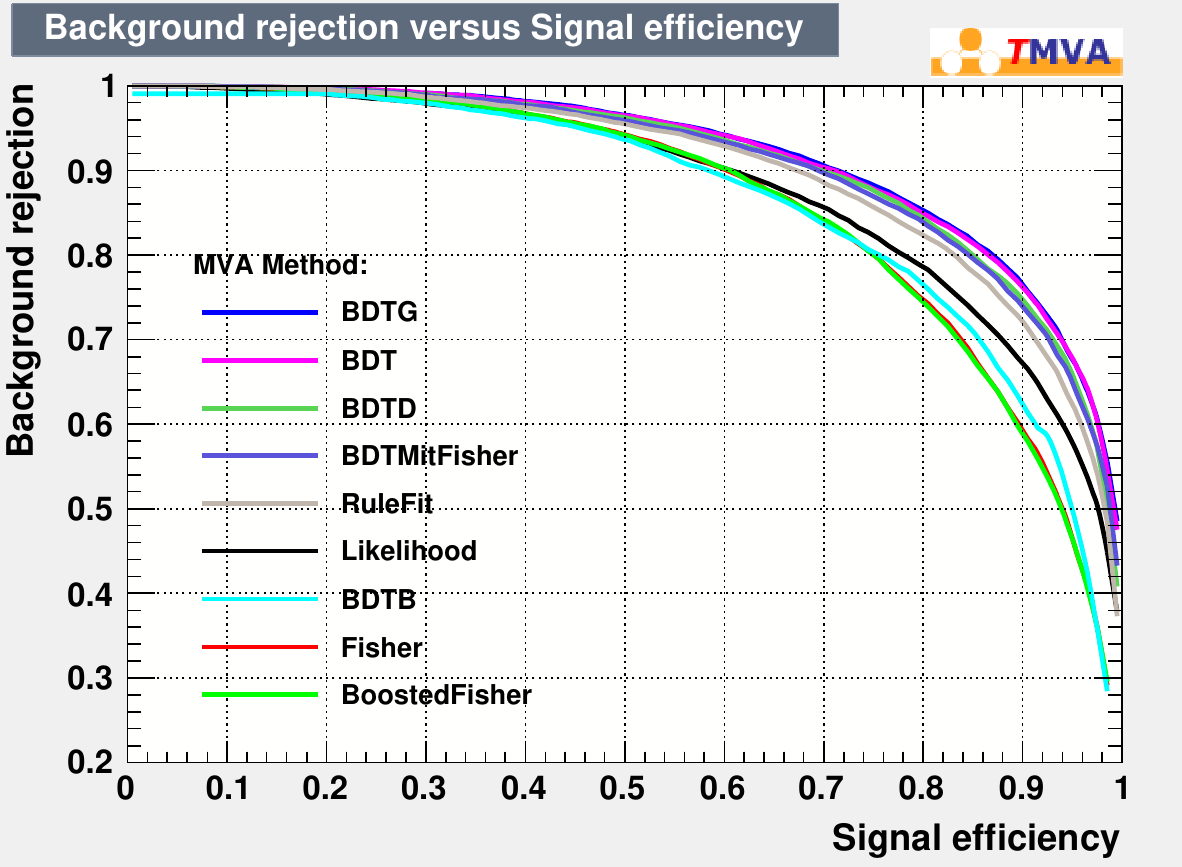}
        \caption{ROC Curve; $t\bar{t}Y_{1^+}$ with $m_{Y_{1^+}}=5$ GeV}
        \label{ROC_spin1minus}
    \end{subfigure}
    \begin{subfigure}[b]{0.48\textwidth}
        \includegraphics[width=\textwidth]{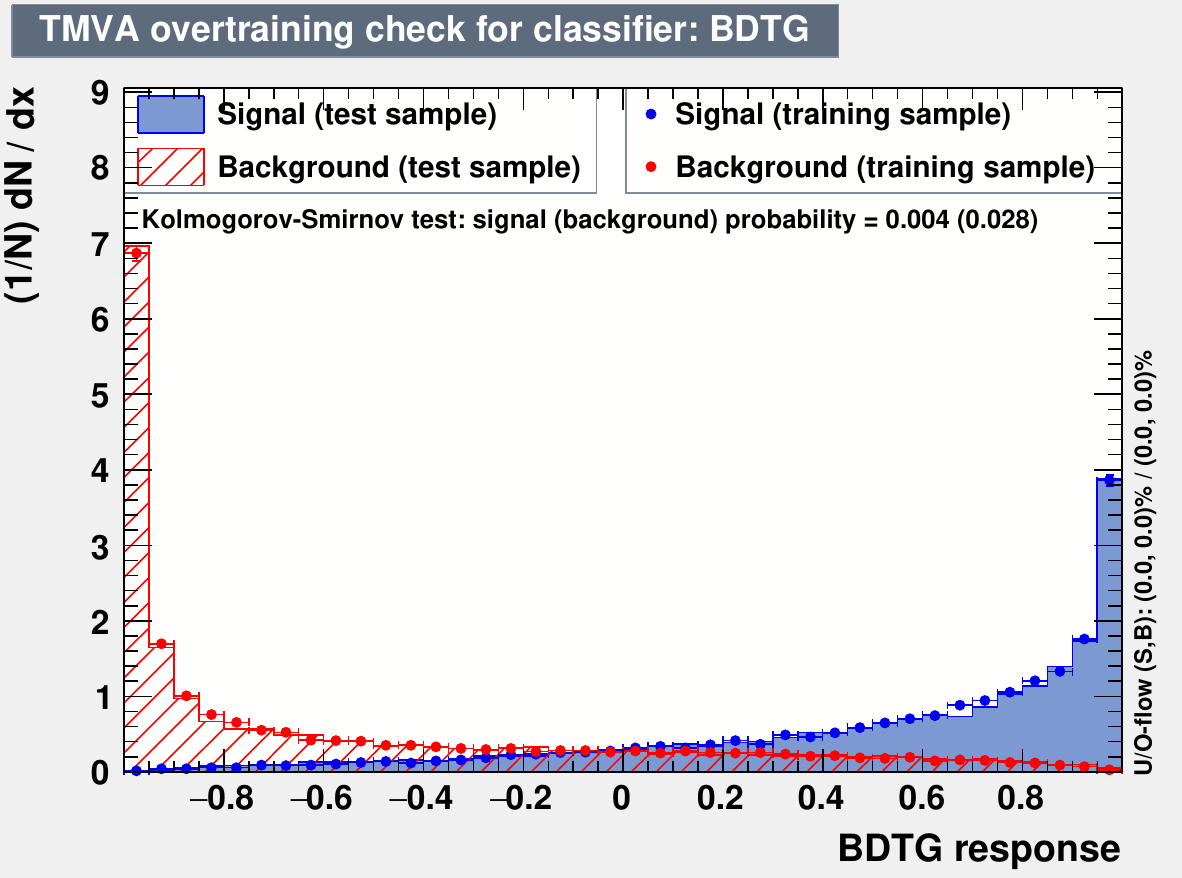}
        \caption{BDTG; $t\bar{t}Y_{1^+}$ with $m_{Y_{1^+}}=5$ GeV}
        \label{BDTG_spin1minus}
    \end{subfigure} \\
	\caption{\textit{Wrong} combination rejection vs \textit{correct} combination acceptance (ROC curve) for different multivariate methods trained from parton-level $t\bar{t}Y_{1^-}$ events with $m_{Y_{1^-}}=0.1$ GeV (\ref{ROC_spin1plus}) and $t\bar{t}Y_{1^+}$ events with $m_{Y_{1^+}}=5$ GeV (\ref{ROC_spin1minus}); distribution of the BDTG discriminant for the ``Signal" and ``Background" in training and test samples from parton-level $t\bar{t}Y_{1^-}$ events with $m_{Y_{1^-}}=0.1$ GeV (\ref{BDTG_spin1plus}) and $t\bar{t}Y_{1^+}$ events with $m_{Y_{1^+}}=5$ GeV (\ref{BDTG_spin1minus}).}
    \label{ROC + BDTG}
	\end{center}
\end{figure}

Several multivariate methods were trained using \texttt{TMVA}, based on the {\it correct} and {\it wrong} combinations distributions introduced above. All individual distributions were combined into a single discriminant classifier for each method. On the left-hand side of Figure~\ref{ROC + BDTG}, we show the Receiver Operating Characteristic (ROC) curves for the vector ($J^P=1^-$) and axial-vector ($J^P=1^+$) mediators, in the top and bottom panels, respectively. From the ROC curves, we observe that the best performance is achieved by a Boosted Decision Tree with Gradient boosting (BDTG). The weights computed by the BDTG method are then used to determine the most probable assignment of jets to the parton-level $b$- and $\bar{b}$-quarks from the $t\bar{t}$ decays, allowing the reconstruction of their four-momenta. The corresponding classifier outputs are shown on the right-hand side of Figure~\ref{ROC + BDTG}, again for the vector and axial-vector mediators in the top and bottom panels, respectively. Comparing both cases, the vector mediator exhibits slightly better performance. As such, most of the results presented in this paper are obtained using the vector mediator analysis. For completeness, we also present results from the axial-vector mediator analysis in Subsection~\ref{subsec:scenario3}. 

To reconstruct the four-momentum of the undetected neutrinos originating from the top quark decays, we impose the following energy-momentum conservation conditions to the events,
\begin{align}
    (p&_{\nu} + p_{\ell^+})^2 = m^2_{W} , \nonumber \\
    (p&_{\bar{\nu}} + p_{\ell^-})^2 = m^2_{W} , \nonumber \\
    (p&_{W^+} + p_b)^2 = m^2_t , \label{eq:etmiss} \\ 
    (p&_{W^-} + p_{\bar{b}})^2 = m^2_{\bar{t}} , \nonumber \\
    p&^x_{\nu} + p^x_{\bar{\nu}} = \slashed{E}^x , \nonumber \\
    p&^y_{\nu} + p^y_{\bar{\nu}} = \slashed{E}^y , \nonumber 
\end{align}
where $p_\varsigma$ ($p_\varsigma^\kappa$) represents the four-momentum of particle $\varsigma$ (its projection along the $\kappa$-axis), and $\slashed{E}^\kappa$ is the missing transverse energy, $\slashed{E}$, projected along the $\kappa$-axis. 
In the first four equations mass constraints are imposed, where neutrinos and charged leptons are required to reconstruct the masses of the $W$ bosons from which they originate. When combined with the corresponding $b$-jets, these should reproduce the top quark masses. As in~\cite{Azevedo:2023xuc}, we assume that the total missing transverse energy is wholly accounted for by the neutrinos (last two equations). 
To determine the best solution for each event, the top quark and $W$ boson masses used in the fit were sampled up to 500 times from two-dimensional probability distribution functions (\textit{p.d.f.}s) obtained from parton-level (with shower effects) $t\bar{t}Y_{1^-}$ and $t\bar{t}Y_{1^+}$ signal events for the vector and axial-vector DM mediator analyses, respectively. The neutrinos four-momenta are obtained by solving the system of equations in Eq.~\ref{eq:etmiss}. If a solution is found, nearby mass values are explored (up to two additional iterations) to check whether a better overall fit can be achieved. Due to the quadratic nature of the mass equations, multiple solutions may exist for a given event. We select the solution that maximizes the likelihood ($L$), constructed from smoothed parton-level (with shower effects) distributions, in particular the \textit{p.d.f.}s for the transverse momenta of the neutrinos, top quarks and $t\bar{t}$ system, $P(p_{T_{\nu}})$, $P(p_{T_{\bar{\nu}}})$, $P(p_{T_t})$, $P(p_{T_{\bar{t}}})$ and $P(p_{T_{t\bar{t}}})$, respectively. This likelihood is defined as
\begin{equation}
    L \propto \frac{1}{p_{T_{\nu}}p_{T_{\bar{\nu}}}} P(p_{T_{\nu}})P(p_{T_{\bar{\nu}}})P(p_{T_{t}})P(p_{T_{\bar{t}}})P(p_{T_{t\bar{t}}}) ,
\end{equation}
where the normalization factor ${1}/{p_{T_{\nu}}p_{T_{\bar{\nu}}}}$ accounts for energy losses due to radiation emission and detector resolution effects, which tend to increase the reconstructed neutrino momenta. 
If no solution is found, the event is discarded. When a solution exists, the reconstructed neutrino kinematics allows the full reconstruction of the $t$ and $\bar{t}$ four-momenta, which is the main goal of this analysis. Further details on the event selection and kinematic reconstruction can be found in~\cite{Azevedo:2020vfw, Azevedo:2020fdl, Azevedo:2023xuc, JoaoLopes:2024MasterThesis}.

\begin{figure}[h!]
	\begin{center}
       \includegraphics[width = 7.5cm]{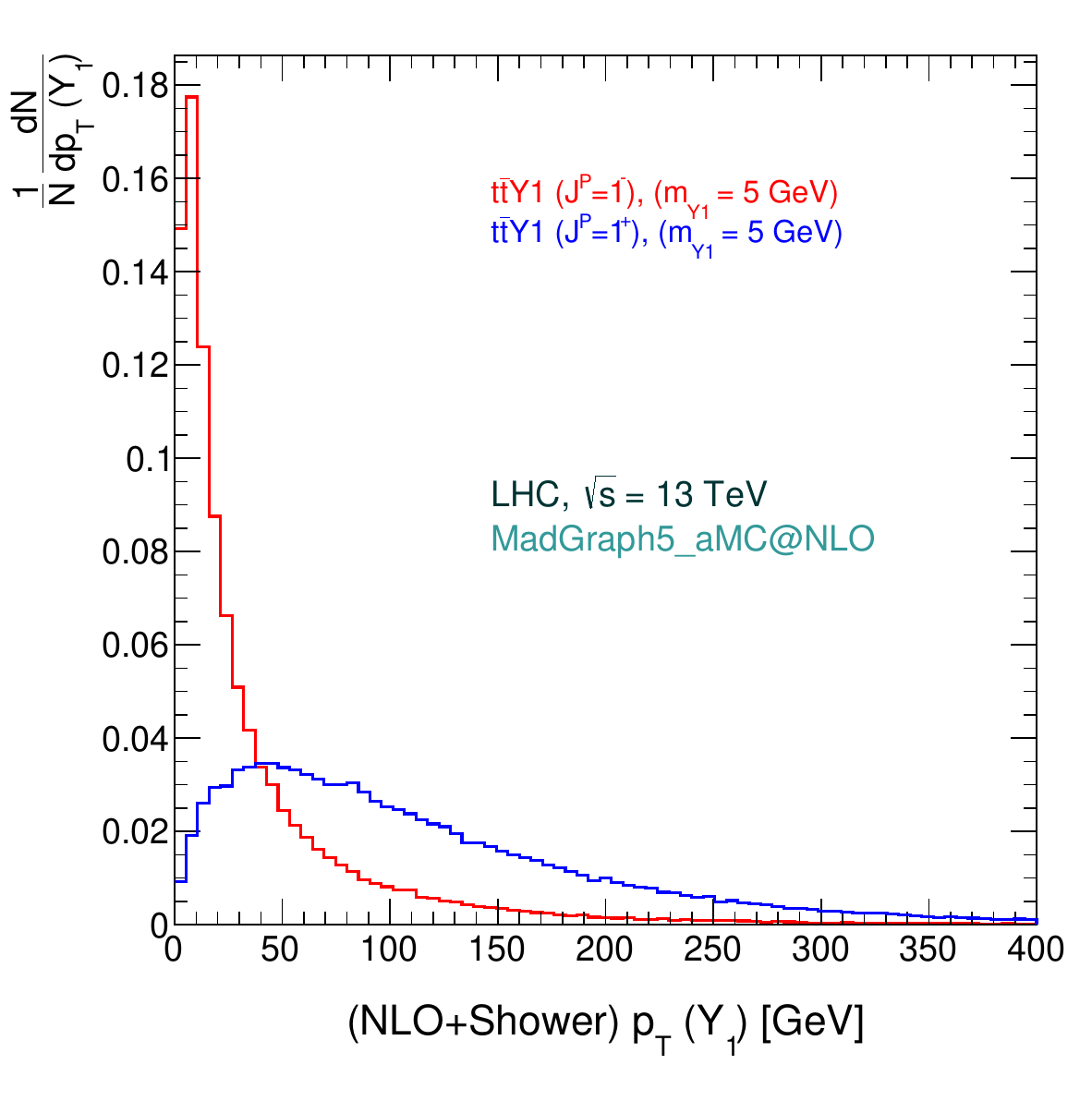}
        \includegraphics[width = 7.5cm]{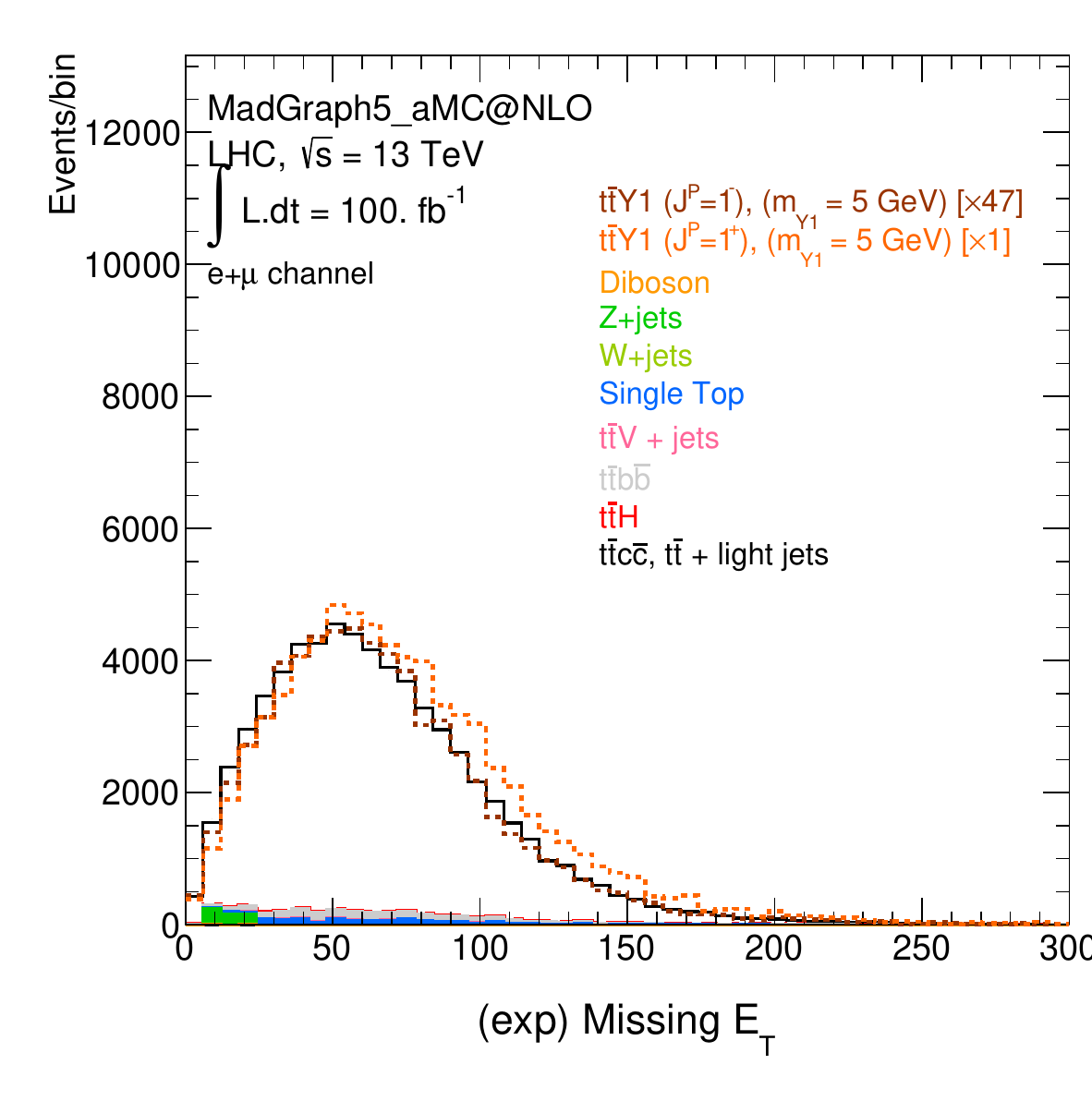}
		\caption{Left: parton level $p_{T}(Y_1)$ distributions with NLO corrections and shower effects (NLO+Shower), for a pure vector (red line) and a pure axial-vector (blue line) mediators with $m_{Y_1}= 5$~GeV. Right: missing transverse energy ($E_T$) distributions of the expected number of events for vector and axial-vector signals with dileptonic final states (dashed curves) together with the SM background processes (full lines), are represented after event selection and kinematic reconstruction (exp), for a reference luminosity of 100~fb$^{-1}$. Scaling factors are applied to the vector and axial-vector signals for convenience.}
		\label{fig:met_tty1}
	\end{center}
\end{figure}

We found that the efficiency of our kinematic reconstruction for $m_{Y_{1}} = 5$~GeV is around 68\% for $t\bar{t}Y_{1^-}$ signal events, and around 48\% for $t\bar{t}Y_{1^+}$ signal events. The former is in good agreement with typical SM $t\bar{t}$ analyses where such kinematic reconstruction is attempted. These efficiencies are smaller than those reported in \cite{Azevedo:2023xuc} (75\% and 72\% for $t\bar{t}Y_{0^+}$ and $t\bar{t}Y_{0^-}$ signal events with $m_{Y_{0}}=0$ GeV, respectively). This difference is partly due to the fact that the production of a $t\bar{t}$ pair in association with a spin-1 mediator modifies the kinematics of the top quarks more significantly than in the spin-0 case. The lower efficiency in the axial-vector case can be understood from its significantly harder transverse momentum spectrum, which features a peak at larger $p_T$ values and a longer high-$p_T$ tail compared to the vector case, whose distribution is strongly peaked at low $p_T$ and rapidly falling (see the left panel of Figure~\ref{fig:met_tty1}). As a result, the assumption that the mediator contribution to the missing transverse energy is negligible becomes less accurate for the axial-vector mediator, making the reconstruction of the neutrino four-momenta more challenging and thus reducing the overall efficiency. This is reflected in the missing transverse energy distribution shown in Figure~\ref{fig:met_tty1} (right panel), which is slightly shifted towards larger values for the axial-vector signal with respect to both the vector case and the dominant $t\bar{t}$ background. Nevertheless, the fact that a significant fraction of events survives the kinematic reconstruction --- even when the analysis does not match the mass, parity, or spin of the DM mediator present in the signal --- indicates that a standard $t\bar{t}$ analysis is unlikely to reject events containing such particles. Therefore, any observed excess in $t\bar{t}$ final states should be interpreted with this possibility in mind.

\begin{figure}[h]
	\begin{center}
		\includegraphics[width = 7.0cm]{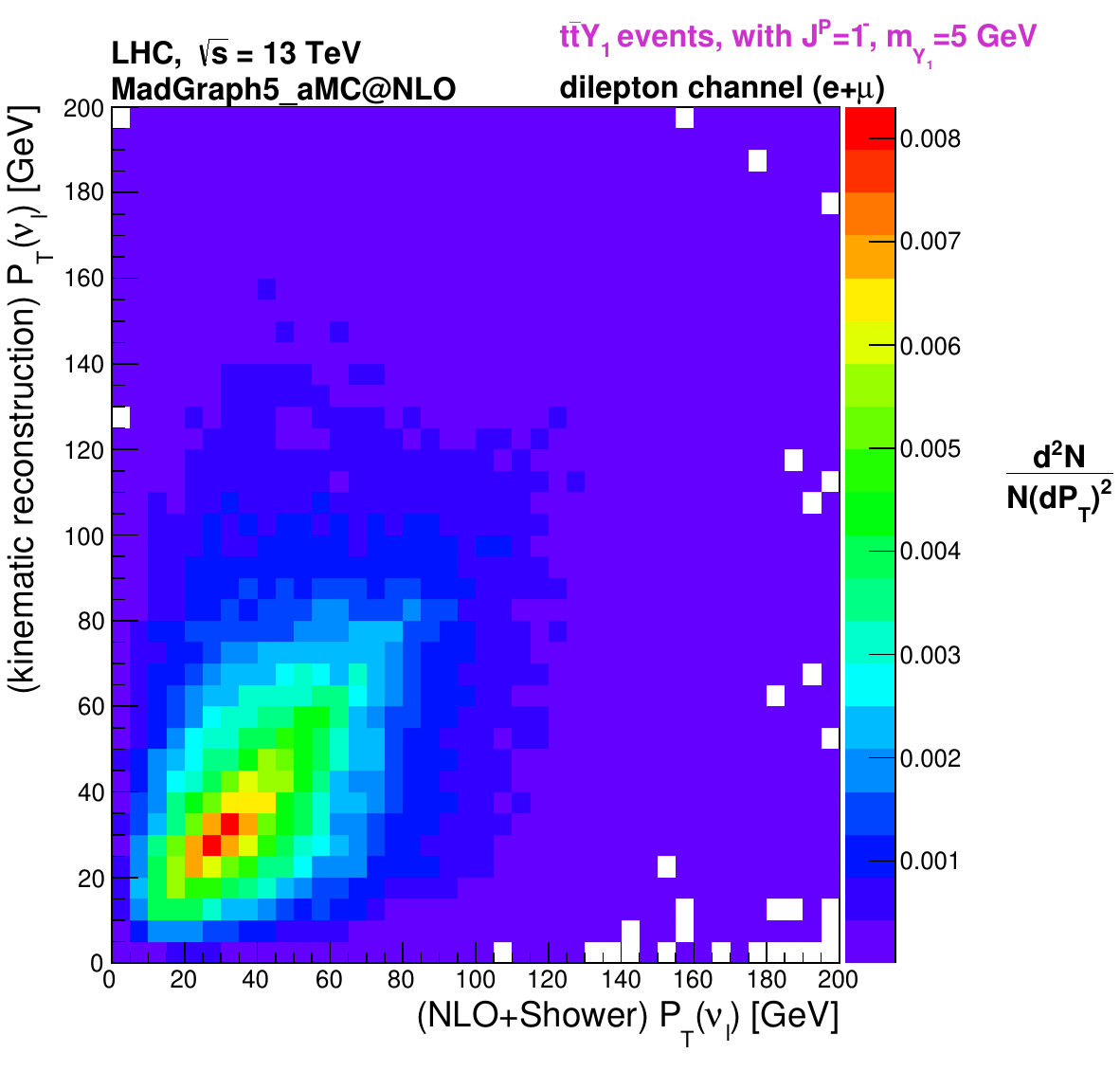}
		\includegraphics[width = 7.0cm]{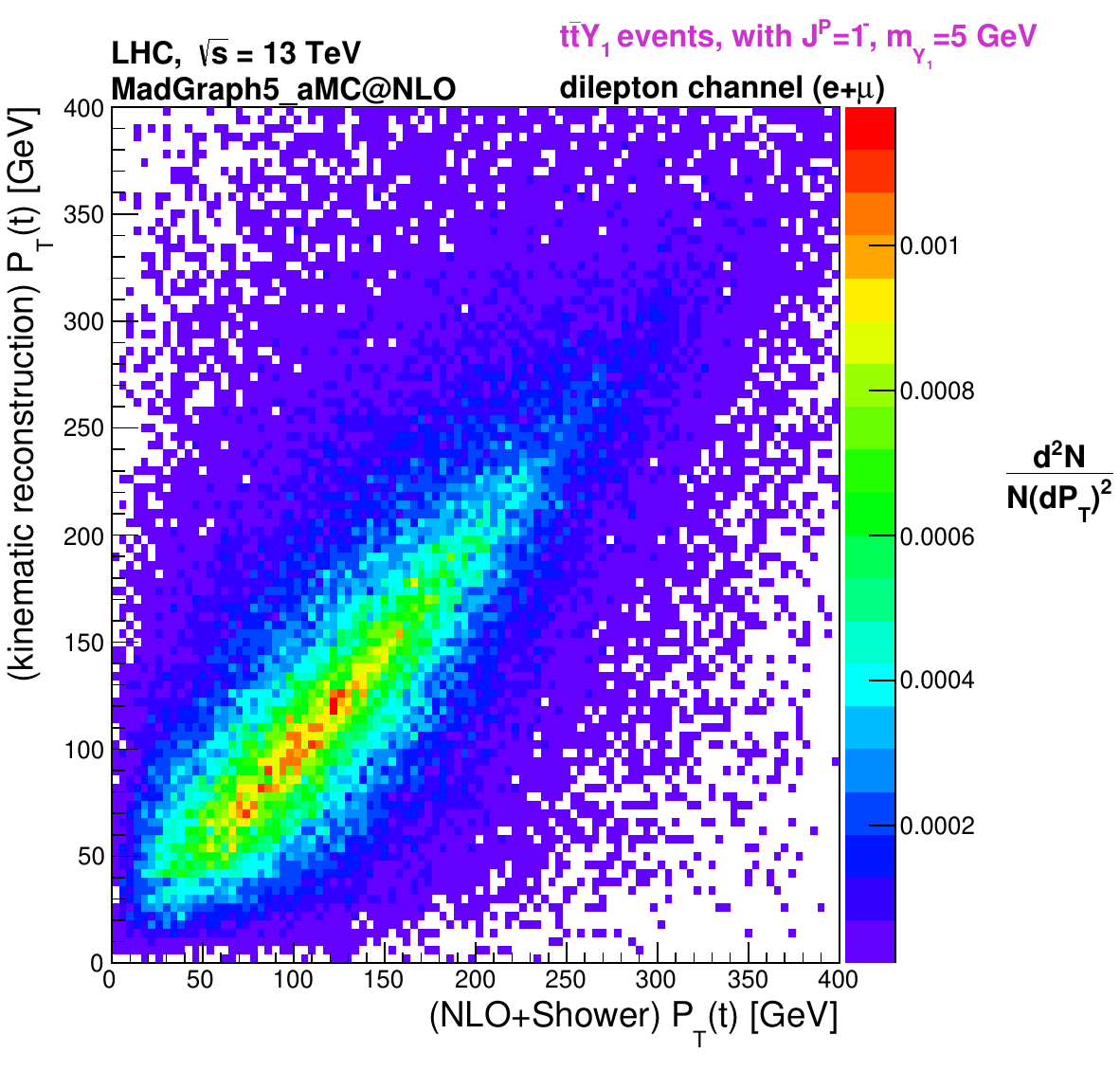} \\
		\includegraphics[width = 7.0cm]{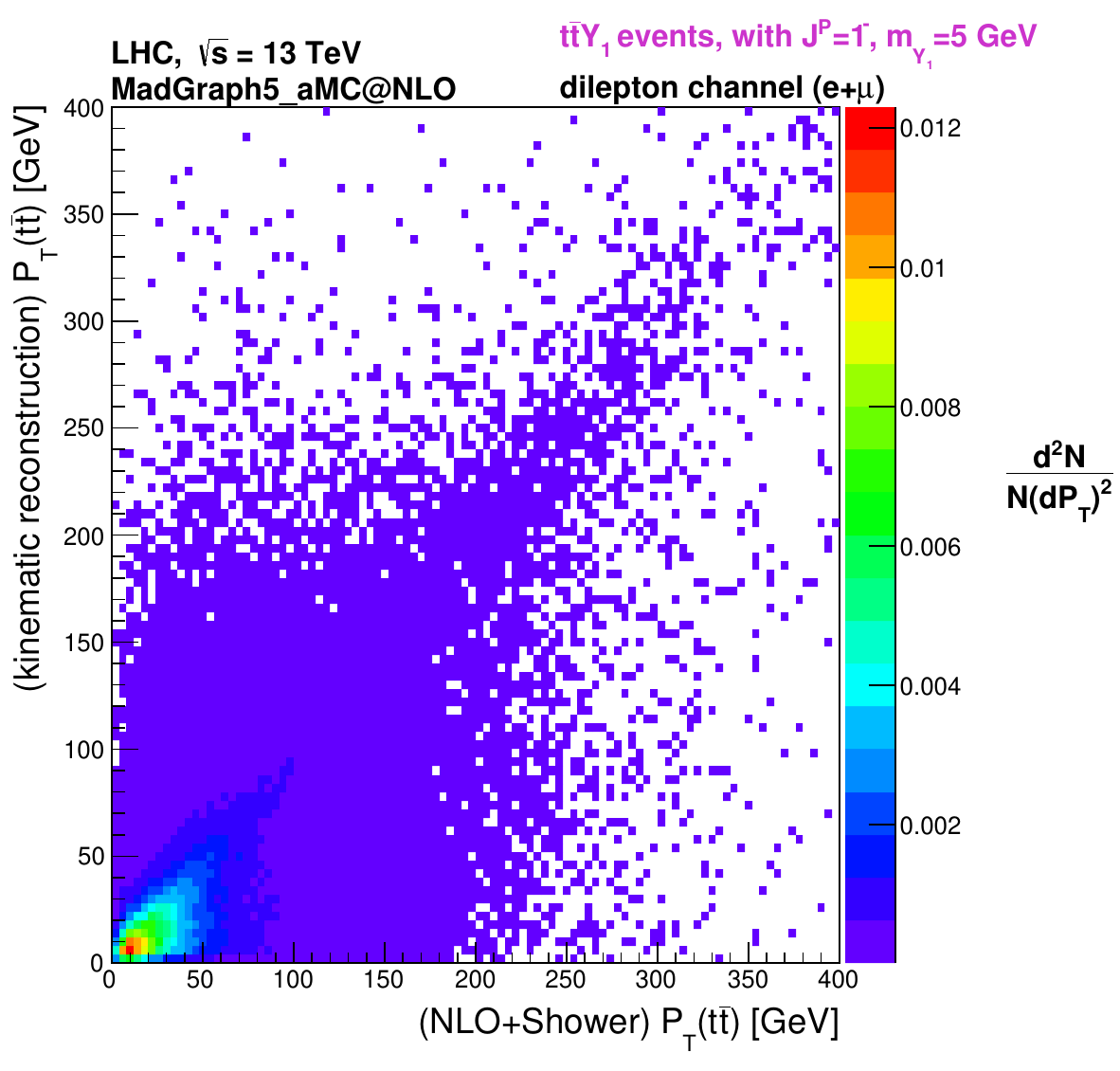}
		\includegraphics[width = 7.0cm]{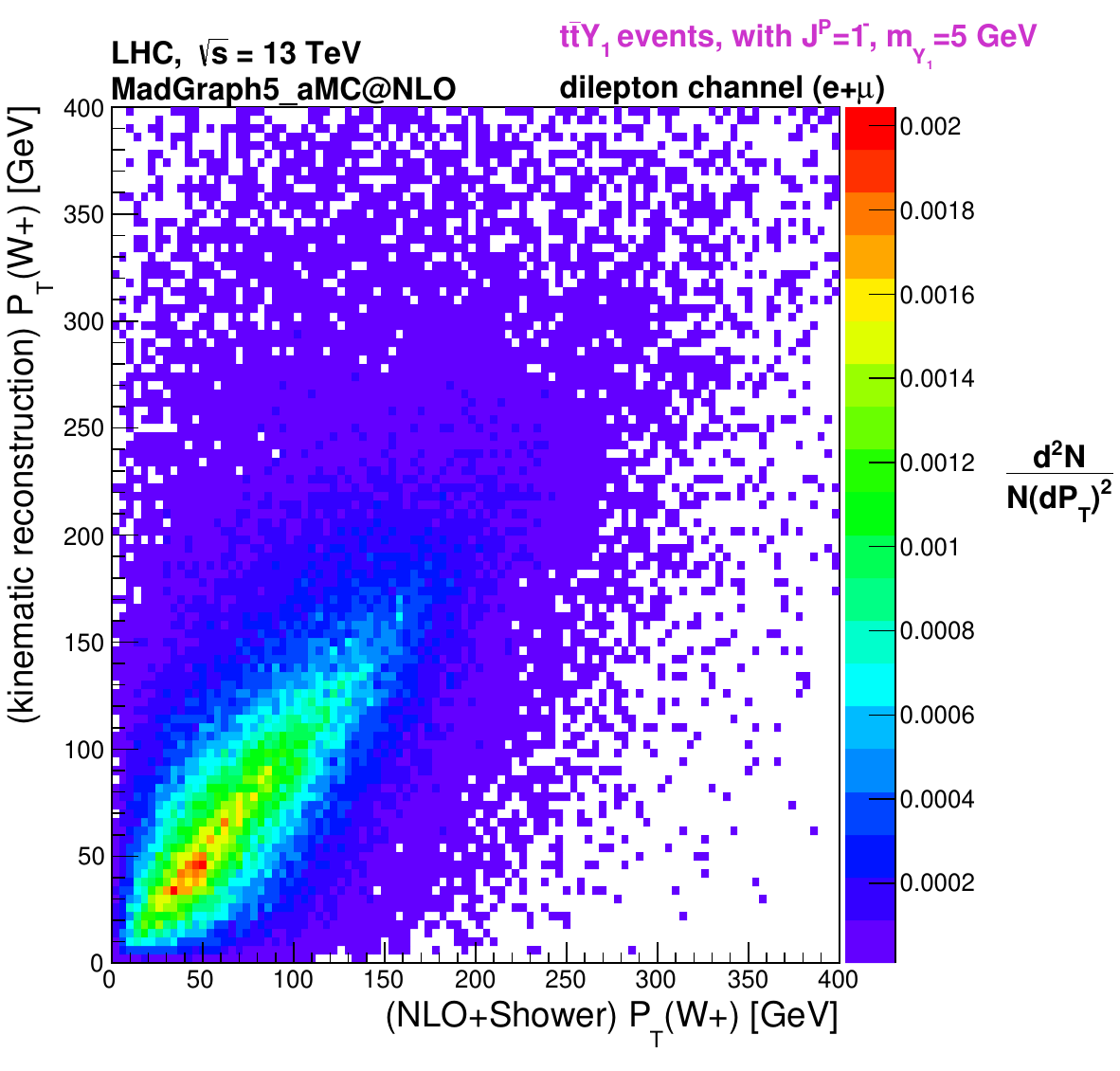} \\
		\caption{Two-Dimensional distributions of $t\bar{t}Y_{1^-}$ signal events with $m_{Y_{1^-}}=5$ GeV: parton-level transverse momentum (NLO+Shower) versus reconstructed transverse momentum (kinematic reconstruction) for several particles (neutrino, top left; $t$, top right; $t\overline{t}$ system, bottom left; W$^{+}$ boson, bottom right).}
		\label{fig:GENvsREC}
	\end{center}
\end{figure}

In Figure~\ref{fig:GENvsREC}, we show the two-dimensional $p_T$ distributions of the neutrino (top left), top quark (top right), $t\bar{t}$ system (bottom left) and $W^+$ boson (bottom right) for the $t\bar{t}Y_{1^-}$ signal events with $m_{Y_{1^-}}=5$ GeV, after applying the vector mediator analysis. We can see that the parton-level with shower effects (NLO+Shower) and the reconstructed kinematics (kinematic reconstruction) are highly correlated for all particles or systems of particles. This implies that, even after experimental effects are taken into account, it is still possible to reconstruct the $t\bar{t}$ system without trying to reconstruct the invisible DM mediator. Similar plots were obtained when the axial-vector mediator analysis was used in the reconstruction, although with slightly worse correlations.

\newpage
\section{Results and Discussion \label{sec:results}}
\hspace{\parindent} 

\begin{figure}[!h]
	\begin{center}
		\includegraphics[width = 7.5cm]{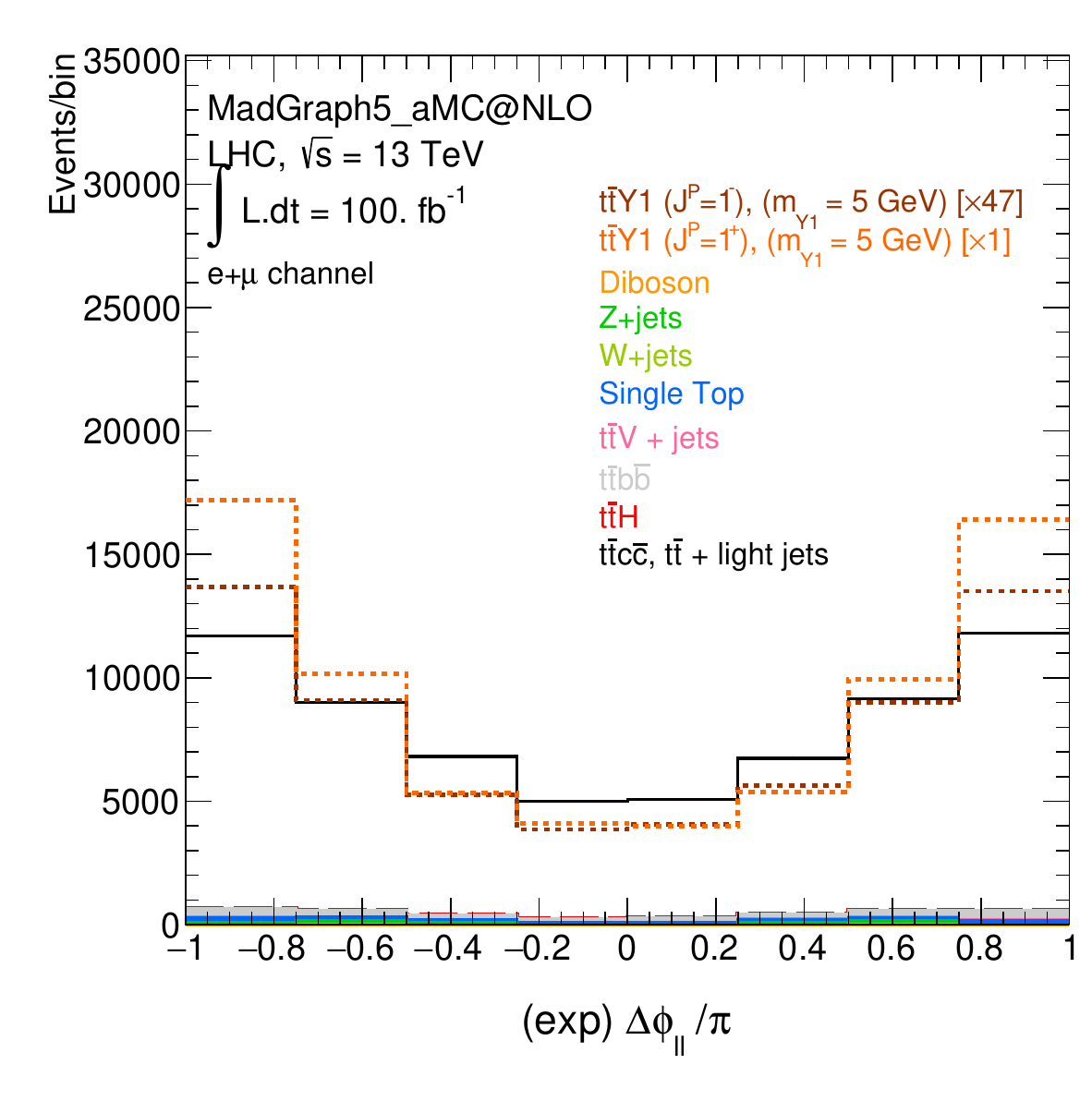}
		\includegraphics[width = 7.5cm]{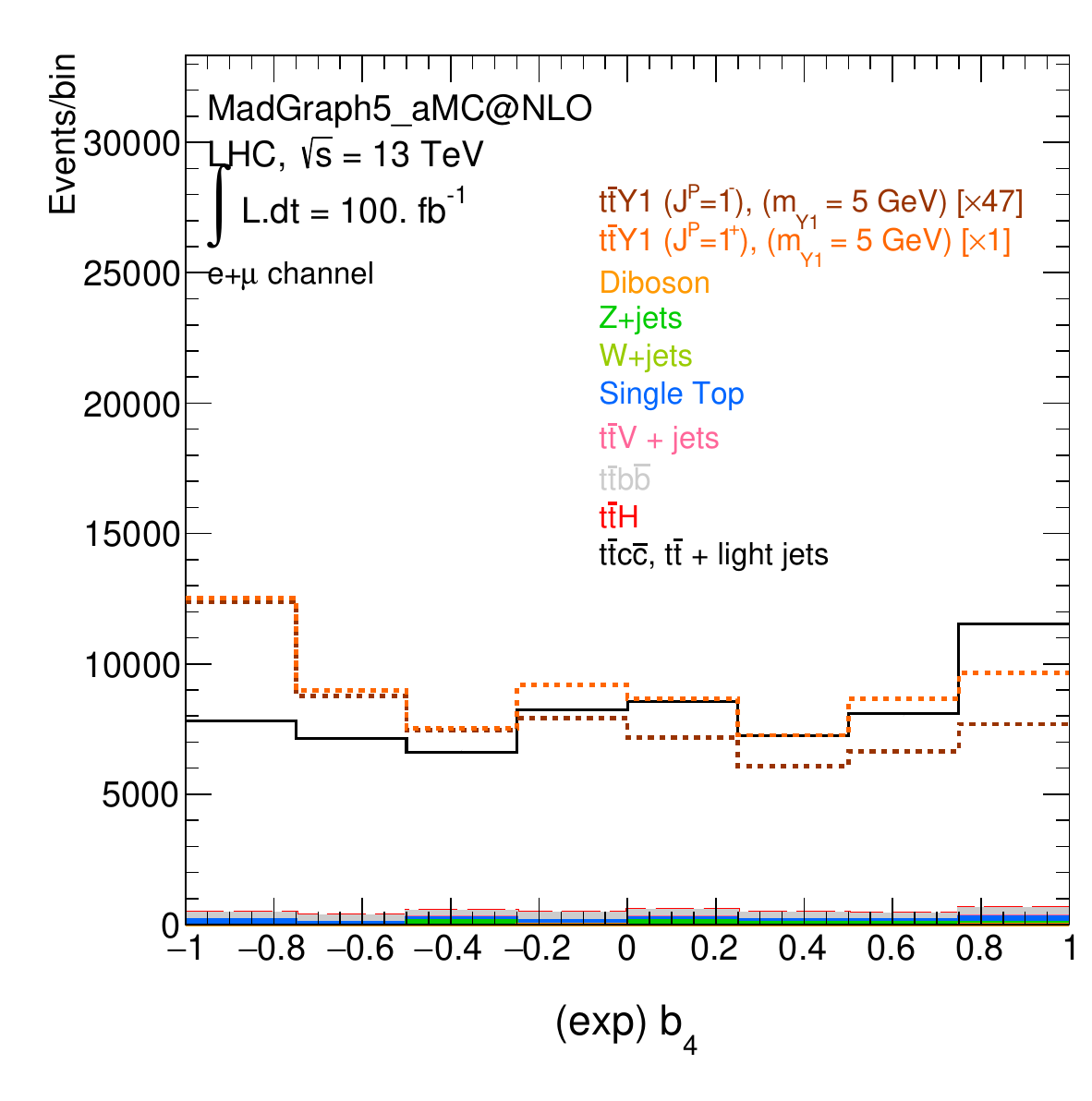}
		\caption{$\Delta \phi_{\ell^+ \ell^-}$ (left) and $b_4$ (right) distributions of the expected number of SM background (full lines), $t\bar{t}Y_{1^-}$ (brown dashed line) and $t\bar{t}Y_{1^+}$ (orange dashed line) events, after event selection and kinematic reconstruction (using the vector mediator analysis), for a reference integrated luminosity of 100 $\text{fb}^{-1}$. The DM mediator mass is $m_{Y_1}=5$ GeV. The $t\bar{t}Y_{1^-}$ signal distribution was scaled by $\times 47$ for convenience.}
		\label{fig:Stackplots1}
	\end{center}
\end{figure}

In Figure~\ref{fig:Stackplots1}, we show the $\Delta \phi_{\ell^+ \ell^-}$ (left) and $b_4$ (right) distributions of the expected number of events for each process after event selection and kinematic reconstruction, for a reference luminosity of 100~fb$^{-1}$. All considered SM backgrounds are represented --- $t\bar{t}$ (plus up to 3~jets) is labeled as ``$t\bar{t}c\bar{c}$, $t\bar{t}$+light jets"; single top quark production via $s$-, $t$- and $Wt$-channels are labeled as ``Single Top"; $W(Z)$ (plus up to 4~jets) and $W$($Z$)$b\bar{b}$ (plus up to 2~jets) are labeled as ``W(Z) + jets"; and $WW, ZZ, WZ$ diboson processes (plus up to 3 jets) as ``Diboson". The labels of the remaining background processes are self-explanatory. The $t\bar{t}Y_1$ vector (brown dashed line) and axial-vector (orange dashed line) signals, with $m_{Y_1}=5$~GeV, are shown as well. The $t\bar{t}Y_{1^-}$ signal is scaled by a factor of 47, for representation purposes only, to allow a better comparison between the SM background and the $t\bar{t}Y_{1^{\pm}}$ signals, in what concerns the distributions shapes. As expected, the dominant SM background arises from $t\bar{t}$ production, as it shares the most similar topology with the $t\bar{t}Y_1$ signal final state. All other background contributions are negligible.

\begin{figure}[H]
	\begin{center}
		\includegraphics[width = 7.5cm]{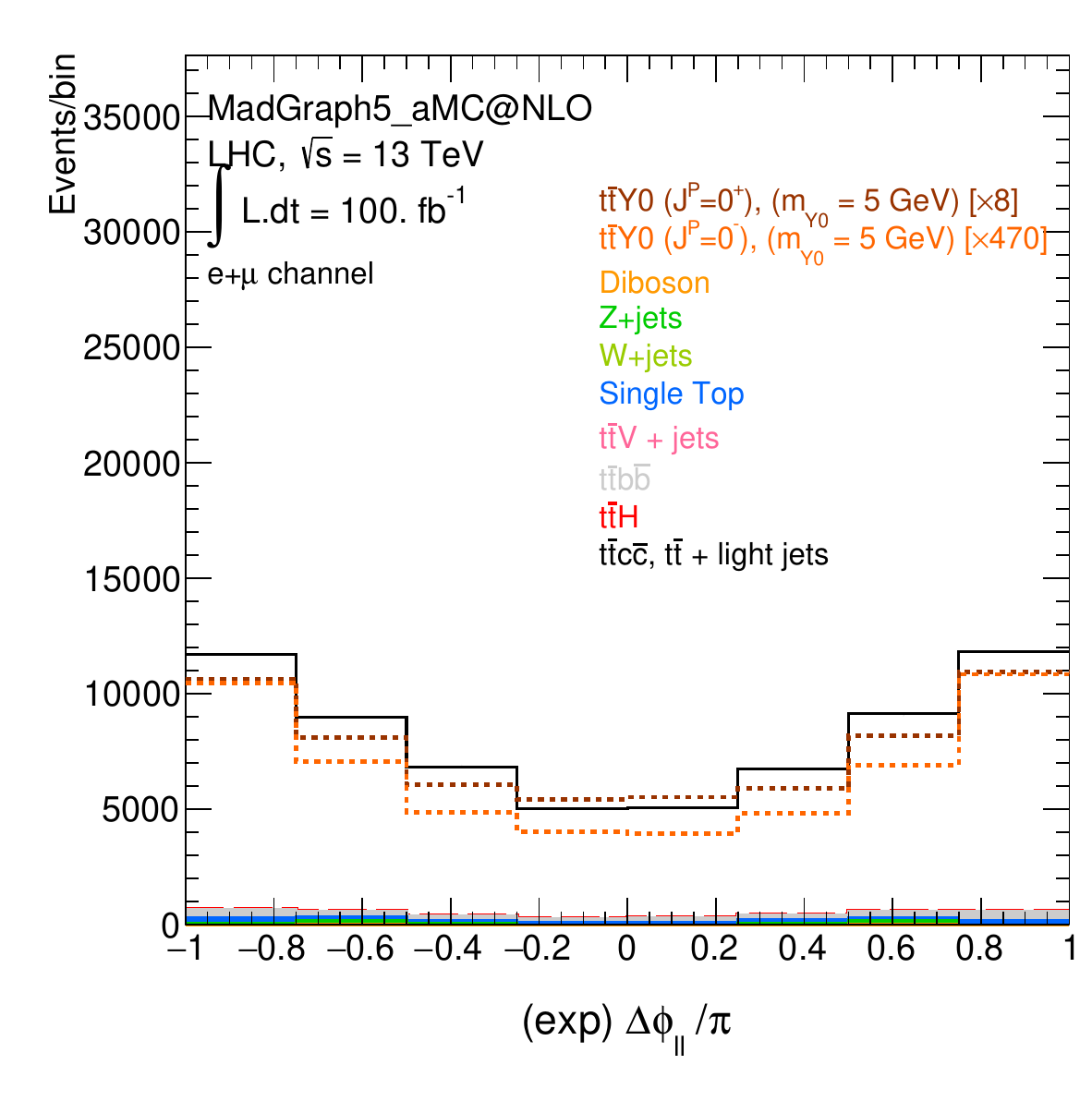}
		\includegraphics[width = 7.5cm]{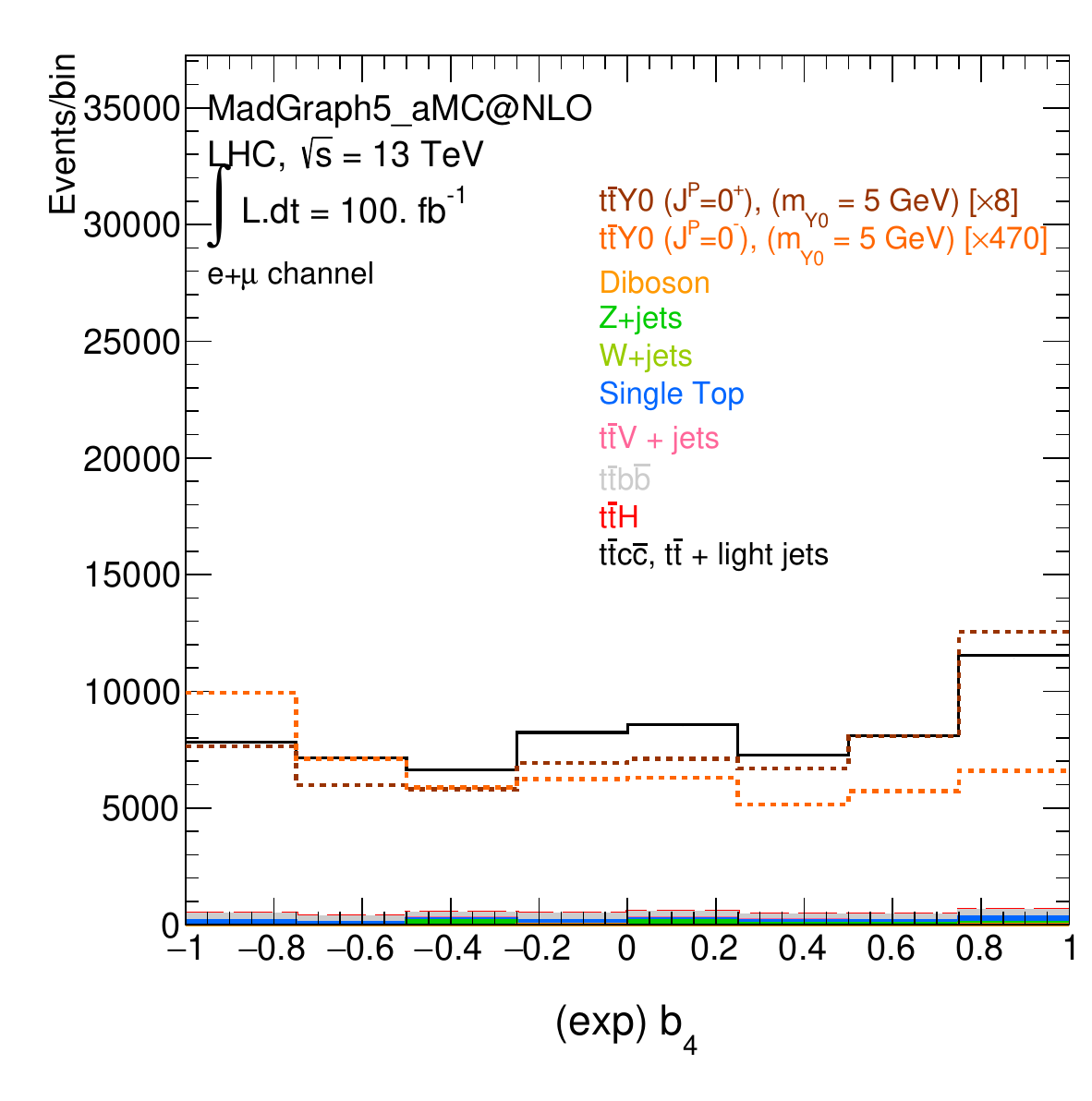}
		\caption{$\Delta \phi_{\ell^+ \ell^-}$ (left) and $b_4$ (right) distributions of the expected number of SM background (full lines), $t\bar{t}Y_{0^+}$ (brown dashed line) and $t\bar{t}Y_{0^-}$ (orange dashed line) events after event selection and kinematic reconstruction (using the vector mediator analysis), for a reference integrated luminosity of 100 $\text{fb}^{-1}$. The DM mediator mass is $m_{Y_0}=5$ GeV. The $t\bar{t}Y_{0^+}$ and $t\bar{t}Y_{0^-}$ signal distributions were respectively scaled by $\times 8$ and $\times 470$ for convenience.}
		\label{fig:Stackplots2}
	\end{center}
\end{figure}

One can observe differences in the shapes of the $t\bar{t}Y_1$ signal distributions. However, the most pronounced differences arise when comparing the SM background with the $t\bar{t}Y_1$ signals, particularly in the $b_4$ distribution. For this observable, the $t\bar{t}Y_1$ signal events tend to populate negative values more than positive ones, in contrast to the SM background, where positive values are more populated. For the $\Delta \phi_{\ell^+ \ell^-}$ distribution, the $t\bar{t}Y_1$ signal events show a stronger preference for extreme values compared to the SM background, which exhibits a relatively flatter profile. 

For completeness, Figure~\ref{fig:Stackplots2} shows the same observables for the $t\bar{t}Y_0$ scalar (brown dashed line) and pseudoscalar (orange dashed line) signals. The $t\bar{t}Y_{0^+}$ and $t\bar{t}Y_{0^-}$ distributions are scaled by factors of 8 and 470, respectively. 
The shapes of these distributions are in agreement with what was observed in Ref.~\cite{Azevedo:2023xuc} (see Fig.~5). 

The $\Delta \phi_{\ell^+ \ell^-}$ and $b_4$ distributions were then used to set confidence level (CL) limits on the exclusion of alternative hypotheses under a given null hypothesis, for different scenarios. The calculation of the CLs follows the prescription described in~\cite{Read:2002hq, Junk:1999kv}. The considered scenarios are:

\begin{itemize}
    \item \textbf{Scenario 1:} Exclusion of the SM plus a combination of pure vector and axial-vector DM mediators as the alternative (signal) hypothesis, against the SM as the null hypothesis.
    \item \textbf{Scenario 2:} Assumption of the SM plus a pure vector DM mediator, $Y_{1^-}$, as the null hypothesis. 
    \begin{itemize}
        \item \textbf{Scenario 2.1:} Exclusion of the SM plus a combination of pure vector and axial-vector DM mediators as the alternative hypothesis.
        \item \textbf{Scenario 2.2:} Exclusion of the SM plus a combination of pure vector and scalar DM mediators as the alternative hypothesis.
        \item \textbf{Scenario 2.3:} Exclusion of the SM plus a combination of pure vector and pseudoscalar DM mediators as the alternative hypothesis.
    \end{itemize}
    \item \textbf{Scenario 3:} Assumption of the SM plus a pure axial-vector DM mediator, $Y_{1^+}$, as the null hypothesis.
    \begin{itemize}
        \item \textbf{Scenario 3.1:} Exclusion of the SM plus a combination of pure axial-vector and vector DM mediators as the alternative hypothesis.
        \item \textbf{Scenario 3.2:} Exclusion of the SM plus a combination of pure axial-vector and scalar DM mediators as the alternative hypothesis.
        \item \textbf{Scenario 3.3:} Exclusion of the SM plus a combination of pure axial-vector and pseudoscalar DM mediators as the alternative hypothesis.
    \end{itemize}
\end{itemize}

The goal of Scenario 1 is to determine exclusion limits on the couplings of the spin-1 DM mediator. The main goal of Scenarios 2 and 3 is to quantify how well a mixed state can be distinguished from a pure spin-1 mediator in the event of a discovery. Furthermore, these scenarios allow us to investigate how well the analysis discriminates between spin-0 and spin-1 DM mediators.

Given that the cross section of the generated $t\bar{t}Y_{1^+}$ signal events, with $g^A_{u_{33}} = g^A_{\mathrm{SM}} = 0.25$, is very large, such a signal would likely already have been observed. Indeed, the axial-vector signal distributions are of the same order of magnitude as the $t\bar{t}$ (plus up to three jets) background shown in Figure~\ref{fig:Stackplots1}. Thus, for Scenario 3, where a pure axial-vector mediator is assumed to have been discovered, we rescale the number of $t\bar{t}Y_{1^+}$ signal events surviving the event selection and reconstruction by a factor $\lambda_r$, such that the signal corresponds to a $5\sigma$ excess over the background. This is equivalent to rescaling the coupling $g^A_{u_{33}} = g^A_{\mathrm{SM}}$ by a factor $\sqrt{\lambda_r}$, such that $g^A_{u_{33}} = g^A_{\mathrm{SM}}\sqrt{\lambda_r}$, with $\lambda_r \approx 0.0174$. 

Finally, an additional bound on the vector coupling $g^V_{u_{33}}$ can be derived from the Higgs invisible branching ratio via the loop-induced decay $h \to Y_1 Y_1$. This is discussed in Appendix~\ref{appA}.

\subsection{Scenario 1}
\hspace{\parindent} 

\begin{figure}[H]
        \hspace*{-5mm}\includegraphics[width = 8.2cm]{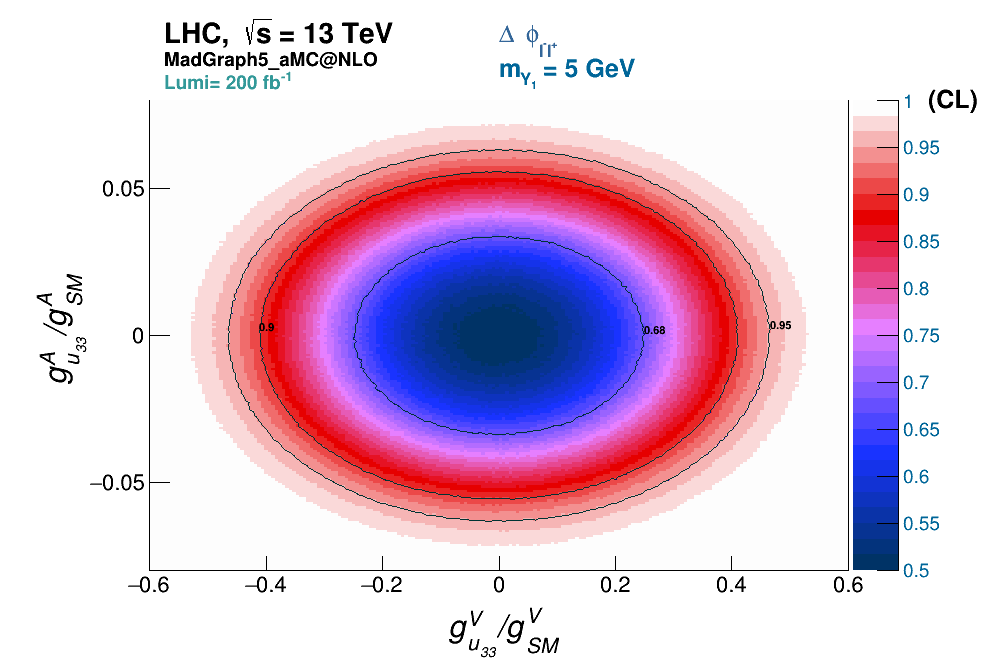}
		\includegraphics[width = 8.2cm]{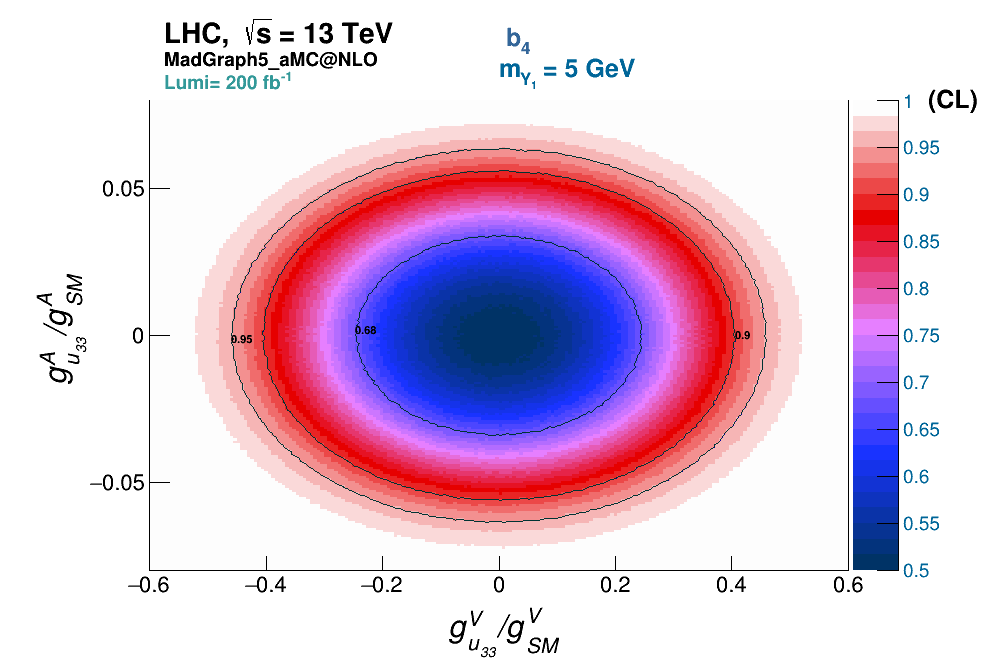}
		\caption{Contour plots of the expected CLs on $g^V_{u_{33}}/g^V_{SM}$ and $g^A_{u_{33}}/g^A_{SM}$ for the exclusion of the SM plus a combination of pure vector and axial-vector
        DM mediators with $m_{Y_1} = 5$ GeV and $g^{V/A}_{SM} = 0.25$ as the alternative hypothesis, assuming the SM as the null hypothesis. The limits are obtained using the $\Delta \phi_{\ell^+ \ell^-}$ (left) and $b_4$ (right) distributions, for an integrated luminosity of $L=200$~fb$^{-1}$.}
		\label{fig:scenario1_200fb}
\end{figure}
\begin{figure}[H]
        \hspace*{-5mm}\includegraphics[width = 8.2cm]{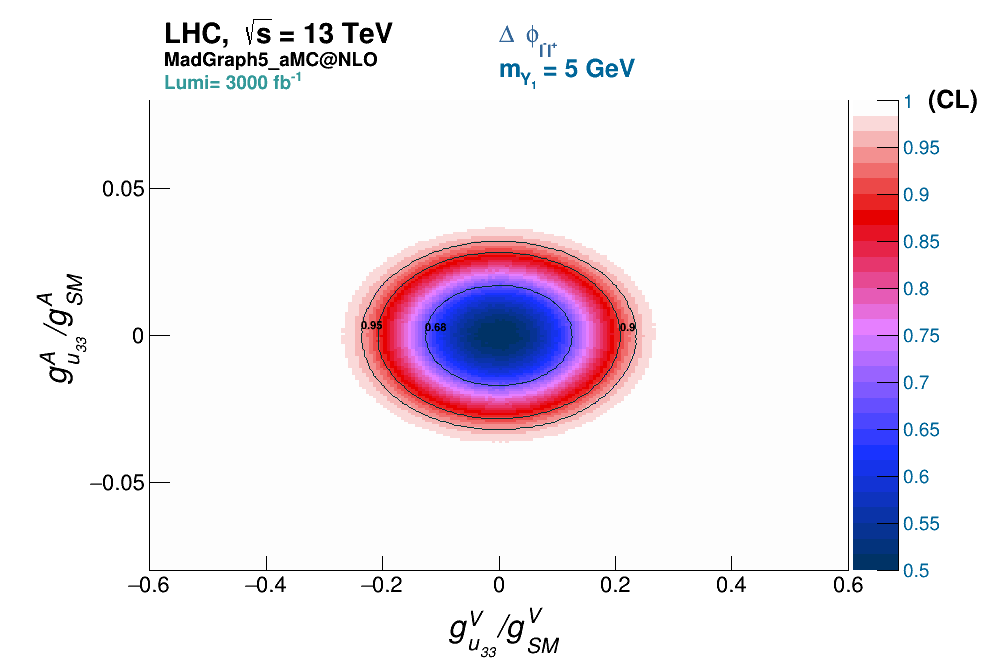}
		\includegraphics[width = 8.2cm]{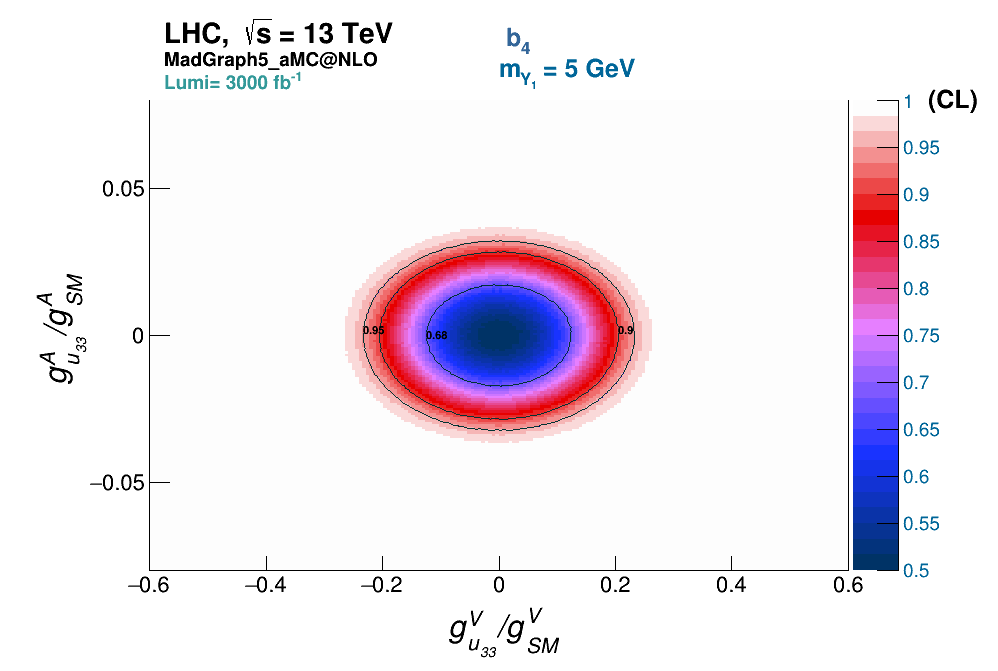}
		\caption{Contour plots of the expected CLs on $g^V_{u_{33}}/g^V_{SM}$ and $g^A_{u_{33}}/g^A_{SM}$ for the exclusion of the SM plus a combination of pure vector and axial-vector
        DM mediators with $m_{Y_1} = 5$ GeV and $g^{V/A}_{SM} = 0.25$ as the alternative hypothesis, assuming the SM as the null hypothesis. The limits are obtained using the $\Delta \phi_{\ell^+ \ell^-}$ (left) and $b_4$ (right) distributions, for an integrated luminosity of $L=3000$~fb$^{-1}$.}
		\label{fig:scenario1_3000fb}
\end{figure}

For Scenario 1, the results are shown in Figures~\ref{fig:scenario1_200fb} and~\ref{fig:scenario1_3000fb}, for integrated luminosities corresponding roughly to the RUN~2 integrated luminosity plus the contribution from the first year of RUN~3, i.e.,  $L\sim 200$~fb$^{-1}$, and to the full luminosity expected at the end of the High-Luminosity phase of the LHC (HL-LHC), i.e., $L\sim 3000$~fb$^{-1}$, respectively. 
The CLs are shown as contour plots in the $(g^i_{u_{33}}/g^i_{SM}, g^j_{u_{33}}/g^j_{SM})$ 2D plane, where $i,j = V, A, S, P$, with $i=V$ and $j=A$ for this scenario. Inspecting Figures~\ref{fig:scenario1_200fb} and~\ref{fig:scenario1_3000fb}, it is clear that the CLs are identical for both observables, as observed in Scenario 1 for the spin-0 DM mediator in Ref.~\cite{Azevedo:2023xuc}. It is also evident that the limits on the axial-vector couplings are significantly stronger than those for the vector mediator, as expected given the much larger cross section in the axial-vector case.

The resulting 68\% and 95\% exclusion limits are presented, for both luminosity values, in Table~\ref{table:exclusion_limits_scenario1_deltaphi} for the $\Delta \phi_{\ell^+ \ell^-}$ observable and in Table~\ref{table:exclusion_limits_scenario1_b4} for the $b_4$ observable. We observe that, for $L=3000$~fb$^{-1}$, the exclusion limits improve by a factor of $\sim 1/2$ relative to those obtained for $L=200$~fb$^{-1}$.

As discussed in Section~\ref{sec:TH}, perturbative unitarity imposes non-trivial constraints on the axial-vector parameter space. In particular, for $m_{Y_1}=5$~GeV and $m_t=172.5$~GeV, Eq.~\ref{eq:uni1} requires $g^A_{u_{33}} \lesssim 0.036$. This constraint is not represented in Figures~\ref{fig:scenario1_200fb} and~\ref{fig:scenario1_3000fb}, since the unitarity exclusion line on the y-axis would be placed at $g^A_{u_{33}}/g^A_{SM} \approx 0.036/0.25 = 0.144$, i.e., outside the bounds of both figures. This means that the exclusion limits set by our analysis, in this scenario, are much stronger than the limits from perturbative unitarity.

\begin{table}[H]
	\renewcommand{\arraystretch}{1.3}
	\begin{center}
		\begin{tabular}{|c|c|cc|cc|}
			
			\hline
			\multicolumn{2}{|c|}{Exclusion Limits}  &    \multicolumn{2}{c|}{$L$ = 200~fb$^{-1}$} & \multicolumn{2}{c|}{$L$ = 3000~fb$^{-1}$} \\
			
			\multicolumn{2}{|c|}{from $\Delta \phi_{\ell^+ \ell^-}$} &  (68\% CL)	& (95\% CL) & (68\% CL) & (95\% CL)  \\ \hline 
			              
			\multicolumn{2}{|c|}{$g^V_{u_{33}}/g^V_{SM} \in$} & [-0.249, 0.249]&   [-0.471, 0.471]& [-0.129, 0.123]&   [-0.237, 0.237]\\               
			  \multicolumn{2}{|c|}{$g^A_{u_{33}}/g^A_{SM} \in$} & [-0.034, 0.034] & [-0.0636, 0.0644] & [-0.0172, 0.0172] & [-0.0324, 0.0324]\\ \hline
		\end{tabular}
		\caption{Exclusion limits on $g^V_{u_{33}}/g^V_{SM}$ and $g^A_{u_{33}}/g^A_{SM}$ for the SM plus a combination of pure vector and axial-vector
        DM mediators hypothesis with $m_{Y_1}=5$ GeV and $g^{V/A}_{SM}=0.25$, assuming the SM as the null hypothesis. The limits are shown for integrated luminosities of 200~fb$^{-1}$ and 3000~fb$^{-1}$, at 68\% and 95\% CLs, using the $\Delta \phi_{\ell^+ \ell^-}$ observable.}
		\label{table:exclusion_limits_scenario1_deltaphi}
	\end{center}
\end{table}

\begin{table}[H]
	\renewcommand{\arraystretch}{1.3}
	\begin{center}
		\begin{tabular}{|c|c|cc|cc|}
			
			\hline
			\multicolumn{2}{|c|}{Exclusion Limits}  &    \multicolumn{2}{c|}{$L$ = 200~fb$^{-1}$} & \multicolumn{2}{c|}{$L$ = 3000~fb$^{-1}$} \\
			
			\multicolumn{2}{|c|}{from $b_4$} &  (68\% CL)	& (95\% CL) & (68\% CL) & (95\% CL)  \\ \hline 
			              
			\multicolumn{2}{|c|}{$g^V_{u_{33}}/g^V_{SM} \in$} & [-0.249, 0.249]&   [-0.465, 0.465] & [-0.123, 0.123] &   [-0.237, -0.237]\\               
			\multicolumn{2}{|c|}{$g^A_{u_{33}}/g^A_{SM} \in$} & [-0.034, 0.0348]& [-0.0644, 0.0644]& [-0.0172, 0.0172]& [-0.0324, 0.0324]\\ \hline          
		\end{tabular}
		\caption{Exclusion limits on $g^V_{u_{33}}/g^V_{SM}$ and $g^A_{u_{33}}/g^A_{SM}$ for the SM plus a combination of pure vector and axial-vector
        DM mediators hypothesis with $m_{Y_1}=5$ GeV and $g^{V/A}_{SM}=0.25$, assuming the SM as the null hypothesis. The limits are shown for integrated luminosities of 200~fb$^{-1}$ and 3000~fb$^{-1}$, at 68\% and 95\% CLs, using the $b_4$ observable.}
		\label{table:exclusion_limits_scenario1_b4}
	\end{center}
\end{table}

\subsection{Scenario 2}
\subsubsection{Scenario 2.1}
\hspace{\parindent} 

\begin{figure}[H]
        \hspace*{-5mm}\includegraphics[width = 8.2cm]{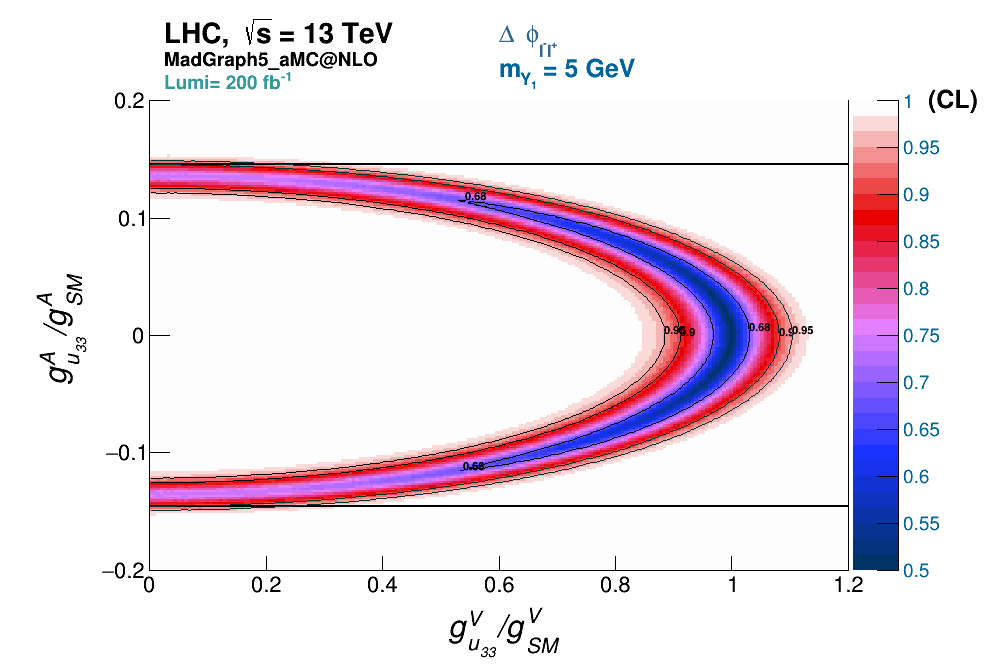}
		\includegraphics[width = 8.2cm]{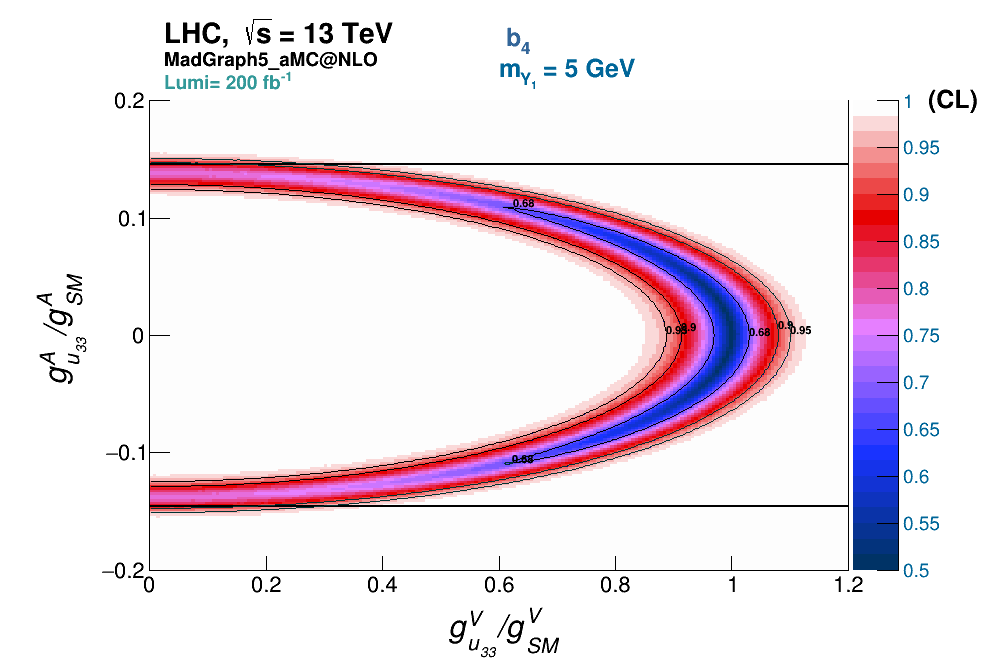}
		\caption{Contour plots of the expected CLs on $g^V_{u_{33}}/g^V_{SM}$ and $g^A_{u_{33}}/g^A_{SM}$ for the exclusion of the SM plus 
        a combination of pure vector and axial-vector
        DM mediators with $m_{Y_1} = 5$ GeV and $g^{V/A}_{SM} = 0.25$ as the alternative hypothesis, assuming the SM plus a pure vector DM mediator as the null hypothesis. The limits are obtained using the $\Delta \phi_{\ell^+ \ell^-}$ (left) and $b_4$ (right) distributions, for an integrated luminosity of $L=200$~fb$^{-1}$. The black lines represent the coupling values under which the perturbative unitarity constraint condition is satisfied for the axial-vector mediator, $g^A_{u_{33}}/g^A_{SM} \approx 0.145$.}
		\label{fig:scenario2.1_V_200fb}
\end{figure}
\begin{figure}[H]
        \hspace*{-5mm}\includegraphics[width = 8.2cm]{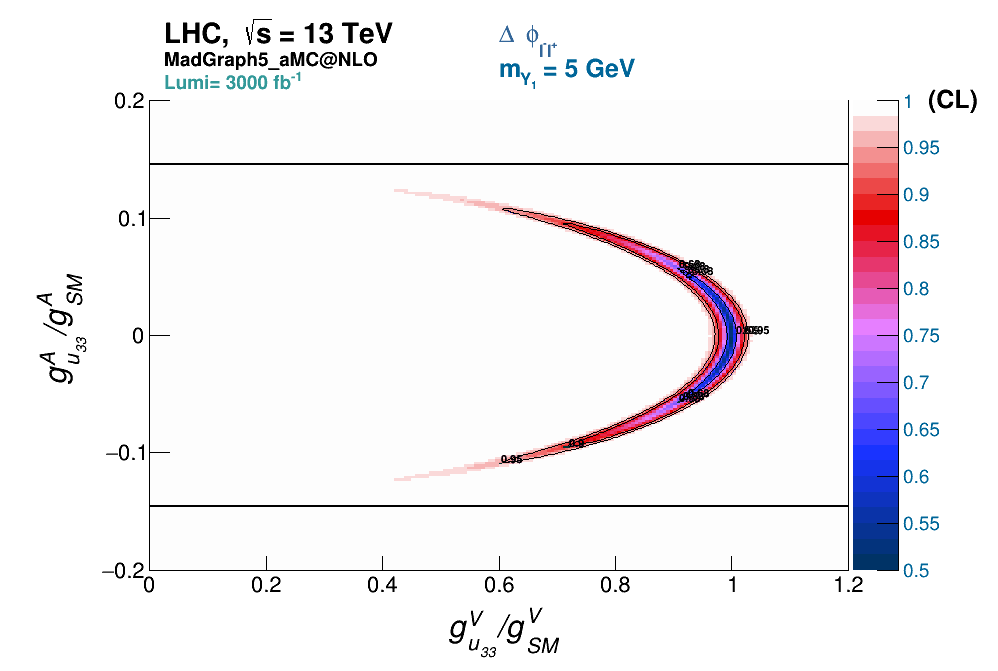}
		\includegraphics[width = 8.2cm]{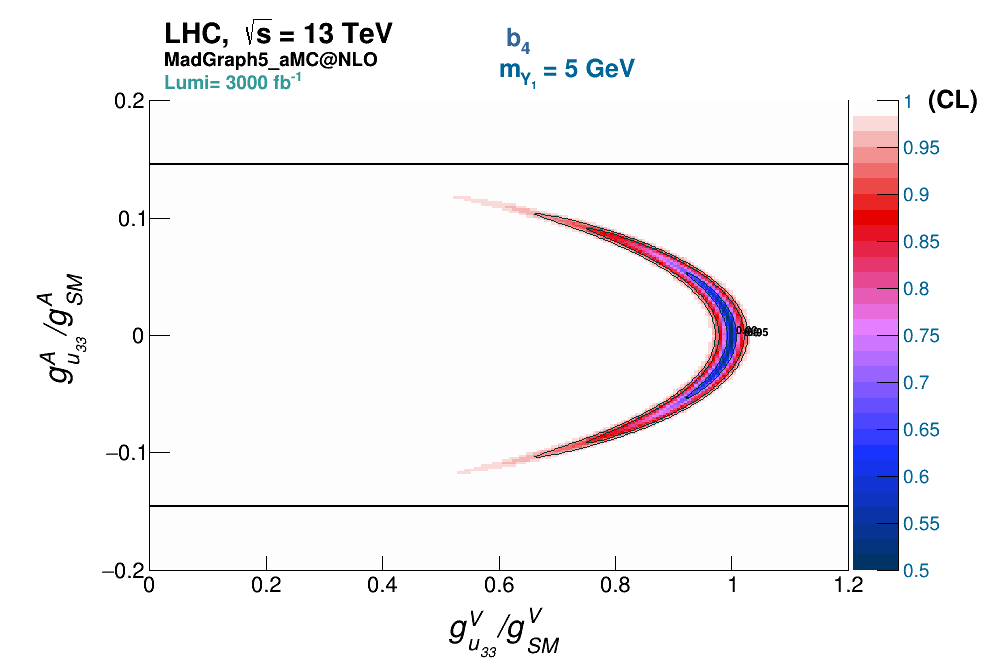}
		\caption{Contour plots of the expected CLs on $g^V_{u_{33}}/g^V_{SM}$ and $g^A_{u_{33}}/g^A_{SM}$ for the exclusion of the SM plus
        a combination of pure vector and axial-vector
        DM mediators with $m_{Y_1} = 5$ GeV and $g^{V/A}_{SM} = 0.25$ as the alternative hypothesis, assuming the SM plus a pure vector DM mediator as the null hypothesis. The limits are obtained using the $\Delta \phi_{\ell^+ \ell^-}$ (left) and $b_4$ (right) distributions, for an integrated luminosity of $L=3000$~fb$^{-1}$. The black lines represent the coupling values under which the perturbative unitarity constraint condition is satisfied for the axial-vector mediator, $g^A_{u_{33}}/g^A_{SM} \approx 0.145$.}
		\label{fig:scenario2.1_V_3000fb}
\end{figure}

The expected CLs for the exclusion of the SM plus a combination of pure vector and axial-vector
DM mediators as the alternative hypothesis, when assuming the discovery of a pure vector DM mediator, are shown in Figures~\ref{fig:scenario2.1_V_200fb} and~\ref{fig:scenario2.1_V_3000fb}, for $L=200$~fb$^{-1}$ and $L=3000$~fb$^{-1}$, respectively. 
In this case, the $b_4$ observable provides the strongest exclusion limits for both luminosities, although the difference is marginal. 

For both observables, the unitarity constraint, pictured in both Figures~\ref{fig:scenario2.1_V_200fb} and~\ref{fig:scenario2.1_V_3000fb} as the black lines corresponding to $|g^A_{u_{33}}/g^A_{SM}| \approx 0.0363/0.25 \approx 0.145$, excludes the highest axial-vector coupling values. For $L=200$~fb$^{-1}$, we observe in Figure~\ref{fig:scenario2.1_V_200fb} that the exclusion limits from perturbative unitary are slightly stronger than the 95\% CL limits from our analysis, for large axial-vector and small vector couplings. For $L=3000$~fb$^{-1}$, the 95\% CL limits from our analysis are the stronger ones, as can be seen in Figure~\ref{fig:scenario2.1_V_3000fb}.

\subsubsection{Scenario 2.2}
\hspace{\parindent}

\begin{figure}[H]
        \hspace*{-5mm}\includegraphics[width = 8.2cm]{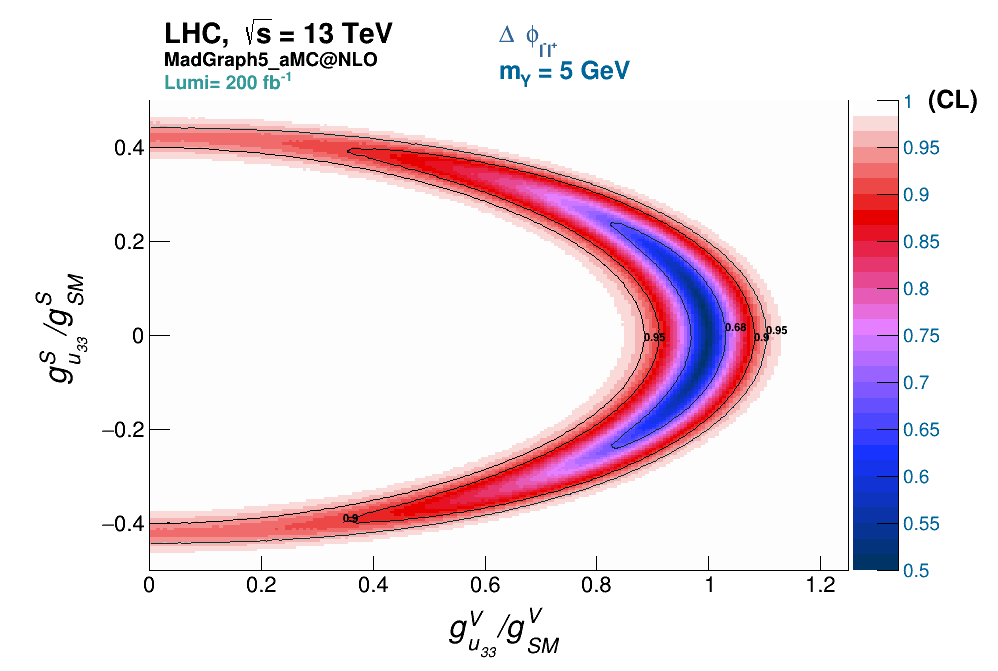}
		\includegraphics[width = 8.2cm]{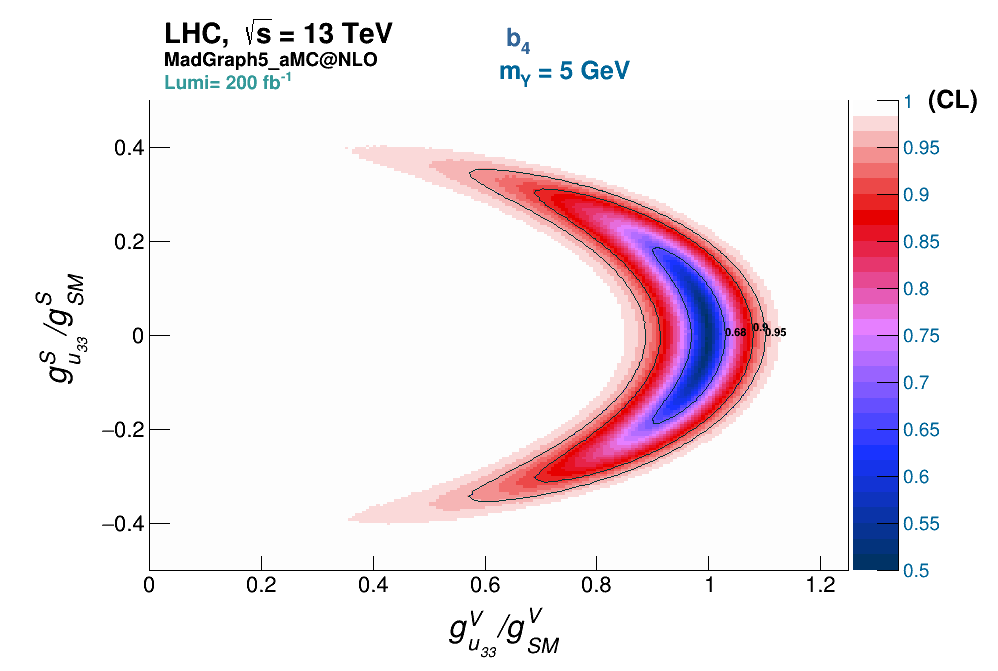}
		\caption{Contour plots of the expected CLs on $g^V_{u_{33}}/g^V_{SM}$ and $g^S_{u_{33}}/g^S_{SM}$ for the exclusion of the SM plus
        a combination of pure vector and scalar
        DM mediators with $m_{Y} = 5$ GeV, $g^{V}_{SM} = 0.25$ and $g^{S}_{SM} = 1$ as the alternative hypothesis, assuming the SM plus a pure vector DM mediator as the null hypothesis. The limits are obtained using the $\Delta \phi_{\ell^+ \ell^-}$ (left) and $b_4$ (right) distributions, for an integrated luminosity of $L=200$~fb$^{-1}$.}
		\label{fig:scenario2.2_V_200fb}
\end{figure}

\begin{figure}[H]
        \hspace*{-5mm}\includegraphics[width = 8.2cm]{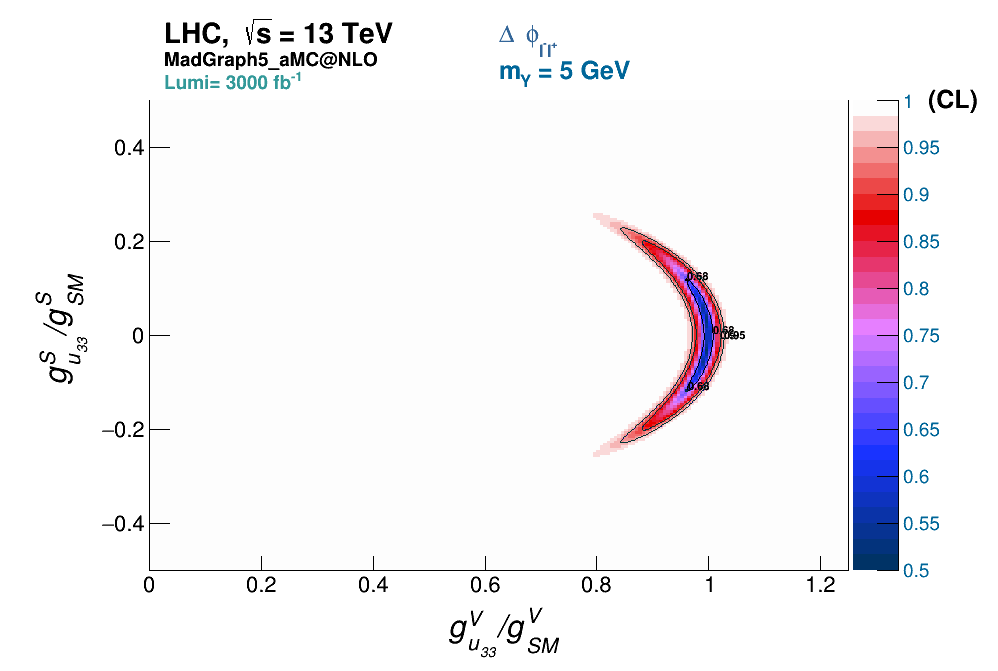}
		\includegraphics[width = 8.2cm]{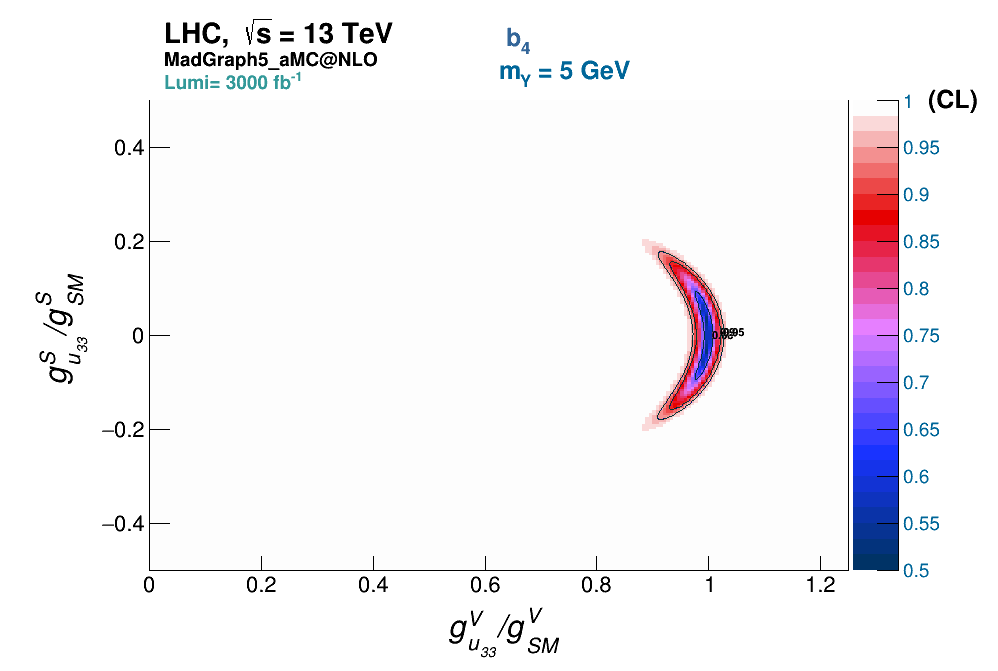}
		\caption{Contour plots of the expected CLs on $g^V_{u_{33}}/g^V_{SM}$ and $g^S_{u_{33}}/g^S_{SM}$ for the exclusion of the SM plus 
        a combination of pure vector and scalar
        DM mediators with $m_{Y} = 5$ GeV, $g^{V}_{SM} = 0.25$ and $g^{S}_{SM} = 1$ as the alternative hypothesis, assuming the SM plus a pure vector DM mediator as the null hypothesis. The limits are obtained using the $\Delta \phi_{\ell^+ \ell^-}$ (left) and $b_4$ (right) distributions, for an integrated luminosity of $L=3000$~fb$^{-1}$.}
		\label{fig:scenario2.2_V_3000fb}
\end{figure}

The expected CLs for the exclusion of the SM plus a
a combination of pure vector and scalar
DM mediators as the alternative hypothesis, when assuming the discovery of a pure vector DM mediator, are shown in Figures~\ref{fig:scenario2.2_V_200fb} and~\ref{fig:scenario2.2_V_3000fb}, for $L=200$~fb$^{-1}$ and $L=3000$~fb$^{-1}$, respectively. 
In this scenario, the $b_4$ observable yields significantly stronger exclusion limits than $\Delta \phi_{\ell^+ \ell^-}$, making it the more suitable of the two observables to probe the existence of a scalar component associated with a vector DM mediator in our analysis.

\subsubsection{Scenario 2.3}
\hspace{\parindent}

\begin{figure}[H]
        \hspace*{-5mm}\includegraphics[width = 8.2cm]{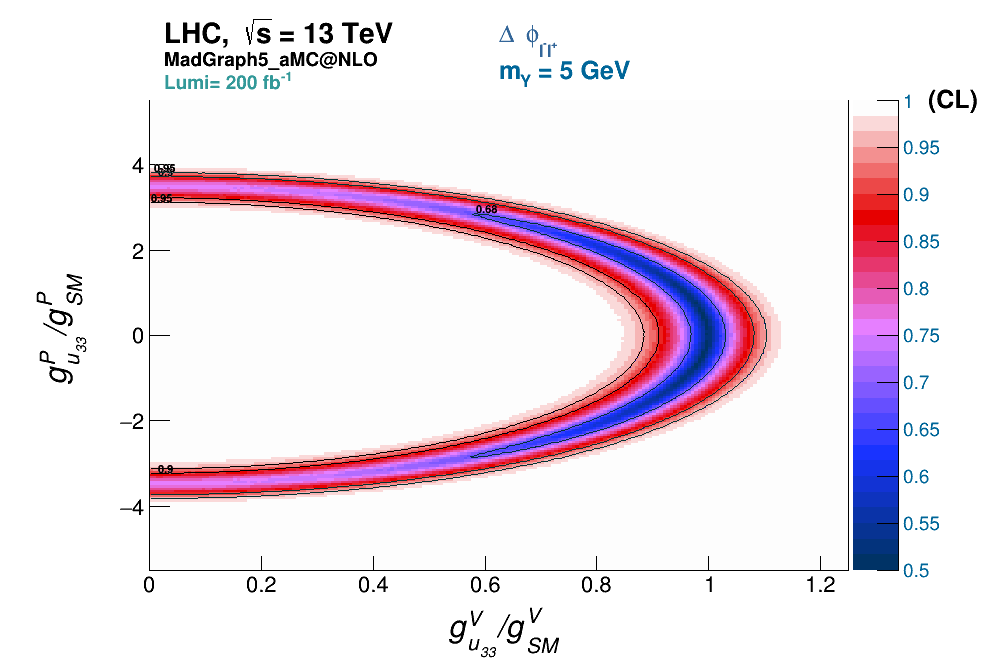}
		\includegraphics[width = 8.2cm]{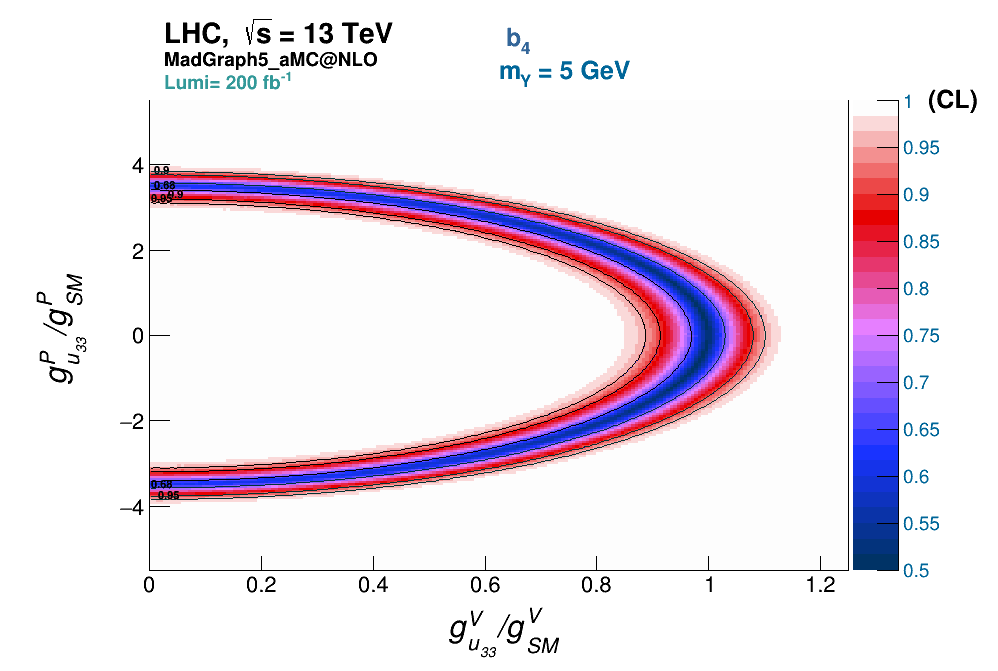}
		\caption{Contour plots of the expected CLs on $g^V_{u_{33}}/g^V_{SM}$ and $g^P_{u_{33}}/g^P_{SM}$ for the exclusion of the SM plus 
        a combination of pure vector and pseudoscalar
        DM mediators with $m_{Y} = 5$ GeV, $g^{V}_{SM} = 0.25$ and $g^{P}_{SM} = 1$ as the alternative hypothesis, assuming the SM plus a pure vector DM mediator as the null hypothesis. The limits are obtained using the $\Delta \phi_{\ell^+ \ell^-}$ (left) and $b_4$ (right) distributions, for an integrated luminosity of $L=200$~fb$^{-1}$.}
		\label{fig:scenario2.3_V_200fb}
\end{figure}

\begin{figure}[H]
        \hspace*{-5mm}\includegraphics[width = 8.2cm]{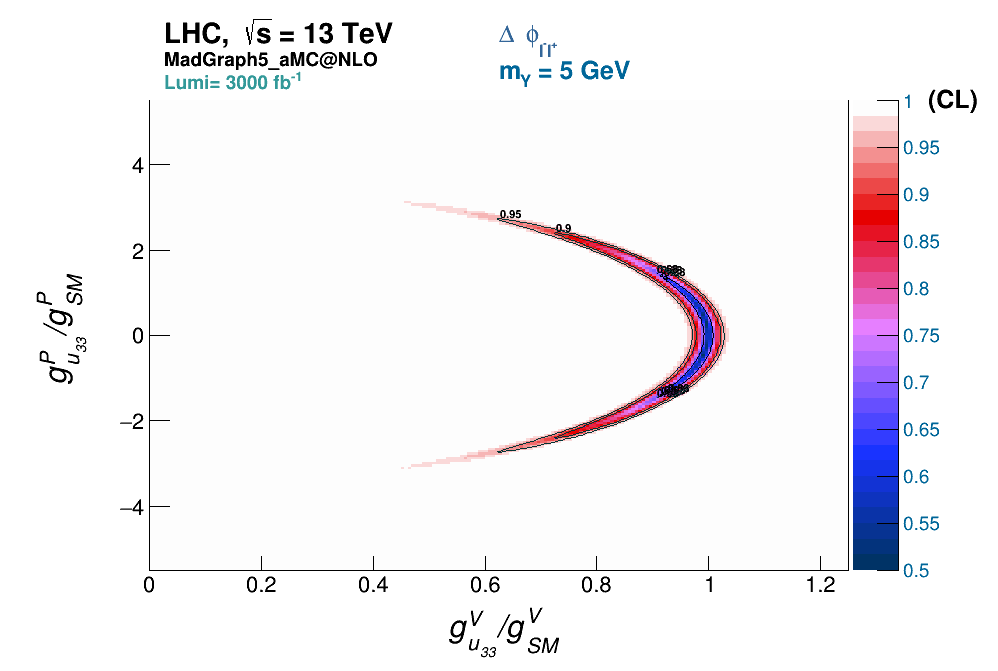}
		\includegraphics[width = 8.2cm]{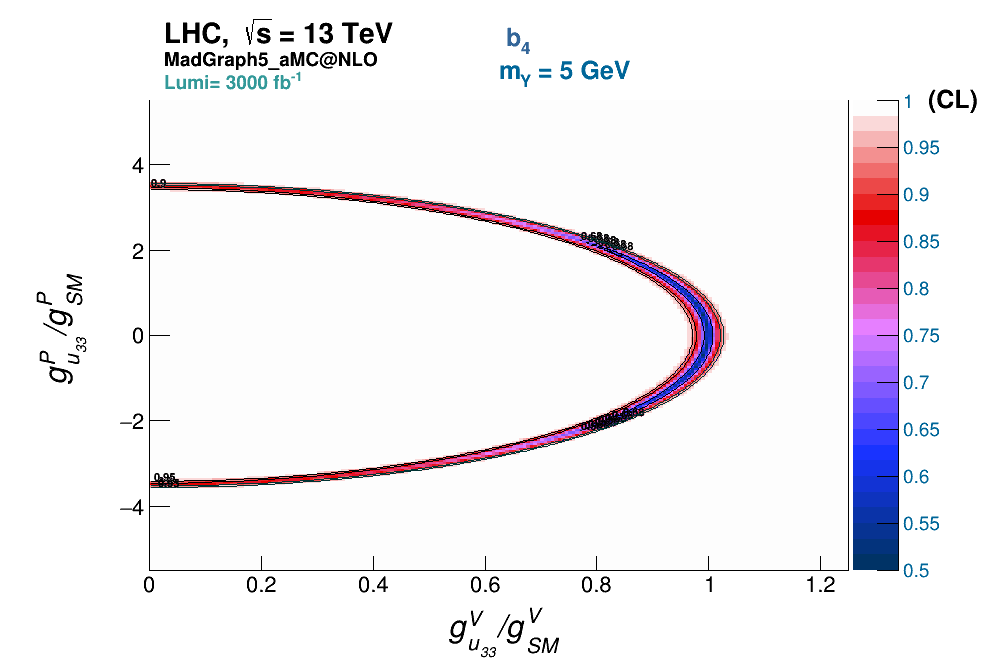}
		\caption{Contour plots of the expected CLs on $g^V_{u_{33}}/g^V_{SM}$ and $g^P_{u_{33}}/g^P_{SM}$ for the exclusion of the SM plus 
        a combination of pure vector and pseudoscalar
        DM mediators with $m_{Y} = 5$ GeV, $g^{V}_{SM} = 0.25$ and $g^{P}_{SM} = 1$ as the alternative hypothesis, assuming the SM plus a pure vector DM mediator as the null hypothesis. The limits are obtained using the $\Delta \phi_{\ell^+ \ell^-}$ (left) and $b_4$ (right) distributions, for an integrated luminosity of $L=3000$~fb$^{-1}$.}
		\label{fig:scenario2.3_V_3000fb}
\end{figure}

The expected CLs for the exclusion of the SM plus 
a combination of pure vector and pseudoscalar
DM mediators as the alternative hypothesis, when assuming the discovery of a pure vector DM mediator, are shown in Figures~\ref{fig:scenario2.3_V_200fb} and~\ref{fig:scenario2.3_V_3000fb}, for $L=200$~fb$^{-1}$ and $L=3000$~fb$^{-1}$, respectively. 

Unlike in the previous cases, here the $\Delta \phi_{\ell^+ \ell^-}$ observable performs better than $b_4$. This behaviour can be understood by considering the distributions shown in the right panels of Figures~\ref{fig:Stackplots1} and~\ref{fig:Stackplots2}. While the scalar and vector mediators exhibit clearly distinct shapes in $b_4$, the pseudoscalar and vector cases display much more similar distributions, both favoring negative values more than positive ones. As a result, the discriminating power of $b_4$ is significantly reduced when attempting to identify a pseudoscalar component associated with a vector mediator. In contrast, since the $\Delta \phi_{\ell^+ \ell^-}$ distributions for the scalar and pseudoscalar mediators are quite similar (see the left panel of Figure~\ref{fig:Stackplots2}), the performance of this observable degrades less severely in the pseudoscalar case. Consequently, in this situation $\Delta \phi_{\ell^+ \ell^-}$ becomes the more effective observable.

The results for $L=200$~fb$^{-1}$ (Fig.~\ref{fig:scenario2.3_V_200fb}) further show that, independently of the observable used, it is not possible to exclude the possibility that the mediator assumed to be a vector is instead a pseudoscalar, as both hypotheses become difficult to distinguish in specific regions of the parameter space. In the extreme case where $g^V_{u_{33}}/g^V_{\mathrm{SM}} = 0$, large values of the pseudoscalar coupling are still allowed. At higher luminosity (see the left panel of Fig.~\ref{fig:scenario2.3_V_3000fb}), the possibility of a vanishing vector coupling is excluded.

These examples highlight the importance of angular observables in improving exclusion limits and, ultimately, in determining the properties of new particles, such as their spin and parity. They also illustrate that the choice of observable is crucial, as their sensitivity strongly depends on the underlying hypothesis being tested.

\subsection{Scenario 3}
\label{subsec:scenario3}
\subsubsection{Scenario 3.1}
\hspace{\parindent}

\begin{figure}[H]
        \hspace*{-5mm}\includegraphics[width = 8.2cm]{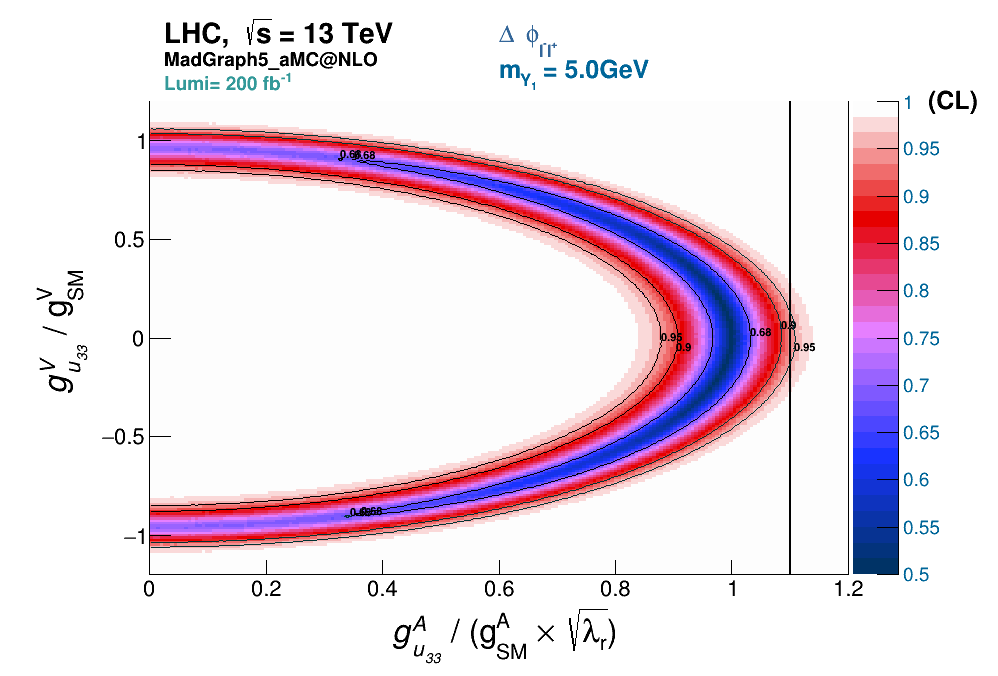}
		\includegraphics[width = 8.2cm]{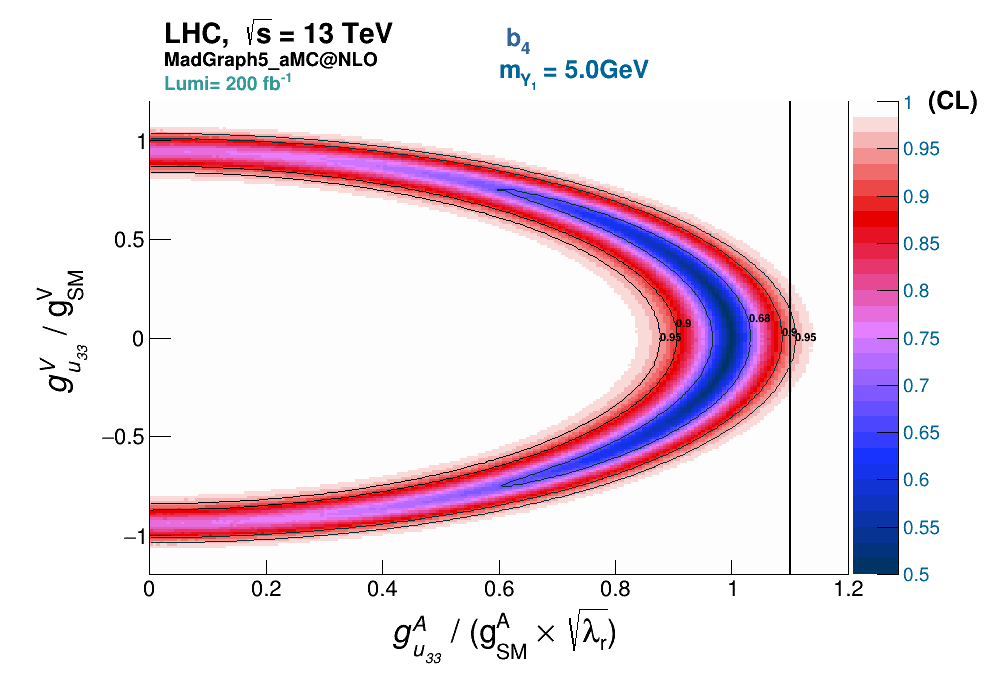}
		\caption{Contour plots of the expected CLs on $g^A_{u_{33}}/(g^A_{SM}\sqrt{\lambda_r})$ and $g^V_{u_{33}}/g^V_{SM}$ for the exclusion of the SM plus a combination of pure axial-vector and vector DM mediators with $m_{Y_1} = 5$ GeV and $g^{V/A}_{SM} = 0.25$ as the alternative hypothesis, assuming the SM plus a pure axial-vector DM mediator as the null hypothesis. The limits are obtained using the $\Delta \phi_{\ell^+ \ell^-}$ (left) and $b_4$ (right) distributions, for an integrated luminosity of $L=200$~fb$^{-1}$. The black line represents the coupling value under which the pertubative unitarity condition is satisfied for the axial-vector mediator, $g^A_{u_{33}}/(g^A_{SM} \sqrt{\lambda_r}) \approx 1.09$.}
		\label{fig:scenario3.1_A_200fb}
\end{figure}

\begin{figure}[H]
        \hspace*{-5mm}\includegraphics[width = 8.2cm]{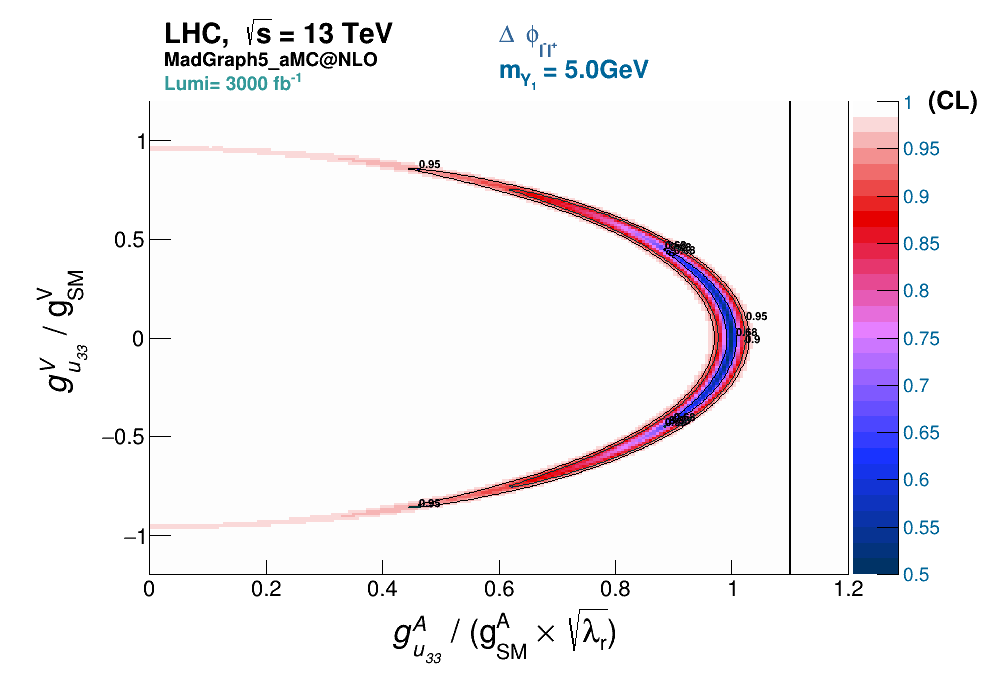}
		\includegraphics[width = 8.2cm]{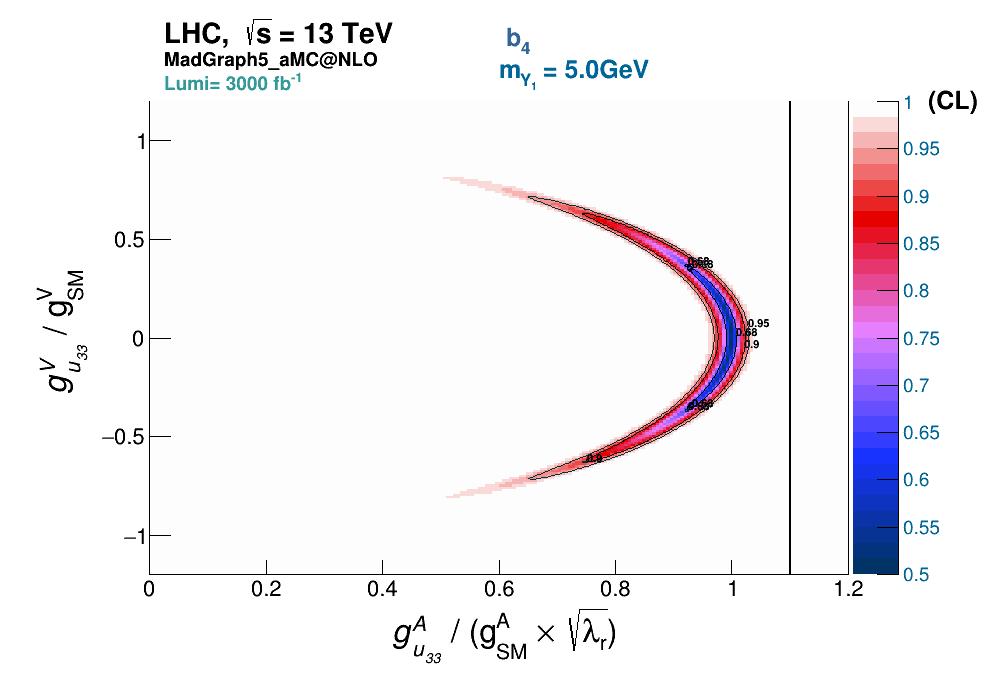}
		\caption{Contour plots of the expected CLs on $g^A_{u_{33}}/(g^A_{SM}\sqrt{\lambda_r})$ and $g^V_{u_{33}}/g^V_{SM}$ for the exclusion of the SM plus a combination of pure axial-vector and vector DM mediators with $m_{Y_1} = 5$ GeV and $g^{V/A}_{SM} = 0.25$ as the alternative hypothesis, assuming the SM plus a pure axial-vector DM mediator as the null hypothesis. The limits are obtained using the $\Delta \phi_{\ell^+ \ell^-}$ (left) and $b_4$ (right) distributions, for an integrated luminosity of $L=3000$~fb$^{-1}$. The black line represents the coupling value under which the pertubative unitarity condition is satisfied for the axial-vector mediator, $g^A_{u_{33}}/(g^A_{SM} \sqrt{\lambda_r}) \approx 1.09$.}
		\label{fig:scenario3.1_A_3000fb}
\end{figure}

Contrary to the previous cases, and as mentioned above, in Scenario~3 all presented results are obtained from the axial-vector analysis. The expected CLs for the exclusion of the SM plus a combination of pure axial-vector and vector DM mediators as the alternative hypothesis, when assuming the discovery of a pure axial-vector DM mediator, are shown in Figures~\ref{fig:scenario3.1_A_200fb} and~\ref{fig:scenario3.1_A_3000fb} for $L=200$~fb$^{-1}$ and $L=3000$~fb$^{-1}$, respectively. 
In this case, the $b_4$ observable provides stronger exclusion limits more clearly than in the corresponding vector case (Scenario~2.1). From the results of Scenarios~2.1 and~3.1, the $b_4$ observable appears to be the most sensitive probe of the parity ($P$) nature of a spin-1 DM mediator within our analysis.

For both observables, for $L=200$~fb$^{-1}$, the unitarity constraint, pictured in Figures~\ref{fig:scenario3.1_A_200fb} and~\ref{fig:scenario3.1_A_3000fb} as a black line corresponding to $g^A_{u_{33}}/(g^A_{SM} \sqrt{\lambda_r}) \approx 0.036/(0.25 \sqrt{0.0174}) \approx 1.09$, excludes the highest axial-vector coupling values. Thus, for this luminosity, the exclusion limits from perturbative unitarity are slightly stronger than the 95\% CL limits from our analysis, for high axial-vector and small vector couplings. This is no longer the case for $L=3000$~fb$^{-1}$, as can be seen in Figure~\ref{fig:scenario3.1_A_3000fb}. These observations remain true for the subsequent scenarios. 


\subsubsection{Scenario 3.2}
\hspace{\parindent}

\begin{figure}[H]
        \hspace*{-5mm}\includegraphics[width = 8.2cm]{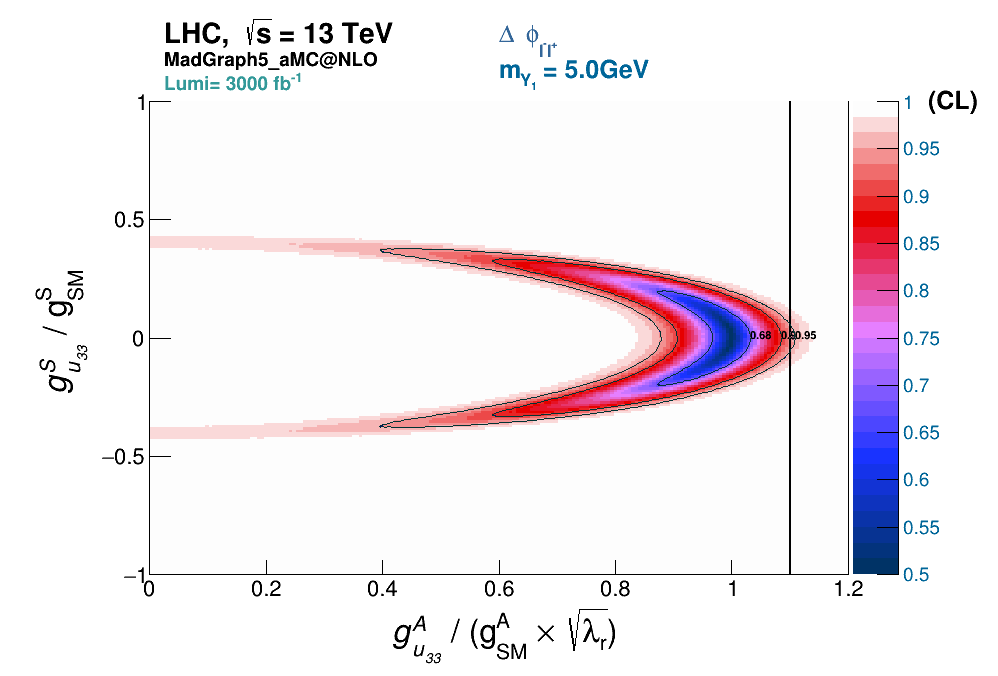}
		\includegraphics[width = 8.2cm]{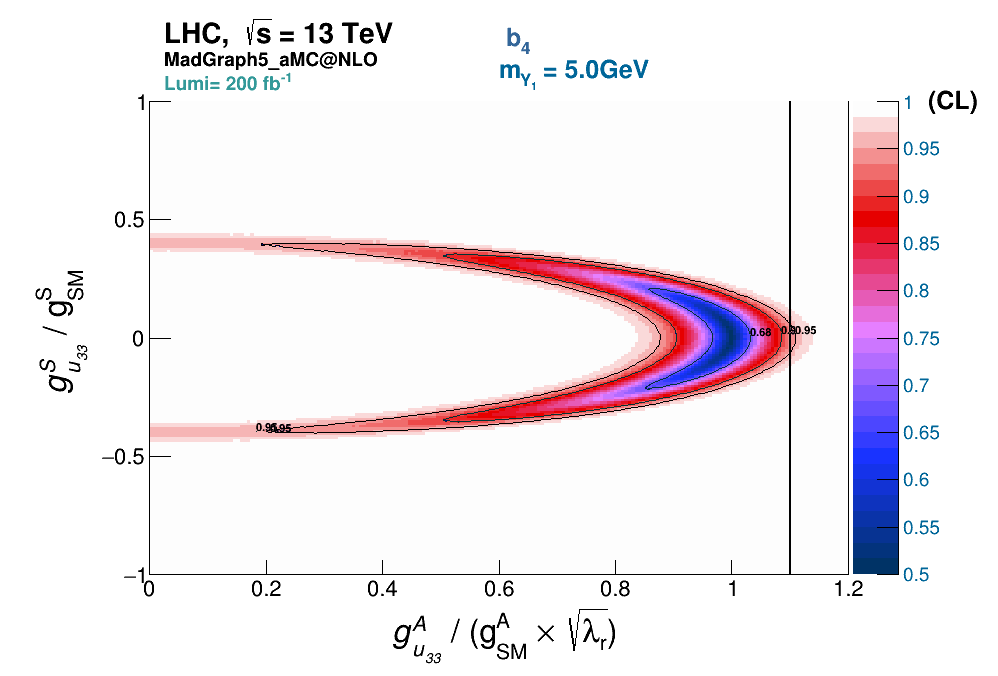}
		\caption{Contour plots of the expected CLs on $g^A_{u_{33}}/(g^A_{SM}\sqrt{\lambda_r})$ and $g^S_{u_{33}}/g^S_{SM}$ for the exclusion of the SM plus a combination of pure axial-vector and scalar DM mediators with $m_{Y} = 5$ GeV, $g^{A}_{SM} = 0.25$ and $g^{S}_{SM} = 1$ as the alternative hypothesis, assuming the SM plus a pure axial-vector DM mediator as the null hypothesis. The limits are obtained using the $\Delta \phi_{\ell^+ \ell^-}$ (left) and $b_4$ (right) distributions, for an integrated luminosity of $L=200$~fb$^{-1}$. The black line represents the coupling value under which the pertubative unitarity condition is satisfied for the axial-vector mediator, $g^A_{u_{33}}/(g^A_{SM} \sqrt{\lambda_r}) \approx 1.09$.}
		\label{fig:scenario3.2_A_200fb}
\end{figure}

\begin{figure}[H]
        \hspace*{-5mm}\includegraphics[width = 8.2cm]{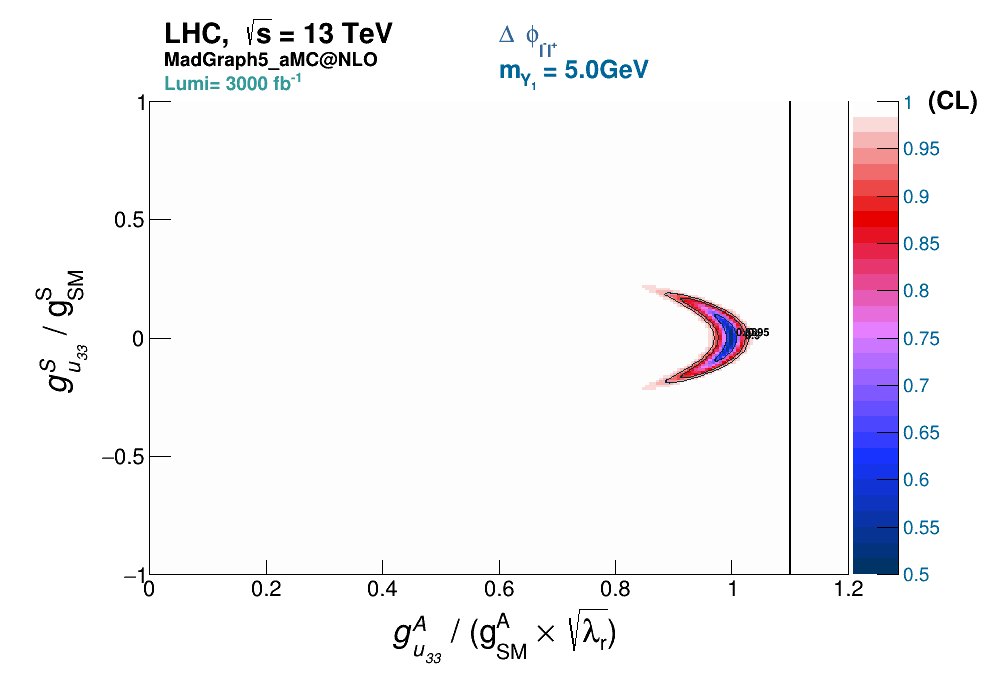}
		\includegraphics[width = 8.2cm]{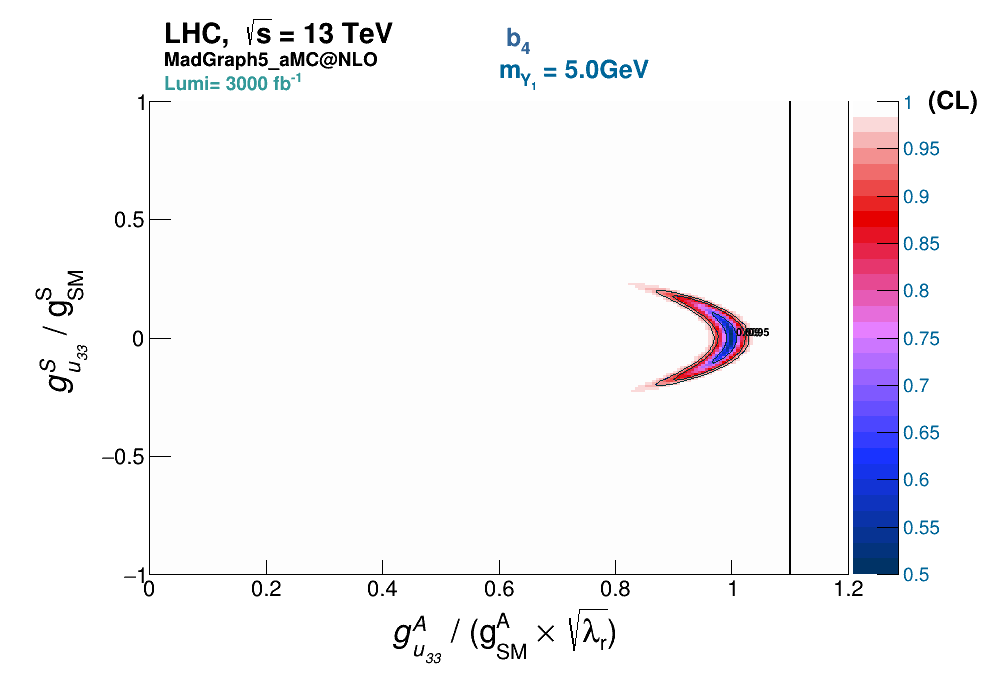}
		\caption{Contour plots of the expected CLs on $g^A_{u_{33}}/(g^A_{SM}\sqrt{\lambda_r})$ and $g^S_{u_{33}}/g^S_{SM}$ for the exclusion of the SM plus a combination of pure axial-vector and scalar DM mediators with $m_{Y} = 5$ GeV, $g^{A}_{SM} = 0.25$ and $g^{S}_{SM} = 1$ as the alternative hypothesis, assuming the SM plus a pure axial-vector DM mediator as the null hypothesis. The limits are obtained using the $\Delta \phi_{\ell^+ \ell^-}$ (left) and $b_4$ (right) distributions, for an integrated luminosity of $L=3000$~fb$^{-1}$. The black line represents the coupling value under which the pertubative unitarity condition is satisfied for the axial-vector mediator, $g^A_{u_{33}}/(g^A_{SM} \sqrt{\lambda_r}) \approx 1.09$.}
		\label{fig:scenario3.2_A_3000fb}
\end{figure}

The expected CLs for the exclusion of the SM plus a combination of pure axial-vector and scalar DM mediators as the alternative hypothesis, when assuming the discovery of a pure axial-vector DM mediator, are shown in Figures~\ref{fig:scenario3.2_A_200fb} and~\ref{fig:scenario3.2_A_3000fb}, for $L=200$~fb$^{-1}$ and $L=3000$~fb$^{-1}$, respectively. 
Here, the differences between the observables are not as evident as in Scenario~2.2, although the 95\% CL limits for the axial-vector couplings are much better for the $\Delta \phi_{\ell^+ \ell^-}$ observable at $L=200$~fb$^{-1}$.

\subsubsection{Scenario 3.3}
\hspace{\parindent}

\begin{figure}[H]
        \hspace*{-5mm}\includegraphics[width = 8.2cm]{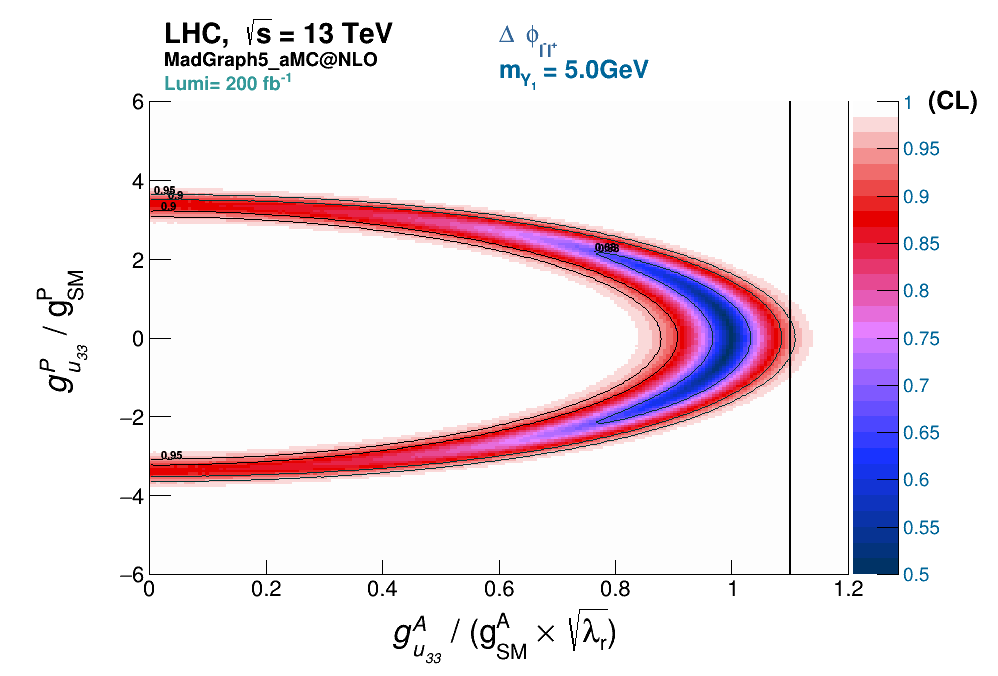}
		\includegraphics[width = 8.2cm]{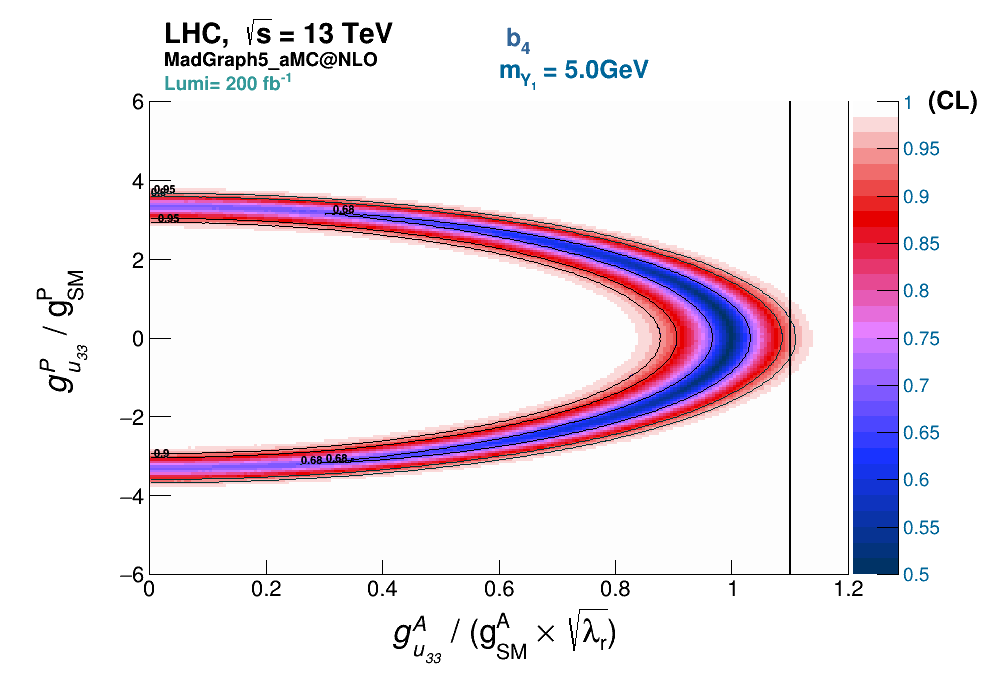}
		\caption{Contour plots of the expected CLs on $g^A_{u_{33}}/(g^A_{SM}\sqrt{\lambda_r})$ and $g^P_{u_{33}}/g^P_{SM}$ for the exclusion of the SM plus a combination of pure axial-vector and pseudoscalar DM mediators with $m_{Y} = 5$ GeV, $g^{A}_{SM} = 0.25$ and $g^{P}_{SM} = 1$ as the alternative hypothesis, assuming the SM plus a pure axial-vector DM mediator as the null hypothesis. The limits are obtained using the $\Delta \phi_{\ell^+ \ell^-}$ (left) and $b_4$ (right) distributions, for an integrated luminosity of $L=200$~fb$^{-1}$. The black line represents the coupling value under which the pertubative unitarity condition is satisfied for the axial-vector mediator, $g^A_{u_{33}}/(g^A_{SM} \sqrt{\lambda_r}) \approx 1.09$.}
		\label{fig:scenario3.3_A_200fb}
\end{figure}

\begin{figure}[H]
        \hspace*{-5mm}\includegraphics[width = 8.2cm]{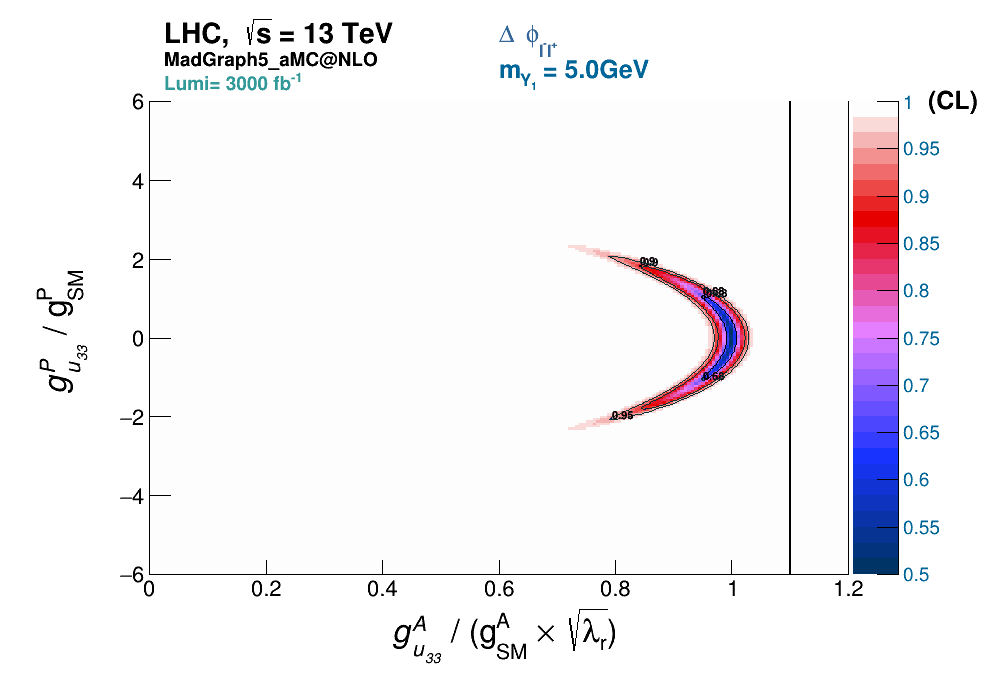}
		\includegraphics[width = 8.2cm]{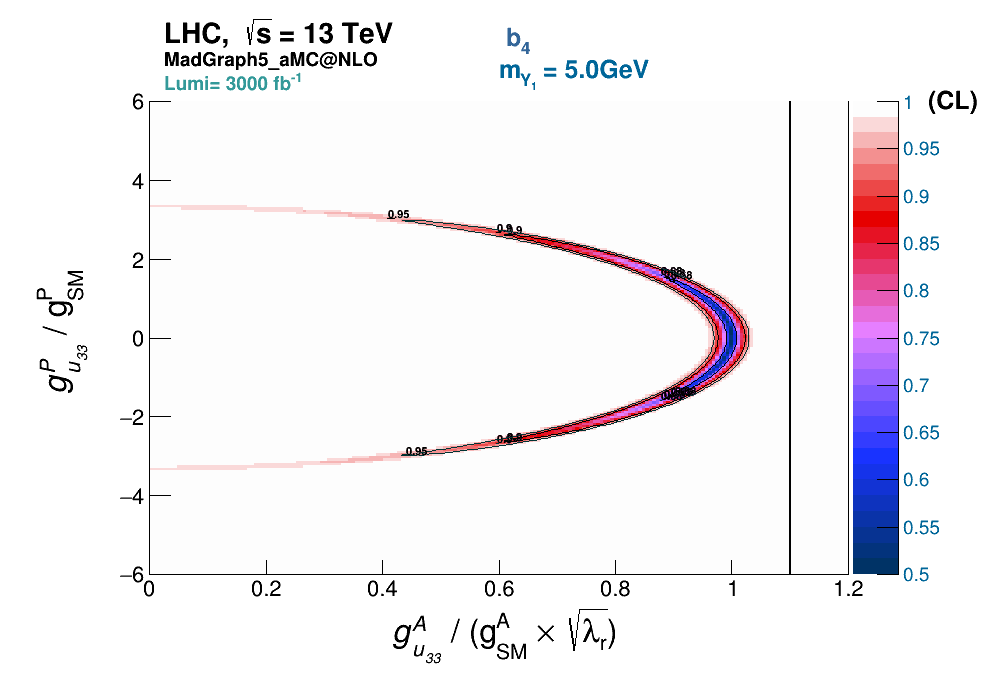}
		\caption{Contour plots of the expected CLs on $g^A_{u_{33}}/(g^A_{SM}\sqrt{\lambda_r})$ and $g^P_{u_{33}}/g^P_{SM}$ for the exclusion of the SM plus a combination of pure axial-vector and pseudoscalar DM mediators with $m_{Y} = 5$ GeV, $g^{A}_{SM} = 0.25$ and $g^{P}_{SM} = 1$ as the alternative hypothesis, assuming the SM plus a pure axial-vector DM mediator as the null hypothesis. The limits are obtained using the $\Delta \phi_{\ell^+ \ell^-}$ (left) and $b_4$ (right) distributions, for an integrated luminosity of $L=3000$~fb$^{-1}$. The black line represents the coupling value under which the pertubative unitarity condition is satisfied for the axial-vector mediator, $g^A_{u_{33}}/(g^A_{SM} \sqrt{\lambda_r}) \approx 1.09$.}
		\label{fig:scenario3.3_A_3000fb}
\end{figure}

The expected CLs for the exclusion of the SM plus a combination of pure axial-vector and pseudoscalar DM mediators as the alternative hypothesis, when assuming the discovery of a pure axial-vector DM mediator, are shown in Figures~\ref{fig:scenario3.3_A_200fb} and~\ref{fig:scenario3.3_A_3000fb}, for $L=200$~fb$^{-1}$ and $L=3000$~fb$^{-1}$, respectively. 
Here, as in Scenario~2.3, the $\Delta \phi_{\ell^+ \ell^-}$ observable provides the strongest exclusion limits. Consequently, the $\Delta \phi_{\ell^+ \ell^-}$ observable is the most sensitive observable to probe the presence of a pseudoscalar component associated with an axial-vector DM mediator, using our analysis.

\section{Conclusions \label{sec:conclusion}}
\hspace{\parindent} 

The recent observations by the ATLAS and CMS collaborations of an enhancement in the $t\bar t$ production rate near threshold have renewed interest in new physics scenarios~\cite{ATLAS:2026nrx, CMS:2025kzt}. Although the light mediators considered in this work are not intended to provide a direct explanation of the observed excess, the measurements highlight the importance of exploring the phenomenology of top-philic states and their possible impact on precision top-quark observables.
 As measurements of the $t\bar t$ threshold region continue to improve, together with the increasing precision of Higgs studies at the LHC and future colliders, the parameter space explored here will become increasingly testable. These searches therefore provide a complementary probe of top-philic dark sectors beyond the SM.

In this paper, we have extended our previous study of invisible scalar mediators produced in association with a $t\bar{t}$ pair at the LHC to the case of spin-1 mediators. Within the framework of simplified DM models, we considered both vector and axial-vector mediators, as well as scalar and pseudoscalar hypotheses, focusing on a light mediator mass of $m_{Y}=5$~GeV. The analysis was performed in the dileptonic $t\bar{t}$ final state, where the full kinematic reconstruction of the $t\bar{t}$ system was achieved without explicitly reconstructing the invisible mediator. We have shown that the kinematic reconstruction remains effective in the presence of a spin-1 mediator. Although its efficiency is reduced when compared to the spin-0 case, a significant fraction of signal events survives the reconstruction, indicating that standard $t\bar{t}$ analyses are sensitive to such scenarios.

We explored the sensitivity of angular observables, in particular $\Delta \phi_{\ell^+ \ell^-}$ and $b_4$, to the presence of new invisible particles and to their spin and CP properties, using these observables to set exclusion limits for several scenarios and integrated luminosities of 200~fb$^{-1}$ and 3000~fb$^{-1}$. We have shown that angular observables play an important role in improving exclusion limits for most scenarios and in probing the properties of new particles, and that their effectiveness depends strongly on the underlying hypothesis being tested. 

At the same time, we have shown that there are regions of parameter space that remain unconstrained, and that different hypotheses can be difficult to distinguish at lower luminosities. Although higher luminosity improves the discriminating power, highlighting the importance of the HL-LHC program, a non-negligible portion of the parameter space remains beyond reach. Given that our kinematic reconstruction is quite efficient regardless of the mediator type, this implies that, in some cases, new invisible particles could be present in $t\bar{t}$ final states without producing clearly distinguishable signatures, and may therefore evade detection within this analysis.

Since we are considering the possibility of both light mediators and light dark matter candidates, it is important to examine the constraints arising from dark matter experiments. This is not always straightforward, as a given simplified model may admit several different ultraviolet completions, leading to different phenomenological implications. For this reason, we discuss the constraints derived from the Higgs invisible width in the Appendix, which provide a useful estimate of the size of the couplings involved. Other constraints, such as those arising from the observed dark matter relic density, have been studied in Ref.~\cite{Albert:2022xla}. 
However, these constraints are not relevant in our case, as the dark matter mass lies below the kinematic threshold for annihilation into top-quark pairs.

Overall, searches in $t\bar{t}$ final states at the LHC provide a powerful framework not only for the discovery of invisible particles, but also for probing their fundamental properties. Nevertheless, identifying such signals in this channel alone can be challenging, emphasizing the need for complementary observables and strategies to search for potential new physics at the LHC.

\vspace{3mm}
\begin{center}
\textbf{Acknowledgements} 
\end{center}
\vspace{-1mm}

The authors thank Igor Ivanov and Thomas Biekötter for useful discussions and suggestions, as well as to Gonçalo Freitas for his participation in the early stages of this project. R.S. and R.C. acknowledge financial support from the Portuguese Foundation for Science and Technology (FCT)  under contracts: UID/PRR2/00618/2025 (https://doi.org/10.54499/UID
/PRR2/00618/2025), UID/PRR/00618/2025 (https://doi.org/
10.54499/UID/PRR/00618/2025), and UID/00618/2025 (https://doi.org/10.54499/UID/00618/2025) and through the project with the reference 2024.03328.CERN. R.C. is additionally supported by FCT with a PhD Grant No.~2020.08221.BD. AO acknowledges financial support by CF-UM-UP through the Strategic Fundings UIDB/04650/2020, UIDP/04650/2020, UID/PRR/04650/2025 and UID/04650/2025 from FCT.

\appendix

\section{Higgs Decay into Vector DM Mediators at One Loop}
\label{appA}
\hspace{\parindent} 

Since both the mediator and the DM particles are very light, one could ask how the Higgs invisible width would impact on the mediator couplings to top quarks. The simplified model could be turned into a full renormalisable model in many different ways. Hence, the constraint is very model dependent. We can however study the particular case of the mediator with spin one and vector coupling to the tops.

In this appendix we study the loop-induced decay of the Higgs boson into pairs of dark-sector mediators within the simplified model in which the mediator couples directly only to the top quark and to dark matter. Since no tree-level Higgs-mediator interaction is present, the decay $h \rightarrow XX$, where ($X$) denotes a scalar, pseudoscalar, vector, or axial-vector mediator, is generated at one loop through a top-quark triangle diagram.
The structure of the resulting amplitudes depends strongly on the Lorentz nature of the mediator. For scalar, pseudoscalar, and axial-vector mediators, the loop amplitudes contain ultraviolet divergences that can be associated with local effective operators sharing the same quantum numbers as the corresponding final states. In contrast, the amplitude for Higgs decay into a pair of vector mediators is found to be finite. This behaviour can be traced to the conservation of the vector current in the fermion loop, which imposes Ward identities on the amplitude and removes the ultraviolet divergent contribution. 

\begin{figure}[H]
	\begin{center}
        \includegraphics[width=5cm]{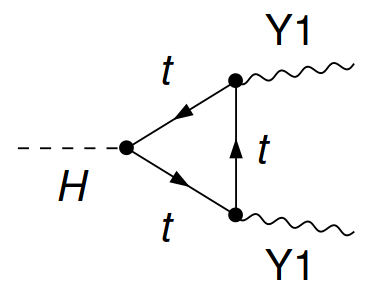}
		\caption{Feynman diagram for the $h \to Y_1 Y_1$ decay.}
		\label{fig:hY1Y1}
	\end{center}
\end{figure}

We will now derive an additional bound on the Yukawa coupling of the pure vector DM mediator, $g^V_{u_{33}}$. As stated, this bound comes from the Higgs invisible branching ratio, BR$(h \to \text{inv})$. The ATLAS collaboration currently sets an upper bound on this quantity of $\text{BR}(h \to \text{inv}) \leq 0.11$~\cite{ATLAS:2023tkt}, and CMS reports $\text{BR}(h \to \text{inv}) \leq 0.15$~\cite{CMS:2023sdw}. Although the Higgs does not have a tree-level coupling to the DM mediator, this decay can occur at the one-loop level, since both of these particles couple to the top quarks. The Feynman diagram for this process is shown in Fig.~\ref{fig:hY1Y1}.

The decay width for $h \to Y_1 Y_1$, in the limit of a massless mediator, is given by
\begin{equation}
    \Gamma(h \to Y_1 Y_1) = \frac{1}{512 \pi^5} \frac{(g^V_{u_{33}})^4 \, (y^t_{33} m_t N_c)^2}{m_h} \left|(m_h^2 - 4m_t^2) C_0(m_h^2, 0, 0, m_t^2, m_t^2, m_t^2) - 2\right|^2 \, ,
    \label{eq:width_hY1Y1}
\end{equation}
where $N_c$ denotes the number of colors and $C_0(m_h^2, 0, 0, m_t^2, m_t^2, m_t^2)$ is the scalar three-point Passarino-Veltman scalar function~\cite{1979NuPhB.160..151P}, given by
\begin{equation}
    C_0(m_h^2, 0, 0, m_t^2, m_t^2, m_t^2) = \frac{1}{2m_h^2} \ln^2\left(\frac{m_h \sqrt{m_h^2 - 4m_t^2} - m_h^2 + 2m_t^2}{2m_t^2}\right) \, .
\end{equation}

After some algebra, one can show that Equation~\ref{eq:width_hY1Y1} reproduces the standard expression for the Higgs decay into two photons, upon the replacement $g^V_{u_{33}} \to Q_t e$ (see, e.g., Eq.~1 of~\cite{Djouadi:1993ji}). For a massive mediator, the full expression becomes more involved and will not be displayed here. However, numerically, it yields very close results to the massless limit for small mediator masses, which is the case in this work since $m_{Y_1} = 5$~GeV.

Applying the bound from Higgs invisible decays, 
\begin{equation}
    \text{BR}(h \to \text{inv}) = \frac{\Gamma(h \to Y_1 Y_1)}{\Gamma(h \to \text{SM}) +  \Gamma(h \to Y_1 Y_1)} \leq 0.11 \, , 
\end{equation}
we obtain $|g^V_{u_{33}}| \leq 1.039$ ($|g^V_{u_{33}}| \leq 1.037$) for $m_{Y_1} = 5$~GeV ($m_{Y_1} = 0$~GeV) and $\Gamma(h \to \text{SM}) = 4.1$~MeV. This limit is weaker than the ones obtained from all the scenarios studied in this paper, in some cases by more than an order of magnitude. For example, in Scenario 1, we find $|g^V_{u_{33}}| \leq 0.237 \times g^V_\text{SM} \approx 0.059$ at 95\% CL and for $L = 3000$~fb$^{-1}$ (see Table~\ref{table:exclusion_limits_scenario1_b4}).

As discussed, the corresponding calculation for scalar, pseudoscalar and axial-vector mediators is not presented here, as their one-loop amplitudes are not finite for the simplified DM models considered in this work. We would need to complete the model in a specific manner, to calculate the process after renormalisation, which goes beyond the scope of this work.

\vspace*{1cm}
\bibliographystyle{h-physrev}
\bibliography{papernovo.bib}

@article{Albert:2022xla,
    author = "Albert, Andreas and others",
    title = "{Displaying dark matter constraints from colliders with varying simplified model parameters}",
    eprint = "2203.12035",
    archivePrefix = "arXiv",
    primaryClass = "hep-ph",
    month = "3",
    year = "2022"
}

@article{Azevedo:2023xuc,
    author = "Azevedo, Duarte and Capucha, Rodrigo and Chaves, Pedro and Martins, Jo{\~a}o Bravo and Onofre, Ant{\'o}nio and Santos, Rui",
    title = "{Search for an invisible scalar in $ t\overline{t} $ final states at the LHC}",
    eprint = "2308.00819",
    archivePrefix = "arXiv",
    primaryClass = "hep-ph",
    doi = "10.1007/JHEP11(2023)125",
    journal = "JHEP",
    volume = "11",
    pages = "125",
    year = "2023"
}

@article{Flacke:2025dwk,
    author = "Flacke, Thomas and Fuks, Benjamin and Kim, Dongchan and Kim, Jinheung and Lee, Seung J. and Munoz-Aillaud, L{\'e}andre",
    title = "{New physics in toponium's shadow?}",
    eprint = "2512.03220",
    archivePrefix = "arXiv",
    primaryClass = "hep-ph",
    reportNumber = "KIAS-Q25020",
    journal = "",
    month = "12",
    year = "2025"
}

@article{CMS:2025kzt,
    author = "Hayrapetyan, Aram and others",
    collaboration = "CMS",
    title = "{Observation of a pseudoscalar excess at the top quark pair production threshold}",
    eprint = "2503.22382",
    archivePrefix = "arXiv",
    primaryClass = "hep-ex",
    reportNumber = "CMS-TOP-24-007, CERN-EP-2025-061",
    doi = "10.1088/1361-6633/adf7d3",
    journal = "Rept. Prog. Phys.",
    volume = "88",
    number = "8",
    pages = "087801",
    year = "2025"
}

@article{Sjostrand:2006za,
    author = "Sjostrand, Torbjorn and Mrenna, Stephen and Skands, Peter Z.",
    title = "{PYTHIA 6.4 Physics and Manual}",
    eprint = "hep-ph/0603175",
    archivePrefix = "arXiv",
    reportNumber = "FERMILAB-PUB-06-052-CD-T, LU-TP-06-13",
    doi = "10.1088/1126-6708/2006/05/026",
    journal = "JHEP",
    volume = "05",
    pages = "026",
    year = "2006"
}

@article{deFavereau:2013fsa,
    author = "de Favereau, J. and Delaere, C. and Demin, P. and Giammanco, A. and Lema\^\i{}tre, V. and Mertens, A. and Selvaggi, M.",
    collaboration = "DELPHES 3",
    title = "{DELPHES 3, A modular framework for fast simulation of a generic collider experiment}",
    eprint = "1307.6346",
    archivePrefix = "arXiv",
    primaryClass = "hep-ex",
    doi = "10.1007/JHEP02(2014)057",
    journal = "JHEP",
    volume = "02",
    pages = "057",
    year = "2014"
}

@article{Azevedo:2020fdl,
    author = "Azevedo, Duarte and Capucha, Rodrigo and Gouveia, Emanuel and Onofre, Ant\'onio and Santos, Rui",
    title = "{Light Higgs searches in $ t\overline{t}\phi $ production at the LHC}",
    eprint = "2012.10730",
    archivePrefix = "arXiv",
    primaryClass = "hep-ph",
    doi = "10.1007/JHEP04(2021)077",
    journal = "JHEP",
    volume = "04",
    pages = "077",
    year = "2021"
}

@article{Azevedo:2020vfw,
    author = "Azevedo, Duarte and Capucha, Rodrigo and Onofre, Ant\'onio and Santos, Rui",
    title = "{Scalar mass dependence of angular variables in $ t\overline{t}\phi$ production}",
    eprint = "2003.09043",
    archivePrefix = "arXiv",
    primaryClass = "hep-ph",
    doi = "10.1007/JHEP06(2020)155",
    journal = "JHEP",
    volume = "06",
    pages = "155",
    year = "2020"
}

@article{Gritsan:2016hjl,
      author         = "Gritsan, Andrei V. and Roentsch, Raoul and Schulze,
                        Markus and Xiao, Meng",
      title          = "{Constraining anomalous Higgs boson couplings to the
                        heavy flavor fermions using matrix element techniques}",
      journal        = "Phys. Rev.",
      volume         = "D94",
      year           = "2016",
      number         = "5",
      pages          = "055023",
      doi            = "10.1103/PhysRevD.94.055023",
      eprint         = "1606.03107",
      archivePrefix  = "arXiv",
      primaryClass   = "hep-ph",
      reportNumber   = "TTP16-020, CERN-TH-2016-135",
      SLACcitation   = "%%CITATION = ARXIV:1606.03107;%%"
}

@article{Ferroglia:2019qjy,
      author         = "Ferroglia, Andrea and Fiolhais, Miguel C. N. and Gouveia,
                        Emanuel and Onofre, António",
      title          = "{Role of the $t{\bar t}h$ rest frame in direct top-quark
                        Yukawa coupling measurements}",
      journal        = "Phys. Rev.",
      volume         = "D100",
      year           = "2019",
      number         = "7",
      pages          = "075034",
      doi            = "10.1103/PhysRevD.100.075034",
      eprint         = "1909.00490",
      archivePrefix  = "arXiv",
      primaryClass   = "hep-ph",
      SLACcitation   = "%%CITATION = ARXIV:1909.00490;%%"
}

@article{Santos:2015dja,
      author         = "Amor dos Santos, S. P. and others",
      title          = "{Angular distributions in $t \overline{t}H(H ? b
                        \overline{b})$ reconstructed events at the LHC}",
      journal        = "Phys. Rev.",
      volume         = "D92",
      year           = "2015",
      number         = "3",
      pages          = "034021",
      doi            = "10.1103/PhysRevD.92.034021",
      eprint         = "1503.07787",
      archivePrefix  = "arXiv",
      primaryClass   = "hep-ph",
      SLACcitation   = "%%CITATION = ARXIV:1503.07787;%%"
}

@article{AmorDosSantos:2017ayi,
      author         = "Amor Dos Santos, S. and others",
      title          = "{Probing the CP nature of the Higgs coupling in $t{\bar
                        t}h$ events at the LHC}",
      journal        = "Phys. Rev.",
      volume         = "D96",
      year           = "2017",
      number         = "1",
      pages          = "013004",
      doi            = "10.1103/PhysRevD.96.013004",
      eprint         = "1704.03565",
      archivePrefix  = "arXiv",
      primaryClass   = "hep-ph",
      SLACcitation   = "%%CITATION = ARXIV:1704.03565;%%"
}

@article{Azevedo:2017qiz,
      author         = "Azevedo, D. and Onofre, A. and Filthaut, F. and Gon\c{c}alo,
                        R.",
      title          = "{CP tests of Higgs couplings in $t\bar{t}h$ semileptonic
                        events at the LHC}",
      journal        = "Phys. Rev.",
      volume         = "D98",
      year           = "2018",
      number         = "3",
      pages          = "033004",
      doi            = "10.1103/PhysRevD.98.033004",
      eprint         = "1711.05292",
      archivePrefix  = "arXiv",
      primaryClass   = "hep-ph",
      SLACcitation   = "%%CITATION = ARXIV:1711.05292;%%"
}

@Article{	  Boudjema:2015nda,
  author	= "Boudjema, Fawzi and Godbole, Rohini M. and Guadagnoli,
		  Diego and Mohan, Kirtimaan A.",
  title		= "{Lab-frame observables for probing the top-Higgs
		  interaction}",
  journal	= "Phys. Rev.",
  volume	= "D92",
  year		= "2015",
  number	= "1",
  pages		= "015019",
  doi		= "10.1103/PhysRevD.92.015019",
  eprint	= "1501.03157",
  archiveprefix	= "arXiv",
  primaryclass	= "hep-ph",
  reportnumber	= "LAPTH-004-15, MSUHEP-150113",
  slaccitation	= "%%CITATION = ARXIV:1501.03157;%%"
}

@Article{	  Gunion:1996xu,
  author	= "Gunion, John F. and He, Xiao-Gang",
  title		= "{Determining the CP nature of a neutral Higgs boson at the
		  LHC}",
  journal	= "Phys. Rev. Lett.",
  volume	= "76",
  year		= "1996",
  pages		= "4468-4471",
  doi		= "10.1103/PhysRevLett.76.4468",
  eprint	= "hep-ph/9602226",
  archiveprefix	= "arXiv",
  primaryclass	= "hep-ph",
  reportnumber	= "UCD-96-05",
  slaccitation	= "%%CITATION = HEP-PH/9602226;%%"
}

@article{Alwall:2011uj,
      author         = "Alwall, Johan and Herquet, Michel and Maltoni, Fabio and
                        Mattelaer, Olivier and Stelzer, Tim",
      title          = "{MadGraph 5 : Going Beyond}",
      journal        = "JHEP",
      volume         = "06",
      year           = "2011",
      pages          = "128",
      doi            = "10.1007/JHEP06(2011)128",
      eprint         = "1106.0522",
      archivePrefix  = "arXiv",
      primaryClass   = "hep-ph",
      reportNumber   = "FERMILAB-PUB-11-448-T",
      SLACcitation   = "%%CITATION = ARXIV:1106.0522;%%"
}

@article{Read:2002hq,
      author         = "Read, Alexander L.",
      title          = "{Presentation of search results: The CL(s) technique}",
      booktitle      = "{Advanced Statistical Techniques in Particle Physics.
                        Proceedings, Conference, Durham, UK, March 18-22, 2002}",
      journal        = "J. Phys.",
      volume         = "G28",
      year           = "2002",
      pages          = "2693-2704",
      doi            = "10.1088/0954-3899/28/10/313",
      note           = "[,11(2002)]",
      SLACcitation   = "%%CITATION = JPAGA,G28,2693;%%"
}

@article{Junk:1999kv,
      author         = "Junk, Thomas",
      title          = "{Confidence level computation for combining searches with
                        small statistics}",
      journal        = "Nucl. Instrum. Meth.",
      volume         = "A434",
      year           = "1999",
      pages          = "435-443",
      doi            = "10.1016/S0168-9002(99)00498-2",
      eprint         = "hep-ex/9902006",
      archivePrefix  = "arXiv",
      primaryClass   = "hep-ex",
      reportNumber   = "CARLETON-OPAL-PHYS-99-01, CERN-EP-99-041",
      SLACcitation   = "%%CITATION = HEP-EX/9902006;%%"
}

@article{hoecker2007tmva,
  title={TMVA-toolkit for multivariate data analysis},
  author={Hoecker, Andreas and Speckmayer, Peter and Stelzer, Joerg and Therhaag, Jan and von Toerne, Eckhard and Voss, Helge and Backes, M and Carli, T and Cohen, O and Christov, A and others},
  journal={arXiv preprint physics/0703039},
  year={2007}
}

@masterthesis{JoaoLopes:2024MasterThesis,
	author	=	"Carreira, João Lopes Rodrigues",
	title	=	"{Search for a light vector dark matter particle in the $t\bar{t}$ final state at the LHC}",
	year	=	"2024"
}

@article{Djouadi:1993ji,
    author = "Djouadi, A. and Spira, M. and Zerwas, P. M.",
    title = "{Two photon decay widths of Higgs particles}",
    eprint = "hep-ph/9305335",
    archivePrefix = "arXiv",
    reportNumber = "DESY-92-170, UDEM-LPN-TH-111",
    doi = "10.1016/0370-2693(93)90564-X",
    journal = "Phys. Lett. B",
    volume = "311",
    pages = "255--260",
    year = "1993"
}

@Article{CMS:2023sdw,
  author        = {Tumasyan, Armen and others},
  title         = {{A search for decays of the Higgs boson to invisible particles in events with a top-antitop quark pair or a vector boson in proton-proton collisions at $\sqrt{s} = 13\,\text {Te}\hspace{-.08em}\text {V} $}},
  journal       = {Eur. Phys. J. C},
  year          = {2023},
  volume        = {83},
  number        = {10},
  pages         = {933},
  archiveprefix = {arXiv},
  collaboration = {CMS},
  doi           = {10.1140/epjc/s10052-023-11952-7},
  eprint        = {2303.01214},
  primaryclass  = {hep-ex},
  reportnumber  = {CMS-HIG-21-007, CERN-EP-2023-004},
}

@Article{ATLAS:2023tkt,
  author        = {Aad, Georges and others},
  title         = {{Combination of searches for invisible decays of the Higgs boson using 139 fb{\ensuremath{-}}1 of proton-proton collision data at s=13 TeV collected with the ATLAS experiment}},
  journal       = {Phys. Lett. B},
  year          = {2023},
  volume        = {842},
  pages         = {137963},
  archiveprefix = {arXiv},
  collaboration = {ATLAS},
  doi           = {10.1016/j.physletb.2023.137963},
  eprint        = {2301.10731},
  primaryclass  = {hep-ex},
  reportnumber  = {CERN-EP-2022-289},
}

@ARTICLE{1979NuPhB.160..151P,
       author = {{Passarino}, G. and {Veltman}, M.},
        title = "{One-loop corrections for e $^{+}$e $^{-}$ annihilation into {\ensuremath{\mu}}$^{+}${\ensuremath{\mu}}$^{-}$ in the Weinberg model}",
      journal = {Nuclear Physics B},
         year = 1979,
        month = nov,
       volume = {160},
       number = {1},
        pages = {151-207},
          doi = {10.1016/0550-3213(79)90234-7},
       adsurl = {https://ui.adsabs.harvard.edu/abs/1979NuPhB.160..151P}
}

@article{ATLAS:2023cbt,
    author = "Aad, Georges and others",
    collaboration = "ATLAS",
    title = "{Probing the CP nature of the top{\textendash}Higgs Yukawa coupling in tt{\textasciimacron}H and tH events with H{\textrightarrow}bb{\textasciimacron} decays using the ATLAS detector at the LHC}",
    eprint = "2303.05974",
    archivePrefix = "arXiv",
    primaryClass = "hep-ex",
    reportNumber = "CERN-EP-2022-208",
    doi = "10.1016/j.physletb.2024.138469",
    journal = "Phys. Lett. B",
    volume = "849",
    pages = "138469",
    year = "2024"
}

@article{Backovic:2015soa,
    author = {Backovi{\'c}, Mihailo and Kr{\"a}mer, Michael and Maltoni, Fabio and Martini, Antony and Mawatari, Kentarou and Pellen, Mathieu},
    title = "{Higher-order QCD predictions for dark matter production at the LHC in simplified models with s-channel mediators}",
    eprint = "1508.05327",
    archivePrefix = "arXiv",
    primaryClass = "hep-ph",
    reportNumber = "MCNET-15-24, CP3-15-25, TTK-15-19",
    doi = "10.1140/epjc/s10052-015-3700-6",
    journal = "Eur. Phys. J. C",
    volume = "75",
    number = "10",
    pages = "482",
    year = "2015"
}

@Article{Bernreuther:1993hq,
  author        = {Bernreuther, Werner and Brandenburg, Arnd},
  title         = {{Tracing CP violation in the production of top quark pairs by multiple TeV proton proton collisions}},
  journal       = {Phys. Rev. D},
  year          = {1994},
  volume        = {49},
  pages         = {4481--4492},
  archiveprefix = {arXiv},
  doi           = {10.1103/PhysRevD.49.4481},
  eprint        = {hep-ph/9312210},
  reportnumber  = {SLAC-PUB-6403, PITHA-93-43},
}

@Article{BhupalDev:2007ftb,
  author        = {Bhupal Dev, P. S. and Djouadi, A. and Godbole, R. M. and Muhlleitner, M. M. and Rindani, S. D.},
  title         = {{Determining the CP properties of the Higgs boson}},
  journal       = {Phys. Rev. Lett.},
  year          = {2008},
  volume        = {100},
  pages         = {051801},
  archiveprefix = {arXiv},
  doi           = {10.1103/PhysRevLett.100.051801},
  eprint        = {0707.2878},
  primaryclass  = {hep-ph},
  reportnumber  = {IISC-CHEP-7-07, CERN-PH-TH-2007-115, LPT-ORSAY-07-49, LAPTH-1194-07},
}

@Article{Khatibi:2014bsa,
  author        = {Khatibi, Sara and Mohammadi Najafabadi, Mojtaba},
  title         = {{Exploring the Anomalous Higgs-top Couplings}},
  journal       = {Phys. Rev. D},
  year          = {2014},
  volume        = {90},
  number        = {7},
  pages         = {074014},
  archiveprefix = {arXiv},
  doi           = {10.1103/PhysRevD.90.074014},
  eprint        = {1409.6553},
  primaryclass  = {hep-ph},
}

@Article{He:2014xla,
  author        = {He, Xiao-Gang and Li, Guan-Nan and Zheng, Ya-Juan},
  title         = {{Probing Higgs boson $CP$ Properties with $t\bar{t}H$ at the LHC and the 100 TeV $pp$ collider}},
  journal       = {Int. J. Mod. Phys. A},
  year          = {2015},
  volume        = {30},
  number        = {25},
  pages         = {1550156},
  archiveprefix = {arXiv},
  doi           = {10.1142/S0217751X15501560},
  eprint        = {1501.00012},
  primaryclass  = {hep-ph},
}

@Article{Dolan:2016qvg,
  author        = {Dolan, Matthew J. and Spannowsky, Michael and Wang, Qi and Yu, Zhao-Huan},
  title         = {{Determining the quantum numbers of simplified models in $t\bar{t}X$ production at the LHC}},
  journal       = {Phys. Rev. D},
  year          = {2016},
  volume        = {94},
  number        = {1},
  pages         = {015025},
  archiveprefix = {arXiv},
  doi           = {10.1103/PhysRevD.94.015025},
  eprint        = {1606.00019},
  primaryclass  = {hep-ph},
  reportnumber  = {IPPP-16-47, DCPT-16-94},
}

@Article{Goncalves:2016qhh,
  author        = {Goncalves, Dorival and Lopez-Val, David},
  title         = {{Pseudoscalar searches with dileptonic tops and jet substructure}},
  journal       = {Phys. Rev. D},
  year          = {2016},
  volume        = {94},
  number        = {9},
  pages         = {095005},
  archiveprefix = {arXiv},
  doi           = {10.1103/PhysRevD.94.095005},
  eprint        = {1607.08614},
  primaryclass  = {hep-ph},
  reportnumber  = {IPPP-16-76, DCPT-16-152},
}

@Article{Buckley:2015ctj,
  author        = {Buckley, Matthew R. and Goncalves, Dorival},
  title         = {{Constraining the Strength and CP Structure of Dark Production at the LHC: the Associated Top-Pair Channel}},
  journal       = {Phys. Rev. D},
  year          = {2016},
  volume        = {93},
  number        = {3},
  pages         = {034003},
  archiveprefix = {arXiv},
  doi           = {10.1103/PhysRevD.93.034003},
  eprint        = {1511.06451},
  primaryclass  = {hep-ph},
  reportnumber  = {IPPP-15-68, DCPT-15-136},
}

@Article{Buckley:2015vsa,
  author        = {Buckley, Matthew R. and Goncalves, Dorival},
  title         = {{Boosting the Direct CP Measurement of the Higgs-Top Coupling}},
  journal       = {Phys. Rev. Lett.},
  year          = {2016},
  volume        = {116},
  number        = {9},
  pages         = {091801},
  archiveprefix = {arXiv},
  doi           = {10.1103/PhysRevLett.116.091801},
  eprint        = {1507.07926},
  primaryclass  = {hep-ph},
  reportnumber  = {IPPP-15-49, DCPT-15-98},
}

@Article{Goncalves:2018agy,
  author        = {Gon\c{c}alves, Dorival and Kong, Kyoungchul and Kim, Jeong Han},
  title         = {{Probing the top-Higgs Yukawa CP structure in dileptonic $ t\overline{t}h $ with M$_{2}$-assisted reconstruction}},
  journal       = {JHEP},
  year          = {2018},
  volume        = {06},
  pages         = {079},
  archiveprefix = {arXiv},
  doi           = {10.1007/JHEP06(2018)079},
  eprint        = {1804.05874},
  primaryclass  = {hep-ph},
  reportnumber  = {PITT-PACC-1807},
}

@Article{Faroughy:2019ird,
  author        = {Faroughy, Darius A. and Kamenik, Jernej F. and Ko\v{s}nik, Nejc and Smolkovi\v{c}, Aleks},
  title         = {{Probing the $CP$ nature of the top quark Yukawa at hadron colliders}},
  journal       = {JHEP},
  year          = {2020},
  volume        = {02},
  pages         = {085},
  archiveprefix = {arXiv},
  doi           = {10.1007/JHEP02(2020)085},
  eprint        = {1909.00007},
  primaryclass  = {hep-ph},
}

@Article{Cao:2020hhb,
  author        = {Cao, Qing-Hong and Xie, Ke-Pan and Zhang, Hao and Zhang, Rui},
  title         = {{A New Observable for Measuring CP Property of Top-Higgs Interaction}},
  journal       = {Chin. Phys. C},
  year          = {2021},
  volume        = {45},
  number        = {2},
  pages         = {023117},
  archiveprefix = {arXiv},
  doi           = {10.1088/1674-1137/abcfac},
  eprint        = {2008.13442},
  primaryclass  = {hep-ph},
}

@Article{Goncalves:2021dcu,
  author        = {Gon\c{c}alves, Dorival and Kim, Jeong Han and Kong, Kyoungchul and Wu, Yongcheng},
  title         = {{Direct Higgs-top CP-phase measurement with $ t\overline{t}h $ at the 14 TeV LHC and 100 TeV FCC}},
  journal       = {JHEP},
  year          = {2022},
  volume        = {01},
  pages         = {158},
  archiveprefix = {arXiv},
  doi           = {10.1007/JHEP01(2022)158},
  eprint        = {2108.01083},
  primaryclass  = {hep-ph},
}

@Article{Barman:2021yfh,
  author        = {Barman, Rahool Kumar and Gon\c{c}alves, Dorival and Kling, Felix},
  title         = {{Machine learning the Higgs boson-top quark CP phase}},
  journal       = {Phys. Rev. D},
  year          = {2022},
  volume        = {105},
  number        = {3},
  pages         = {035023},
  archiveprefix = {arXiv},
  doi           = {10.1103/PhysRevD.105.035023},
  eprint        = {2110.07635},
  primaryclass  = {hep-ph},
  reportnumber  = {DESY 21-161},
}

@Article{Bahl:2020wee,
  author        = {Bahl, Henning and Bechtle, Philip and Heinemeyer, Sven and Katzy, Judith and Klingl, Tobias and Peters, Krisztian and Saimpert, Matthias and Stefaniak, Tim and Weiglein, Georg},
  title         = {{Indirect $\mathcal{CP}$ probes of the Higgs-top-quark interaction: current LHC constraints and future opportunities}},
  journal       = {JHEP},
  year          = {2020},
  volume        = {11},
  pages         = {127},
  archiveprefix = {arXiv},
  doi           = {10.1007/JHEP11(2020)127},
  eprint        = {2007.08542},
  primaryclass  = {hep-ph},
  reportnumber  = {DESY-20-102},
}

@Article{Bahl:2021dnc,
  author        = {Bahl, Henning and Brass, Simon},
  title         = {{Constraining $ \mathcal{CP} $-violation in the Higgs-top-quark interaction using machine-learning-based inference}},
  journal       = {JHEP},
  year          = {2022},
  volume        = {03},
  pages         = {017},
  archiveprefix = {arXiv},
  doi           = {10.1007/JHEP03(2022)017},
  eprint        = {2110.10177},
  primaryclass  = {hep-ph},
  reportnumber  = {DESY 21-160},
}

@article{Artoisenet:2012st,
    author = "Artoisenet, Pierre and Frederix, Rikkert and Mattelaer, Olivier and Rietkerk, Robbert",
    title = "{Automatic spin-entangled decays of heavy resonances in Monte Carlo simulations}",
    eprint = "1212.3460",
    archivePrefix = "arXiv",
    primaryClass = "hep-ph",
    reportNumber = "NIKHEF-2012-021, CERN-PH-TH-2012-329",
    doi = "10.1007/JHEP03(2013)015",
    journal = "JHEP",
    volume = "03",
    pages = "015",
    year = "2013"
}

@Article{Ellis:2013yxa,
  author        = {Ellis, John and Hwang, Dae Sung and Sakurai, Kazuki and Takeuchi, Michihisa},
  title         = {{Disentangling Higgs-Top Couplings in Associated Production}},
  journal       = {JHEP},
  year          = {2014},
  volume        = {04},
  pages         = {004},
  archiveprefix = {arXiv},
  doi           = {10.1007/JHEP04(2014)004},
  eprint        = {1312.5736},
  primaryclass  = {hep-ph},
  reportnumber  = {KCL-PH-TH-2013-47, LCTS-2013-35, CERN-PH-TH-2013-312},
}

@Article{Mileo:2016mxg,
  author        = {Mileo, Nicolas and Kiers, Ken and Szynkman, Alejandro and Crane, Daniel and Gegner, Ethan},
  title         = {{Pseudoscalar top-Higgs coupling: exploration of CP-odd observables to resolve the sign ambiguity}},
  journal       = {JHEP},
  year          = {2016},
  volume        = {07},
  pages         = {056},
  archiveprefix = {arXiv},
  doi           = {10.1007/JHEP07(2016)056},
  eprint        = {1603.03632},
  primaryclass  = {hep-ph},
}

@Article{Bortolato:2020zcg,
  author        = {Bortolato, Bla\v{z} and Kamenik, Jernej F. and Ko\v{s}nik, Nejc and Smolkovi\v{c}, Aleks},
  title         = {{Optimized probes of $CP$ -odd effects in the $t \bar t h$ process at hadron colliders}},
  journal       = {Nucl. Phys. B},
  year          = {2021},
  volume        = {964},
  pages         = {115328},
  archiveprefix = {arXiv},
  doi           = {10.1016/j.nuclphysb.2021.115328},
  eprint        = {2006.13110},
  primaryclass  = {hep-ph},
}

@Article{Demartin:2014fia,
  author        = {Demartin, Federico and Maltoni, Fabio and Mawatari, Kentarou and Page, Ben and Zaro, Marco},
  title         = {{Higgs characterisation at NLO in QCD: CP properties of the top-quark Yukawa interaction}},
  journal       = {Eur. Phys. J. C},
  year          = {2014},
  volume        = {74},
  number        = {9},
  pages         = {3065},
  archiveprefix = {arXiv},
  doi           = {10.1140/epjc/s10052-014-3065-2},
  eprint        = {1407.5089},
  primaryclass  = {hep-ph},
  reportnumber  = {CP3-14-59, LPN14-096, MCNET-14-21},
}

@Article{Frederix:2011zi,
  author        = {Frederix, Rikkert and Frixione, Stefano and Hirschi, Valentin and Maltoni, Fabio and Pittau, Roberto and Torrielli, Paolo},
  title         = {{Scalar and pseudoscalar Higgs production in association with a top\textendash{}antitop pair}},
  journal       = {Phys. Lett. B},
  year          = {2011},
  volume        = {701},
  pages         = {427--433},
  archiveprefix = {arXiv},
  doi           = {10.1016/j.physletb.2011.06.012},
  eprint        = {1104.5613},
  primaryclass  = {hep-ph},
  reportnumber  = {CERN-PH-TH-2011-091, CP3-11-17, ZU-TH-07-11},
}

@Article{Li:2017dyz,
  author        = {Li, Jinmian and Si, Zong-guo and Wu, Lei and Yue, Jason},
  title         = {{Central-edge asymmetry as a probe of Higgs-top coupling in $t\bar{t}h$ production at the LHC}},
  journal       = {Phys. Lett. B},
  year          = {2018},
  volume        = {779},
  pages         = {72--76},
  archiveprefix = {arXiv},
  doi           = {10.1016/j.physletb.2018.02.009},
  eprint        = {1701.00224},
  primaryclass  = {hep-ph},
}

@Article{Kobakhidze:2014gqa,
  author        = {Kobakhidze, Archil and Wu, Lei and Yue, Jason},
  title         = {{Anomalous Top-Higgs Couplings and Top Polarisation in Single Top and Higgs Associated Production at the LHC}},
  journal       = {JHEP},
  year          = {2014},
  volume        = {10},
  pages         = {100},
  archiveprefix = {arXiv},
  doi           = {10.1007/JHEP10(2014)100},
  eprint        = {1406.1961},
  primaryclass  = {hep-ph},
}

@Article{Bramante:2014gda,
  author        = {Bramante, Joseph and Delgado, Antonio and Martin, Adam},
  title         = {{Cornering a hyper Higgs boson: Angular kinematics for boosted Higgs bosons with top pairs}},
  journal       = {Phys. Rev. D},
  year          = {2014},
  volume        = {89},
  number        = {9},
  pages         = {093006},
  archiveprefix = {arXiv},
  doi           = {10.1103/PhysRevD.89.093006},
  eprint        = {1402.5985},
  primaryclass  = {hep-ph},
}

@article{Rosenberg:2000wb,
    author = "Rosenberg, L. J. and van Bibber, K. A.",
    title = "{Searches for invisible axions}",
    doi = "10.1016/S0370-1573(99)00045-9",
    journal = "Phys. Rept.",
    volume = "325",
    pages = "1--39",
    year = "2000"
}

@article{Schumann:2019eaa,
    author = "Schumann, Marc",
    title = "{Direct Detection of WIMP Dark Matter: Concepts and Status}",
    eprint = "1903.03026",
    archivePrefix = "arXiv",
    primaryClass = "astro-ph.CO",
    doi = "10.1088/1361-6471/ab2ea5",
    journal = "J. Phys. G",
    volume = "46",
    number = "10",
    pages = "103003",
    year = "2019"
}

@Article{Azevedo:2022jnd,
  author        = {Azevedo, Duarte and Capucha, Rodrigo and Onofre, Ant\'onio and Santos, Rui},
  title         = {{CP-violation, asymmetries and interferences in $ t\overline{t}\phi $}},
  journal       = {JHEP},
  year          = {2022},
  volume        = {09},
  pages         = {246},
  archiveprefix = {arXiv},
  doi           = {10.1007/JHEP09(2022)246},
  eprint        = {2208.04271},
  primaryclass  = {hep-ph},
}

@article{Boveia:2016mrp,
    author = "Boveia, Antonio and others",
    editor = "Buchmueller, Oliver and Doglioni, Caterina and Hahn, Kristian and Haisch, Ulrich and Kahlhoefer, Felix and Mangano, Michelangelo and McCabe, Christopher and Tait, Tim M. P.",
    title = "{Recommendations on presenting LHC searches for missing transverse energy signals using simplified $s$-channel models of dark matter}",
    eprint = "1603.04156",
    archivePrefix = "arXiv",
    primaryClass = "hep-ex",
    reportNumber = "CERN-LPCC-2016-001",
    doi = "10.1016/j.dark.2019.100365",
    journal = "Phys. Dark Univ.",
    volume = "27",
    pages = "100365",
    year = "2020"
}

@article{ParticleDataGroup:2024cfk,
    author = "Navas, S. and others",
    collaboration = "Particle Data Group",
    title = "{Review of particle physics}",
    doi = "10.1103/PhysRevD.110.030001",
    journal = "Phys. Rev. D",
    volume = "110",
    number = "3",
    pages = "030001",
    year = "2024"
}

@article{Kahlhoefer:2015bea,
    author = "Kahlhoefer, Felix and Schmidt-Hoberg, Kai and Schwetz, Thomas and Vogl, Stefan",
    title = "{Implications of unitarity and gauge invariance for simplified dark matter models}",
    eprint = "1510.02110",
    archivePrefix = "arXiv",
    primaryClass = "hep-ph",
    reportNumber = "DESY-15-182",
    doi = "10.1007/JHEP02(2016)016",
    journal = "JHEP",
    volume = "02",
    pages = "016",
    year = "2016"
}

@article{ATLAS:2026nrx,
    collaboration = "ATLAS",
    title = "{Observation of a cross-section enhancement near the $t\bar{t}$ production threshold in $\sqrt{s} = 13${\,}TeV pp collisions with the ATLAS detector}",
    doi = "10.1088/1361-6633/ae60a0",
    journal = "Rept. Prog. Phys.",
    volume = "89",
    number = "5",
    pages = "057801",
    year = "2026"
}

\end{document}